\documentclass[prb,aps,twocolumn,footinbib,floatfix,10pt,longbibliography,superscriptaddress]{revtex4-1}
\usepackage[english]{babel}
\usepackage{breakcites}
\usepackage{listings}
\usepackage[utf8]{inputenc}
\usepackage[T1]{fontenc}
\usepackage{textcomp}
\usepackage{gensymb}

\usepackage{graphicx}
\usepackage{braket}
\usepackage{color}
\usepackage{epstopdf}
\usepackage{mathtools}
\usepackage{amsmath}
\usepackage{amssymb}
\usepackage{float}
\usepackage{comment}
\usepackage{amsmath}

\usepackage{amsfonts}
\usepackage{bm}
\usepackage{bbold}

\usepackage{color}

\graphicspath{{./figs/}}
\usepackage{gensymb}
\usepackage[colorinlistoftodos,prependcaption]{todonotes}
\usepackage{xargs}
\newcommandx{\unsure}[2][1=]{\todo[linecolor=red,backgroundcolor=red!25,bordercolor=red,#1]{#2}}
\newcommandx{\change}[2][1=]{\todo[linecolor=blue,backgroundcolor=blue!25,bordercolor=blue,#1]{#2}}
\newcommandx{\info}[2][1=]{\todo[linecolor=OliveGreen,backgroundcolor=OliveGreen!25,bordercolor=OliveGreen,#1]{#2}}
\newcommandx{\improvement}[2][1=]{\todo[linecolor=Plum,backgroundcolor=Plum!25,bordercolor=Plum,#1]{#2}}
\newcommandx{\thiswillnotshow}[2][1=]{\todo[disable,#1]{#2}}
\newcommandx{\greencom}[2][1=]
{\todo[inline, color=green!40,#1]{#2}}
\newcommandx{\bluecom}[2][1=]
{\todo[inline, color=blue!40,#1]{#2}}

\definecolor{winered}{rgb}{0.5,0,0}
\hypersetup
{
colorlinks=true,
linkcolor=winered,
urlcolor={winered},
filecolor={winered},
citecolor={winered},
allcolors={winered}
}

\usepackage{letltxmacro}
\LetLtxMacro{\ORIGselectlanguage}{\selectlanguage}
\makeatletter
\DeclareRobustCommand{\selectlanguage}[1]{%
  \@ifundefined{alias@\string#1}
    {\ORIGselectlanguage{#1}}
    {\begingroup\edef\x{\endgroup
       \noexpand\ORIGselectlanguage{\@nameuse{alias@#1}}}\x}%
}
\newcommand{\definelanguagealias}[2]{%
  \@namedef{alias@#1}{#2}%
}
\makeatother

\definelanguagealias{en}{english}
\definelanguagealias{EN}{english}

\begin{document}

\title{
Efficient near-field to far-field transformations for
 quasinormal modes of optical cavities and plasmonic resonators}
\author{Juanjuan Ren}
\email{jr180@queensu.ca}
\affiliation{\hspace{0pt}Department of Physics, Engineering Physics, and Astronomy, Queen's University, Kingston, Ontario K7L 3N6, Canada\hspace{0pt}}
\author{Sebastian~Franke}
\affiliation{Technische Universit\"at Berlin, Institut f\"ur Theoretische Physik,
Nichtlineare Optik und Quantenelektronik, Hardenbergstra{\ss}e 36, 10623 Berlin, Germany}
 \author{Andreas Knorr}
 \affiliation{Technische Universit\"at Berlin, Institut f\"ur Theoretische Physik,
 Nichtlineare Optik und Quantenelektronik, Hardenbergstra{\ss}e 36, 10623 Berlin, Germany}
   \author{Marten Richter}
 \affiliation{Technische Universit\"at Berlin, Institut f\"ur Theoretische Physik,
 Nichtlineare Optik und Quantenelektronik, Hardenbergstra{\ss}e 36, 10623 Berlin, Germany}
\author{Stephen Hughes}
\affiliation{\hspace{0pt}Department of Physics, Engineering Physics, and Astronomy, Queen's University, Kingston, Ontario K7L 3N6, Canada\hspace{0pt}}

\date{\today}

\begin{abstract}
We describe an efficient near-field to far-field transformation for optical quasinormal modes, which are the dissipative modes
of open cavities and plasmonic resonators with complex eigenfrequencies.
As an application of the theory, we show how one can
 compute the reservoir modes (or regularized quasinormal modes) outside the resonator, which are essential to use in both classical and quantum optics.
We subsequently demonstrate how to efficiently compute the
quantum optical parameters necessary in the theory of
quantized quasinormal modes
[Franke {\it et al.},
 Phys. Rev. Lett. {\bf 122}, 213901 (2019)].
To confirm the accuracy of our technique, we directly compare
with a Dyson equation approach
currently used in the literature (in regimes where this is possible),
and demonstrate several order of magnitude improvement
for the calculation run times.
We also introduce an efficient pole approximation for
computing the quantized quasinormal mode parameters, since they require
an integration over a range of frequencies.
 Using this approach,
 we show how to compute  regularized quasinormal modes
 and quantum optical parameters for a full 3D metal dimer
 in under one minute on a standard desktop computer.
Our technique is exemplified by
studying the quasinormal modes
of metal dimers and
and a hybrid structure consisting of a
gold dimer on top of a photonic crystal beam.
In the latter example, we show how to compute the quantum optical parameters
that describe   a pronounced Fano resonance, using structural
geometries that cannot practically be solved using
a Dyson equation approach.
All calculations for the spontaneous emission
rates are confirmed with full-dipole calculations
in Maxwell's equations and are shown to be in excellent agreement.


\end{abstract}

\maketitle

\section{Introduction}
\label{Sec1}

Open-cavity photonic structure are widely used to enhance light-matter interaction at the nanoscale \cite{vahala_optical_2003,chang1996optical}, especially plasmonic cavities \cite{bergman_surface_2003,maier_plasmonics:_2007,noginov_demonstration_2009,novotny_antennas_2011,chang_single-photon_2007,andersen_strongly_2011,jacob_plasmonics_2011,tame_quantum_2013,berini_surface_2012}, which allow one to
enhance light-matter interactions without any fundamental
bounds from diffraction---giving rise to
high field electromagnetic hots spots. These optical hot spots have been used
to strongly couple single molecules at room temperature and probe
molecular optomechanics in the regime
of surface enhanced Raman spectroscopy (SERS)~\cite{Chikkaraddy2016,Benz2016}.

In optical cavity physics, and especially
cavity-QED (quantum electrodynamics),
a few ``mode'' description for these cavities is of great benefit.
A cavity mode description not only helps to explain the underlying physics of light-matter enhancement, but it can quantify the separation of radiative and nonradiative decay processes, and allow design insights into the important figures of merit. For closed systems without absorption, the cavity modes can be  described by normal modes with real eigenfrequencies and infinite lifetimes \cite{morse_methods_1954}.
This is often a good approximation for
high $Q$ resonances (where $Q$ is the quality factor), but is still ambiguous in general~\cite{kristensen_generalized_2012,kristensen_modes_2014}.
However, most---if not all---systems are dissipative via radiation decay or/and absorption, and thus it is highly desirable to work with the
correct dissipative modes.

One of the most powerful approaches to this problem is to use  quasinormal modes (QNMs) \cite{lai_time-independent_1990,leung_completeness_1994,leung_time-independent_1994,leung_completeness_1996,lee_dyadic_1999,kristensen_generalized_2012,sauvan_theory_2013,kristensen_modes_2014,bai_efficient_2013-1,PhysRevA.98.043806,lalanne_light_2018,1910.05412},
which are open cavity modes
with complex eigenfrequency $\tilde \omega_c$, with a finite
cavity lifetime $\tau_c=2\pi/(-{\rm Im}(\tilde\omega_c)$).
 The key advantage of such
a discrete modal approach is that
often only a few QNMs are needed, and frequently just one QNM,
which can be used
to give an accurate description of light-matter interactions over a wide range of positions and frequencies~\cite{lee_dyadic_1999,muljarov_brillouin-wigner_2010,kristensen_generalized_2012,sauvan_theory_2013,kristensen_modes_2014,lalanne_light_2018}. In the semiclassical regime, one can rigorously compute a generalized effective mode volume
 \cite{leung_time-independent_1994,kristensen_generalized_2012}, the photon Green's function \cite{leung_completeness_1994, ge_quasinormal_2014}, and  the enhanced spontaneous emission rate (related to generalized Purcell factors) \cite{ge_quasinormal_2014,2017PRA_hybrid} in system-reservoir theory of quantum optics.
Moreover, it
has been recognized
that quantization of QNMs would be a significant improvement in quantum optics/plasmonics theories \cite{fernandez-dominguez_plasmon-enhanced_2018}.
Some progress has been made
 for one-dimensional dielectric structures \cite{ho_second_1998,severini_second_2004}, but this approach  does not lead to Fock states, typically used to expand multiphoton quantum field states.
 Recently, Franke {\it et al.} introduced a
 quantization for leaky optical cavities and plasmonic resonators based on QNMs \cite{franke_quantization_2018}, which allows one to rigorously study multi-photon problems for
 open-cavity resonators, including dielectrics and metals.
 However, as input to the quantization theory, one requires
 the QNMs and reservoir modes (``regularized QNMs''~\cite{ge_quasinormal_2014}) outside the cavity region.

The QNMs represent a highly accurate description to the total field for positions inside the resonator but is not a good description for fields far outside, where one needs the continuous reservoir fields~\cite{ge_quasinormal_2014}.
However, a  special feature of the QNM field, $\tilde{\mathbf{f}}_{\mu}(\mathbf{r})$, is that they diverge exponentially outside the resonator, originating from the complex resonance frequency and the Silver-M\"uller radiation condition \cite{martin2006multiple}.
The locations for this divergent behavior depend mainly on the quality factor of
optical cavity structure, and the shape of the mode,
e.g.,
the spatial divergence may begin a few microns away
from the resonator
for typical low $Q$ plasmonic cavity modes \cite{kamandar_dezfouli_regularized_2018}.
At these locations, the QNMs are no longer a good representation of the fields
outside the cavity, and generally one should only use the
QNMs inside the cavity region; outside the cavity system, one needs the reservoir modes
or regularized modes
which become a function of continuous frequency.
For example, one can obtain regularized fields outside the resonator
from the solution inside by using a Dyson equation \cite{ge_quasinormal_2014},
\begin{equation}\label{Dysoneq}
\tilde{\mathbf{F}}_{\mu}(\mathbf{R},\omega) \approx \int_{V}d\mathbf{r}\mathbf{G}^{\rm B}(\mathbf{R},\mathbf{r,\omega})\Delta\epsilon(\mathbf{r},\omega)\tilde{\mathbf{f}}_{\mu}(\mathbf{r}),
\end{equation}
where $\mathbf{G}^{\rm B}(\mathbf{R},\mathbf{r,\omega})$
is the Green's function for the background medium
and $\Delta\epsilon$ is the change in dielectric constant from the spatially-dependent resonator.
In this way, one can obtain the fields everywhere, by only using the QNMs within the structure;
other contributions can be included as needed, such as
background contributions from evanescent modes if very near a metal surface \cite{ge_quasinormal_2014}.
Here we consider
${\bf R}$ to be outside the resonator, and ${\bf r}$ inside. We also note that
the $\tilde{\mathbf{F}}_{\mu}(\mathbf{R},\omega)$ are continuous in frequency (and these are expected to be accurate within the QNM spectral region of interest), while the
$\tilde{\mathbf{f}}_{\mu}(\mathbf{r})$ is associated with the QNM complex frequency.

These regularized fields $\tilde{\mathbf{F}}_{\mu}$  have been shown to be highly accurate
for obtaining the Purcell factor outside the resonator,
and they properly converge in the far field~\cite{ge_quasinormal_2014}.  Recently, it was also shown how these fields $\tilde{\mathbf{F}}_{\mu}$
are required for QNM quantization of arbitrary media~\cite{franke_quantization_2018}. While the Dyson approach works in principle, the computation can be tedious and impractical. For example,
for the quantization scheme, one needs to integrate such fields from the outside region over a closed surface
that surrounds the resonator, and over a wide range of frequencies;
this approach requires significant computational memory and is extremely time consuming, especially for complex nanostructures. Indeed, even for simple metal dimer structures, computing $\tilde{\mathbf{F}}_{\mu}$ can take weeks on a high performance desktop computer, as we will also demontrate in this paper with several concrete examples. For more complex cavity structures, such as
dimers on top of photonic crystal (PC) cavities~\cite{2017PRA_hybrid}, the general Dyson approach to obtain $\tilde{\mathbf{F}}_{\mu}$ is numerically intractable.
Given the importance of using these frequency-continuous fields,
$\tilde{\mathbf{F}}_{\mu}$, especially
for connecting to observables,  and for their use
in quantized QNM theories, there is now an urgent need to develop
more efficient and insightful way to obtain these QNM
reservoir fields.

In this paper we present an efficient
solution to this problem.
We define a fictitious boundary surrounding the cavity that radiates
to the far field through the
 appropriate surface currents flowing over the boundary;
the sources inside a domain are replaced with sources on the surface of this domain, and the fields inside the domain can be chosen zero (field equivalence principle).
We take advantage of this principle to introduce an efficient near field to far field (NF2FF) transformation for QNMs to obtain regularized  fields
that give  the correct far field radiation flow.
Moreover, the same transformation can be used to obtain QNMs in the far field, by either projecting in real frequency space (regularized QNM) or
complex frequency space (divergent QNMs).
For practical use in quantized QNM theories,
we also show how the far fields
 can easily be computed via decomposing the near fields into spherical (3D) or cylindrical (2D) waves,
 which can then be propagated separately
to the very far field regime \cite{Neartofar_1992,schneider_understanding_nodate}.
The near fields we use as input are the QNMs fields at a surface close to the
resonator.
With these numerically computed QNMs, obtained
for arbitrarily shaped 3D resonators, we demonstrate how one  can then  carry out NF2FF transformation with real frequencies and
show how the results accurately converge.

The rest of our paper is organized as follows: In Sec.~\ref{Sec2}, we introduce all the main theory needed in this paper. In \ref{subSec2.1}, the core QNMs theory is presented, including the QNM Green's function expansion, classical Purcell factors and classical $\beta$ factors. Due to the divergent behaviour of the QNM fields, one current solution---the Dyson equation approach---is introduced in \ref{subSec2.2}, which can be used to calculate the regularized (i.e., non-divergent) fields outside the resonator. This approach works, but is complicated and time consuming, especially for hybrid structures
(i.e., a combined material created from dielectric and metal
cavity parts). In \ref{subSec2.3}, we introduce an alternative way to obtain these normalized fields using a
NF2FF transformation. As further motivation to why we need these fields,
the basic background of a recently developed quantized QNM theory \cite{franke_quantization_2018,Hughes_SPS_2019} is shown in Sec. \ref{subSec2.4}, where we show how the regularized fields in addition to the QNMs are needed for the ``quantum $S$ parameters''; these matrix elements relate to the commutation rules
for quantization of the QNMs, and are only Kronecker  delta symbols
in the limit of no loss \cite{franke_quantization_2018,Hughes_SPS_2019}.
In \ref{subSec2.5}, we introduce an efficient pole approximation, which simplifies the required integration over frequency for obtaining the $S$ factors, and we give analytical solutions
for single and coupled QNM structures.

Using the above theory, various numerical examples are shown in
Sec.~\ref{Sec3}-\ref{Sec4}.
In \ref{Sec3}, we concentrate on single QNM results, and explore
metal dimer gap modes, with different material losses,
and investigate the resonance features, including the quality factor and complex eigenfrequency
of the dominant localized plasmon mode; the field distribution are described in \ref{subSec3.1}.  In \ref{subSec3.2}, the regularized field $\tilde{\mathbf{F}}(\mathbf{R},\omega)$ are first obtained from Dyson approach,  and compared with our newly developed NF2FF, where
 we show excellent agreement as the fields evolve to the far field. As an example application
 for classical optics, we show $\tilde{\mathbf{F}}(\mathbf{R},\omega)$ at several far field surfaces, which are the fields that experiments can detect directly. Sections~\ref{subSec3.3} and \ref{subSec3.4} show  detailed calculation for quantum parameters, including both non-radiative and radiative contributions, including detailed numerical convergence study.
 Section \ref{subSec3.5} summarizes the computational run times to calculate the radiative contribution for both Dyson approach and NF2FF transformation, which
 is shown to  {\it reduce the calculation run times from
 days-weeks to under $1$ minute} using a standard workstation implemented
 in Matlab.
 In Sec.~\ref{Sec4}, we
 study a complex coupled QNM system, in a regime where the modes strongly overlap
 and the Purcell factors exceed 1 million.
Specifically, we  study a hybrid structure consisting of a gold ellipsoid dimer and a high-Q PC cavity, where two QNMs are overlapping in the frequency region of interest. The interfering
modes  yield  a striking Fano-like resonance, which we show can be
well explain using both the classical and quantum theory. In the latter case, the calculation using a Dyson to a NF2FF approach
would require years of computational time,
but are calculated here in minutes.
The complex details of the Fano resonance feature
are fully obtained using the quantized QNM approach in the bad cavity limit, without
any fitting parameters,
Finally, we present our conclusions
 in Sec.~\ref{Sec5}.

\section{Theory}\label{Sec2}

\subsection{Quasinormal modes, Green's function expansions, classical Purcell factors and beta factors}\label{subSec2.1}

The QNMs, $\tilde{\mathbf{f}}_{{\mu}}\left(\mathbf{r}\right)$, are solutions to the Helmholtz equation,
\begin{equation}\label{smallf}
\boldsymbol{\nabla}\times\boldsymbol{\nabla}\times\tilde{\mathbf{f}}_{{\mu}}\left(\mathbf{r}\right)-\left(\dfrac{\tilde{\omega}_{{\mu}}}{c}\right)^{2}\epsilon\left(\mathbf{r},\tilde{\omega}_\mu\right)\,\tilde{\mathbf{f}}_{{\mu}}\left(\mathbf{r}\right)=0,
\end{equation}
subject to open boundary conditions,
i.e.,  the Silver-M\"uller radiation condition \cite{Kristensen2015}.
Here $\epsilon(\mathbf{r},\tilde{\omega}_{\mu})$  is the dielectric constant and $\tilde{\omega}_{{\mu}}= \omega_{{\mu}}-i\gamma_{{\mu}}$ the complex eigenfrequency
with  quality factor  $Q_{\mu}=\omega_{\mu}/2\gamma_{\mu}$.
Once normalized, the QNMs can be used to construct the transverse Green's function through~\cite{leung_completeness_1994,ge_quasinormal_2014}
\begin{equation}
\mathbf{G}\left(\mathbf{r},\mathbf{r}_{0},\omega\right)= \sum_{\mu} A_{\mu}\left(\omega\right)\,\tilde{\mathbf{f}}_{\mu}\left(\mathbf{r}\right)\tilde{\mathbf{f}}_{\mu}\left(\mathbf{r}_{0}\right),\label{eq:GFwithSUM}
\end{equation}
for locations near (or within) the scattering geometry with volume V, where the QNMs can form a complete basis~\cite{leung_time-independent_1994,leung_completeness_1996}.
The photon Green's function, ${\bf G}(\mathbf{r},\mathbf{r}_{0},\omega)$, fulfills  the equation:
\begin{equation}
\nabla\times\nabla\times{\bf G}(\mathbf{r},\mathbf{r}_{0},\omega)-\frac{\omega^2}{c^2}\epsilon(\mathbf{r},\omega){\bf G}(\mathbf{r},\mathbf{r}_{0},\omega)=\frac{\omega^2}{c^2}\mathbf{1}\delta(\mathbf{r}-\mathbf{r}_{0}),\label{eq:GreenHelmholtz}
\end{equation}
with corresponding radiation conditions, where $\mathbf{1}$ is a unit tensor and $c$ is light speed in vacuum.

Although there are several forms for
 $A_{\mu}(\omega)$, which are related by a sum relationship~\cite{Kristensen_coupled_modes_2017,lee_dyadic_1999}, below we use
\begin{equation}\label{A1}
    A_{\mu}(\omega)=\frac{\omega}{2(\tilde{\omega}_{\mu}-\omega)}.
\end{equation}
However, practically, when we limit the expansion to just a few modes, we
use a slightly different form as an approximation:
\begin{equation}\label{A1b}
    A_{\mu}(\omega) \approx \frac{\omega}{2(\tilde{\omega}_{\mu}-\omega)}
    {\rm Rect}\left (\frac{\omega- \omega_\mu - \omega^{\rm cut}_\mu}{\omega- \omega_\mu}\right),
\end{equation}
 where we now also include the
top-hat or rectangular function:
${\rm Rect}(t)=1$, if $|t|<\frac{1}{2}$, else ${\rm Rect}(t)=0$.
Later we show that
a practical value for the cut-off is
${\omega}^{\rm cut}_\mu=14\gamma_\mu$
when we also compare
with an efficient pole approximation,
to evaluate the integrations over frequency.

We first consider a single QNM, $\mu=\rm c$, so the Green's function can be written as
\begin{equation}
\mathbf{G}_{\rm c}\left(\mathbf{r},\mathbf{r}_{0},\omega\right) \approx A_{\rm c}(\omega)\,\tilde{\mathbf{f}}_{\rm c}\left(\mathbf{r}\right)\tilde{\mathbf{f}}_{\rm c}\left(\mathbf{r}_{0}\right),
\label{eq:GF_QNM}
\end{equation}
where again this holds only nearby the cavity region.
This QNM expansion of Green's function can easily be used to compute the spontaneous emission (SE) rate and Purcell factor.
For example, if one considers a quantum dipole emitter with dipole moment $\mathbf{d}$ (=d$\mathbf{n}_{\rm d}$) at location $\mathbf{r}_{0}$, then the SE rate is
~\cite{kristensen_modes_2014}
\begin{equation}
    \Gamma(\mathbf{r}_{0},\omega)=\frac{2}{\hbar\epsilon_{0}}\mathbf{d}\cdot{\rm Im}\{\mathbf{G}_{\rm c}(\mathbf{r}_{0},\bf{r}_{0},\omega)\}\cdot\mathbf{d}.
\end{equation}
If the emitter is in a homogeneous medium,
then
\begin{align}
\begin{split}
    \Gamma_{0}(\mathbf{r}_{0},\omega)=&\frac{2}{\hbar\epsilon_{0}}\mathbf{d}\cdot{\rm Im}\{\mathbf{G}^{\rm B}(\mathbf{r}_{0},{\bf r}_{0},\omega)\}\cdot\mathbf{d}\\
    =&\frac{\omega^{3}n_{\rm B}{\rm d}^{2}}{3\pi\epsilon_{0}\hbar c^{3}},
\end{split}
\end{align}
where Im$\{\mathbf{G}^{\rm B}({\bf r}_0,{\bf r}_0,\omega)\}=(\omega^3n_{\rm B}/6\pi c^3)\mathbf{1}$,
and  $n_{\rm B}$ is the background refractive index.
Thus the generalized Purcell factor is \cite{Anger2006,kristensen_modes_2014}
\begin{align}\label{QNMpurcell}
\begin{split}
    F_{{\rm P}}^{\rm QNM}({\bf r}_0,\omega) &=1+\frac{\Gamma(\mathbf{r}_{0},\omega)}{\Gamma_{0}(\mathbf{r}_{0},\omega)}\\
    &=1+\frac{\mathbf{n}_{\rm d}\cdot{\rm Im}\{\mathbf{G}_{\rm c}\left(\mathbf{r}_0,\mathbf{r}_0,\omega\right)\}\cdot\mathbf{n}_{\rm d}}{\mathbf{n}_{\rm d}\cdot{\rm Im}\{\mathbf{G}^{\rm B}\left(\mathbf{r}_0,\mathbf{r}_0,\omega\right)\}\cdot\mathbf{n}_{\rm d}}\\
    &=1+\frac{6\pi c^{3}}{\omega^{3}n_{\rm B}}\,\mathbf{n}_{\rm d}\cdot{\rm Im}\{\mathbf{G}_{\rm c}\left(\mathbf{r}_0,\mathbf{r}_0,\omega\right)\}\cdot\mathbf{n}_{\rm d}.
\end{split}
\end{align}
Note that  we have added the extra factor of $1$, which can be derived  from a Dyson equation scattering problem for dipole located outside the resonator (essentially the contribution from the homogeneous radiation modes) \cite{ge_quasinormal_2014}. The actual QNM contribution here is thus the modification to unity.

One can also use the QNMs to calculate the modal nonradiative decay rate of the same dipole emitter \cite{2017PRA_hybrid, Anger2006}, from
\begin{equation}
    \Gamma^{\rm nrad}(\mathbf{r}_{0},\omega)=\frac{2}{\hbar\omega\epsilon_{0}}\int_{\rm V}{\rm Re}\Big\{\mathbf{j(r)\cdot E^{*}(r)}\Big\}d\mathbf{r},
\end{equation}
where $\mathbf{E(r)}=\mathbf{G}_{\rm c}(\mathbf{r},\mathbf{r}_{0},\omega)\cdot\frac{\mathbf{d}}{\epsilon_{0}}$ is the field of the dipole emitter, and $\mathbf{j(r)}=\epsilon_{0}\omega{\rm Im}\{\epsilon(\mathbf{r})\}\mathbf{E(r)}$ represents the dipole induced current density inside metal.
Therefore, the nonradiative and radiative $\beta$ factor can be defined as
\begin{equation}\label{betanradQNMsingle}
    \beta^{\rm nrad}_{\rm QNM}(\mathbf{r}_{0},\omega)=\frac{\Gamma^{\rm nrad}(\mathbf{r}_{0},\omega)}{\Gamma(\mathbf{r}_{0},\omega)},
\end{equation}
and
\begin{equation}\label{betaradQNMsingle}
    \beta^{\rm rad}_{\rm QNM}(\mathbf{r}_{0},\omega)=1-\beta^{\rm nrad}_{\rm QNM}(\mathbf{r}_{0},\omega)=1-\frac{\Gamma^{\rm nrad}(\mathbf{r}_{0},\omega)}{\Gamma(\mathbf{r}_{0},\omega)}.
\end{equation}
Impoartantly, these modal beta factors are associated with the QNM of interest, and define the probability that
an emitted photon through the QNM will decay radiatively ($\beta^{\rm rad}_{\rm QNM}$)
or decay into heating ($\beta^{\rm nrad}_{\rm QNM}$).

 If several QNMs contribute in the
 spectral region of interest,
 then we rewrite Eq. \eqref{eq:GFwithSUM} as
 \begin{align}
 \begin{split}
 \mathbf{G}\left(\mathbf{r},\mathbf{r}_{0},\omega\right)&=
 \sum_{\mu}\mathbf{G}_{\mu}\left(\mathbf{r},\mathbf{r}_{0},\omega\right),
 \end{split}
 \end{align}
 where $\mathbf{G}_{\mu}\left(\mathbf{r},\mathbf{r}_{0},\omega\right)=A_{\mu}\left(\omega\right)\,\tilde{\mathbf{f}}_{\mu}\left(\mathbf{r}\right)\tilde{\mathbf{f}}_{\mu}\left(\mathbf{r}_{0}\right)$.
Thus, the total decay rate of a dipole  emitter is
 \begin{align}
 \begin{split}
 \Gamma_{\rm total}(\mathbf{r}_{0},\omega)&=\frac{2}{\hbar\epsilon_{0}}\mathbf{d}\cdot{\rm Im}\{\mathbf{G}(\mathbf{r}_{0},\bf{r}_{0},\omega)\}\cdot\mathbf{d},\\
 &=\sum_{\mu}\frac{2}{\hbar\epsilon_{0}}\mathbf{d}\cdot{\rm Im}\{\mathbf{G}_{\mu}(\mathbf{r}_{0},\bf{r}_{0},\omega)\}\cdot\mathbf{d},\\
 \end{split}
 \end{align}
 and the total generalized Purcell factor is
 \begin{align}\label{QNMpurcelltotal}
 \begin{split}
     &F_{{\rm total}}^{\rm QNM}({\bf r}_0,\omega) =1+\frac{\Gamma_{\rm total}(\mathbf{r}_{0},\omega)}{\Gamma_{0}(\mathbf{r}_{0},\omega)}\\
     & \ \ \ =1+\frac{\sum_{\mu}\mathbf{n}_{\rm d}\cdot{\rm Im}\{\mathbf{G}_{\mu}\left(\mathbf{r}_0,\mathbf{r}_0,\omega\right)\}\cdot\mathbf{n}_{\rm d}}{\mathbf{n}_{\rm d}\cdot{\rm Im}\{\mathbf{G}^{\rm B}\left(\mathbf{r}_0,\mathbf{r}_0,\omega\right)\}\cdot\mathbf{n}_{\rm d}}\\
     & \ \ \ =1+\sum_{\mu}\frac{6\pi c^{3}}{\omega^{3}n_{\rm B}}\,\mathbf{n}_{\rm d}\cdot{\rm Im}\{\mathbf{G}_{\mu}\left(\mathbf{r}_0,\mathbf{r}_0,\omega\right)\}\cdot\mathbf{n}_{\rm d}.
 \end{split}
 \end{align}

 It is important to note that the total QNM decay rates
 contain both radiative and nonradiative
 contributions.
 The nonradiative decay rate \cite{2017PRA_hybrid} is
 \begin{equation}
     \Gamma^{\rm nrad}_{\rm total}(\mathbf{r}_{0},\omega)=\frac{2}{\hbar\omega\epsilon_{0}}\int_{\rm V}{\rm Re}\Big\{\mathbf{j_{\rm total}(r)\cdot E_{\rm total}^{*}(r)}\Big\}d\mathbf{r},
 \end{equation}
 where $\mathbf{E_{\rm total}(r)}=\sum_{\mu}\mathbf{G}_{\mu}(\mathbf{r},\mathbf{r}_{0},\omega)\cdot\frac{\mathbf{d}}{\epsilon_{0}}$ is the total field of the dipole emitter, and $\mathbf{j_{\rm total}(r)}=\epsilon_{0}\omega{\rm Im}\{\epsilon(\mathbf{r})\}\mathbf{E_{\rm total}(r)}$ represents the dipole induced total current density inside metal.
Thus, the total nonradiative and total radiative $\beta$ factor can be defined as
 \begin{equation}\label{betanradQNMs}
     \beta^{\rm nrad}_{\rm total}(\mathbf{r}_{0},\omega)=\frac{\Gamma_{\rm total}^{\rm nrad}(\mathbf{r}_{0},\omega)}{\Gamma_{\rm total}(\mathbf{r}_{0},\omega)},
 \end{equation}
\begin{align}\label{betaradQNMs}
\begin{split}
\beta^{\rm rad}_{\rm total}(\mathbf{r}_{0},\omega)&=1-\beta^{\rm nrad}_{\rm total}(\mathbf{r}_{0},\omega)\\
&=1-\frac{\Gamma_{\rm total}^{\rm nrad}(\mathbf{r}_{0},\omega)}{\Gamma_{\rm total}(\mathbf{r}_{0},\omega)}.
\end{split}
\end{align}
In the limit of a single mode, these beta factors
define the single QNM beta factors.

\subsection{Regularized QNM fields, \texorpdfstring{${\tilde {\bf F}_{\rm c}({\bf R},\omega)}$}{Lg},  from the Dyson equation}\label{subSec2.2}

As mentioned in the introduction,
one critical feature of the QNM field, $\tilde{\mathbf{f}}_{\mu}({\bf r})$, is that it diverges at locations outside  the resonator~\cite{ge_quasinormal_2014,PhysRevB.98.085418},
and is not convenient, neither classically nor in quantum optics.
One way to rectify this problem, for arbitrarily shaped resonators, is  to employ the Dyson equation to reconstruct a regularized QNM field outside the resonator,
given by Eq.~\eqref{Dysoneq}, which we repeat here for clarity:
$\tilde{\mathbf{F}}_{\mu}(\mathbf{R},\omega)=\int_{V}d\mathbf{r}\mathbf{G}^{\rm B}(\mathbf{R},\mathbf{r,\omega})\Delta\epsilon(\mathbf{r},\omega)\tilde{\mathbf{f}}_{\mu}(\mathbf{r})$.
In this way, one is only using the QNM {\it within} the resonator.
Note that one can also use the same Dyson equation to compute the actual QNMs outside, by replacing the real frequency with the complex QNM pole frequency $\tilde{\omega}_{\mu}$~\cite{PhysRevB.98.085418,kristensen_generalized_2012}:
\begin{equation}\label{f}
\tilde{\mathbf{f}}_{\mu}(\mathbf{R})=\int_{V}d\mathbf{r}\,\mathbf{G}^{\rm B}(\mathbf{R},\mathbf{r,\tilde{\omega}_{\mu}})\Delta\epsilon(\mathbf{r},\tilde{\omega}_{\mu})\tilde{\mathbf{f}}_{\mu}(\mathbf{r}),
\end{equation}
which in certain cases can considerably simplify the computational process of obtaining the QNMs over a wide spatial range.

Although mathematically intuitive, and convenient
for obtaining the renormalized fields
as certain locations,
the Dyson approach to obtain
a range of spatial points (e.g., ${\bf R}$ on a surface) is far from trivial, and can be very
 complex and time consuming.
This is because a full
3D spatial integral is required for every single
spatial point (${\bf R}$) and also for a single frequency.
For many problems, especially in quantum optics,
one requires a full surface of points, and fields that are
computed over a wide range
of frequencies \cite{franke_quantization_2018, Hughes_SPS_2019}.
To exemplify the computational complexity for nanostructures,
if the 3D spatial grid size of $0.2$ ($0.5$) nm is chosen in the volume integral, then it will take  approximately $\approx8.5-10$ ($0.6-0.7$) minutes to obtain $\tilde{\mathbf{F}}_{\mu}$ at a
single point $\mathbf{R}$ and single frequency;
if the grid size is $0.1$ nm (typical for metal nanoparticles), it will take $\approx 75{-}80$ minutes, also for a single point. These numbers are for a
high performance  workstation with 256\,GB RAM, using Matlab.
Obtaining these points over a closed surface
 (which is required to  compute the quantum optical parameters, as discussed below) becomes extremely time demanding, and easily the most difficult part of the numerical calculation.
Thus there is a pressing need to develop a more efficient way to obtain these regularized QNM fields, since they are required as input to quantized QNMs, and form the only reliable way to obtain meaningful modal fields outside the resonator.

\subsection{Regularized QNM fields from a Near-Field to Far-Field transformation}\label{subSec2.3}

\begin{figure}[bht]
  \centering
  \includegraphics[width=0.99\columnwidth]{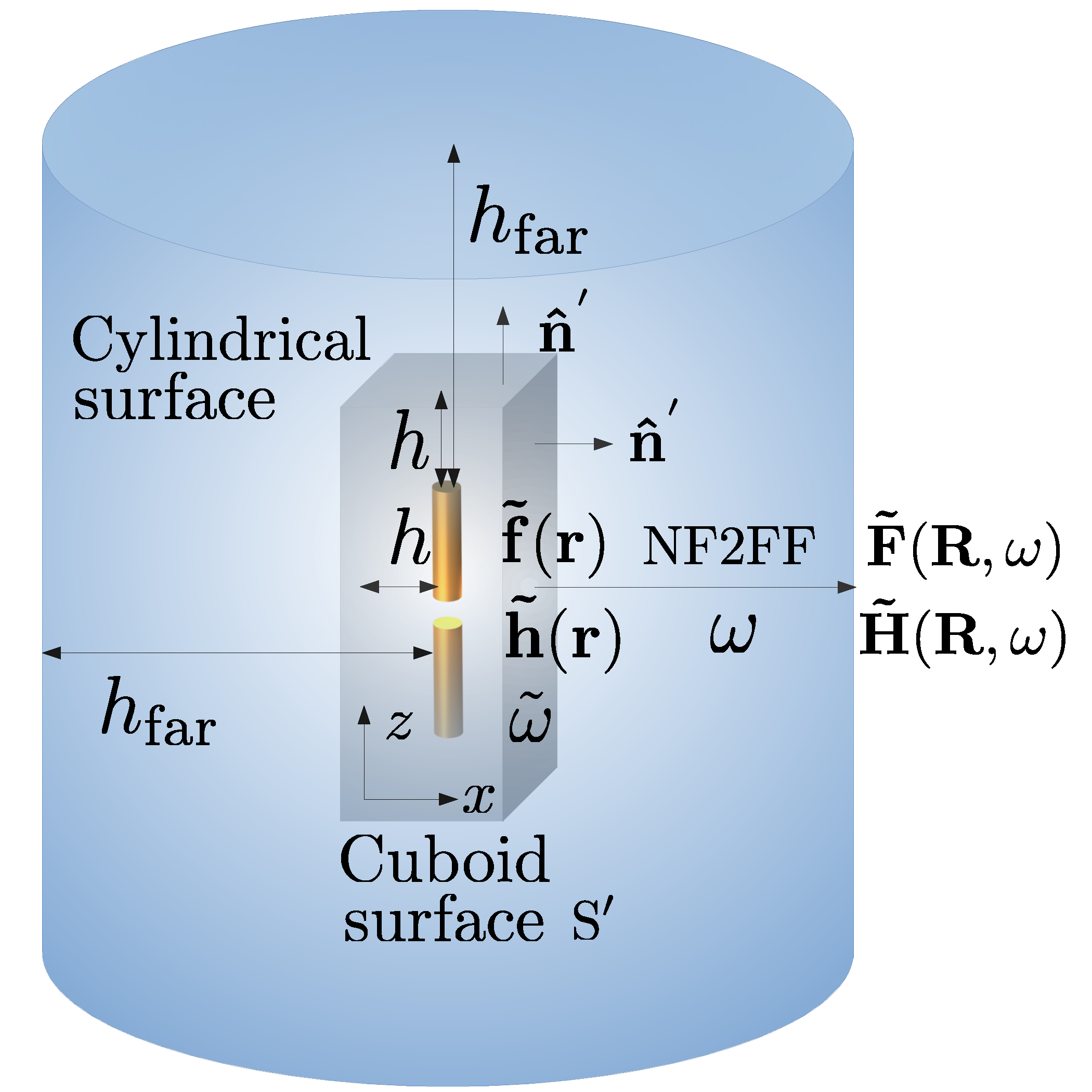}
  \caption{Schematic of NF2FF transformation. The origin of the coordinate axis is at the center of Au dimer gap. A cuboid surface and a cylindrical surface are chosen as near field surface and far field surface separately. The
  labels $h$ and $h_{\rm far}$ represent the distances between them and the surface of the resonator (Au dimer). $\mathbf{\hat{n}}^{'}$ is a unit vector normal to near field surface, pointing outward. Using this transformation with real frequency $\omega$, one could obtain ${\mathbf{ \tilde{F}_{\rm c}}(\mathbf{R},\omega)}$ and ${\mathbf{ \tilde{H}_{\rm c}}(\mathbf{R},\omega)}$ at a far field surface from ${\mathbf{ \tilde{f}_{\rm c}}(\mathbf{r})}$ and ${\mathbf{ \tilde{h}_{\rm c}}(\mathbf{r})}$ obtained from the near field surfaces.
  }\label{neartofar}
\end{figure}

An alternative method to obtain $\tilde{\mathbf{F}}(\mathbf{R},\omega)$ is to perform
a NF2FF transformation~\cite{Neartofar_1992}, using the QNM field $\tilde{\mathbf{f}}$ at a surface close to the resonator.
Near field to far field transformations are frequently exploited in antenna theory to obtain the far field radiation, but without the computational burden of including such fields
in the main calculation (e.g., the near field solution requires
a full numerical simulation, but the propagation to the far field can be done analytically). We adopt such an idea here to transform QNM fields into the desired regularized fields outside the scattering geometry.
As shown in Fig.~\ref{neartofar}, we choose a cuboid surface close to resonator, as the near field ($\mathbf{\tilde{f}}$) surface, and transform to a cylindrical surface away from resonator as far field ($\mathbf{\tilde{F}}$) surface.
The distances to the near and far field surface from the surface of the shown Au dimer are $h$ and $h_{\rm far}$.
In this way, using a field equivalence principle~\cite{FieldEquiv}, i.e., the sources inside a domain are replaced with sources on the surface of this domain, and the fields inside the domain is zero, we can use the QNM with a complex frequency and transform it to a real frequency field, which has the correct far field behavior; note, the near fields
are not appropriate for the evaluation of the total
contribution from the radiative reservoir fields, which should be evaluated in the far field, and in real frequency space.


The electric surface current $\tilde{\mathbf{J}}_{\rm S'}(\mathbf{r}')$ and magnetic surface current $\tilde{\mathbf{M}}_{\rm S'}(\mathbf{r}')$ on a near field surface $\rm S'$, close to resonator, are given by~\cite{Neartofar_1992}
\begin{align}
\tilde{\mathbf{J}}_{\rm S'}(\mathbf{r}')=\mathbf{\hat{n}'\times\tilde{\bf{h}}(\mathbf{r}')},\\
\tilde{\mathbf{M}}_{\rm S'}(\mathbf{r}')=-\mathbf{\hat{n}'\times\tilde{\bf{f}}(\mathbf{r}')},
\end{align}
where
\begin{equation}
    \mathbf{\tilde{h}_{\mu}(r')}=\frac{1}{i\tilde{\omega}_{\mu}\mu_{0}}\nabla\times\mathbf{\tilde{f}_{\mu}(r')},
\end{equation}
are the magnetic QNMs and
$\mathbf{\hat{n}'}$ is normal to the surface $\rm S'$, pointing outward.
Assuming the resonator is placed in a homogeneous medium with dielectric constant $\epsilon_{\rm B}=n_{\rm B}^{2}$, the QNMs fields $\tilde{\bf f}\equiv\tilde{\bf f}_{\mu}$, and magnetic QNMs are $\tilde{\bf h}\equiv\tilde{\bf h}_{\mu}$,
the vector potentials at $\mathbf{R}$ generated by the electric and magnetic currents  at
some area element $\rm dS'$, are
\begin{align}
\tilde{\mathbf{X}}(\mathbf{R},\omega)=\frac{\epsilon_{\rm B}\epsilon_{0}}{4\pi}\frac{e^{ik|\mathbf{R-r'}|}}{|\mathbf{R-r'}|}\tilde{\mathbf{M}}_{\rm S'}(\mathbf{r}'){\rm dS'},\\
\tilde{\mathbf{A}}(\mathbf{R},\omega)=\frac{\mu_{0}}{4\pi}\frac{e^{ik|\mathbf{R-r'}|}}{|\mathbf{R-r'}|}\tilde{\mathbf{J}}_{\rm S'}(\mathbf{r}'){\rm dS'}.
\end{align}
Thus, integrating the contributions from all the point sources distributed over
the surface  $\rm S'$, we obtain the total vector potential fields~\cite{Neartofar_1992}
\begin{align}
\tilde{\mathbf{X}}(\mathbf{R},\omega)=\frac{\epsilon_{\rm B}\epsilon_{0}}{4\pi}\oint_{S'}\frac{e^{ik|\mathbf{R-r'}|}}{|\mathbf{R-r'}|}\tilde{\mathbf{M}}_{\rm S'}(\mathbf{r}'){\rm dS'},\\
\tilde{\mathbf{A}}(\mathbf{R,\omega})=\frac{\mu_{0}}{4\pi}\oint_{\rm S'}\frac{e^{ik|\mathbf{R-r'}|}}{|\mathbf{R-r'}|}\tilde{\mathbf{J}}_{\rm S'}(\mathbf{r}'){\rm dS'}.
\end{align}
Subsequently,  the regularized electric QNMs $\tilde{\bf{F}}$ and magnetic QNMs $\tilde{\bf{H}}$ in the far field are obtained from~\cite{FieldEquiv,Neartofar_1992}
\begin{align}
\begin{split}\label{bigFnear2far}
\tilde{\bf{F}}(\bf{R},\omega)  =& i\omega\bigg[\tilde{\mathbf{A}}(\mathbf{R},\omega)+\frac{c^{2}}{(\omega n_{\rm B})^{2}}\mathbf{\nabla\Big(\nabla\cdot \tilde{\mathbf{A}}(\mathbf{R,\omega})\Big)}\bigg]\\
&-\frac{1}{\epsilon_{\rm B}\epsilon_{0}}\mathbf{\nabla \times \tilde{\mathbf{X}}(\mathbf{R},\omega)},
\end{split}
\end{align}
\begin{align}
\begin{split}\label{bigHnear2far}
\tilde{\bf{H}}(\bf{R},\omega)  =& i\omega\bigg[\tilde{\mathbf{X}}(\mathbf{R},\omega)+\frac{c^{2}}{(\omega n_{\rm B})^{2}}\mathbf {\nabla\Big(\nabla \cdot \tilde{\mathbf{X}}(\mathbf{R},\omega)\Big)}\bigg]\\
&+\frac{1}{\mu_{0}}\mathbf{\nabla \times\tilde{\mathbf{A}}(\mathbf{R,\omega})}.
\end{split}
\end{align}

Mathematically,
it is also useful to consider the
fields in the very far spatial domain ($\mathbf{R}
\rightarrow \mathbf{R}_\infty$), since one can perform the far field approximation to obtain the simplified form for the reservoir field expression:
\begin{align}
\begin{split}
\tilde{\mathbf{F}}(\mathbf{R}_\infty,\omega)\approx i\omega\Bigg( &\tilde{\mathbf{A}}(\mathbf{R}_\infty,\omega) -\hat{\mathbf{R}}_\infty\left(\hat{\mathbf{R}}_\infty\cdot\tilde{\mathbf{A}}(\mathbf{R}_\infty,\omega)\right)  \\
& -\eta^{\rm B}_\alpha(\omega)\hat{\mathbf{R}}_\infty\times\tilde{\mathbf{X}}(\mathbf{R}_\infty,\omega) \Bigg)\label{VeryfarFieldNF2FF},
\end{split}
\end{align}
with $\eta^{\rm B}_\alpha(\omega) = \sqrt{\mu_0/\epsilon_0\epsilon^{\rm B}_\alpha(\omega)}$, $\hat{\mathbf{R}}=\mathbf{R}/|\mathbf{R}|$ and the approximated vector potentials
\begin{gather}
\tilde{\mathbf{A}}(\mathbf{R}_\infty,\omega)\approx\mu_0\frac{e^{ik|\mathbf{R}_\infty|}}{4\pi|\mathbf{R}_\infty|}\oint_{S'} {\rm d}S' e^{-ik\hat{\mathbf{R}}_\infty\cdot\mathbf{r}_s'}\tilde{\mathbf{J}}(\mathbf{r}_s')\label{AVecApprox},\\
\tilde{\mathbf{X}}(\mathbf{R}_\infty,\omega)\approx\epsilon_0\epsilon^{\rm B}\frac{e^{ik|\mathbf{R}_\infty|}}{4\pi|\mathbf{R}_\infty|}\oint_{S'} {\rm d}S' e^{-ik\hat{\mathbf{R}}_\infty\cdot\mathbf{r}_s'}\tilde{\mathbf{M}}(\mathbf{r}_s')\label{XVecApprox}.
\end{gather}
As we will show below, these considerably simplify the
problem of having to integrate the fields over a closed surface,
which is required to obtain the radiative coupling parameters
for quantized QNM theory~\cite{franke_quantization_2018}.

\subsection{Quantized quasinormal mode parameters:
quantum mechanical ``\texorpdfstring{$S$}{Lg} factors'' for Fock space normalization}\label{subSec2.4}

To connect the developed NF2FF to the basic formalism of a recently developed quantized QNM theory \cite{franke_quantization_2018,Hughes_SPS_2019},
the basic theory of the quantized QNM
approach is briefly recapitulated. In particular, we will show what QNM fields are required for the formalism and the related calculation of the quantum parameters required to represent the fields.

As shown in Refs.~\onlinecite{franke_quantization_2018,Hughes_SPS_2019}, combining a well established quantization approach for general absorptive and spatial inhomogeneous media\cite{dung_three-dimensional_1998,suttorp_fano_2004} with the QNM Green function, Eq.~\eqref{eq:GFwithSUM}, one can derive an expansion of the medium-assisted electric field operator $\hat{\mathbf{E}}(\mathbf{r}_{\rm s})$  at position $\mathbf{r}_{\rm s}$ (system region):
\begin{equation}
 \hat{\mathbf{E}}(\mathbf{r}_{\rm s})=\sum_{\mu}i\sqrt{\frac{\hbar\omega_{\mu} }{2\epsilon_0}}\, \tilde{\mathbf{f}}^{s}_{\mu}(\mathbf{r}_{\rm s}) \hat{a}_{\mu} + \text{H.a.},\label{eq:Esymm_multi}
 \end{equation}
 with the symmetrized QNM functions,
  \begin{equation}
 \tilde{\mathbf{f}}^{s}_{\mu}(\mathbf{r}_{\rm s})=\sum_{\nu}(\mathbf{S}^{\frac{1}{2}})_{\nu\mu}\sqrt{\omega_{\nu}/\omega_{\mu}}\tilde{\mathbf{f}}_{\nu}(\mathbf{r}_{\rm s}),
 \end{equation}
where $\hat{a}_{\mu}$ and $\hat{a}_{\mu}^\dagger$ are suitable annihilation and creation operators to obtain plasmon/photon Fock states for the symmetrized QNMs.

The quantum $S$ factors, which become a photon coupling matrix $S_{\mu\eta}$
if more than one QNM is considered,
are given via
\begin{align}\label{eq:S-definition_QNMs}
 \begin{split}
 S_{\mu\eta} &=\int_0^{\infty}\!{\rm d}\omega \frac{2A_{\mu}(\omega)A^{\ast}_{\eta}(\omega)}{\pi\sqrt{\omega_\mu\omega_\eta}}\left[S_{\mu\eta}^{\rm nrad}(\omega){+}S_{\mu\eta}^{\rm rad}(\omega)\right], \\
 &\equiv S_{\mu\eta}^{\rm nrad} + S_{\mu\eta}^{\rm rad},
 \end{split}
 \end{align}
 where
 \begin{equation}\label{SnradfullwMulti}
 S_{\mu\eta}^{\rm nrad}(\omega)=\int_{V} {\rm d}\mathbf{r}\,\epsilon_I(\mathbf{r},\omega)\,
 \tilde {\mathbf{f}}_{\mu}({\bf r})\cdot\tilde {\mathbf{f}}^{\ast}_{\eta}({\bf r}),
 \end{equation}
 accounts for absorption due to the metallic losses, and
 \begin{align}\label{SradfullwMulti}
 S_{\mu\eta}^{\rm rad}(\omega)\!=\!\frac{1}{2\epsilon_0\omega }\int_{S_{\rm V}}{\rm d}A_{\rm s}\hat{\mathbf{n}}_{\rm s}\!\cdot\!\big(\tilde{\mathbf{F}}_{\mu}(\mathbf{s},\omega)\!\times\!\tilde{\mathbf{H}}_{\eta}^{\ast}(\mathbf{s},\omega) \!+\!  \underset{(\mu\leftrightarrow \eta)}{{\rm H.c.}}\big),
 \end{align}
describes radiation leaving the system through the surface $S_V$ with the normal vector $\hat{\mathbf{n}}_{\mathbf{s}}$ pointing outward from the resonator volume $V$, and $\tilde{\mathbf{H}}(\mathbf{s},\omega){=}1/(i\mu_0\omega)\nabla
\times\tilde{\mathbf{F}}(\mathbf{s},\omega)$ is the QNM magnetic
field. Furthermore, by choosing  $S_V$ in the very far field at $S_{\rm \infty}$, we can apply the Silver-M\"uller radiation condition $\hat{\mathbf{n}}_s\times{\mathbf{H}}(\mathbf{s},\omega)\rightarrow -n_{\rm B}c\epsilon_0\tilde{\mathbf{F}}(\mathbf{s},\omega)$ and in addition use the approximated version of the NF2FF results (Eqs.\eqref{VeryfarFieldNF2FF}-\eqref{XVecApprox}); we then arrive at an approximated formula for the radiative contribution %
\begin{align}
S^{\rm rad}_{\mu\eta}(\omega)  \approx \frac{n_{\rm B}c}{\omega}\int_{S_{\infty}}dA_{\rm s}\mathbf{\tilde{F}}_\mu(\mathbf{s}_\infty,\omega)\cdot\mathbf{\tilde{F}}_\eta^*(\mathbf{s}_\infty,\omega). \label{SradSM}
\end{align}
Choosing $S_{\rm \infty}$ as a sphere and transforming into spherical coordinates, leads to a further simplification
\begin{equation}
S^{\rm rad}_{\mu\eta}(\omega) = \frac{n_{\rm B}c}{\omega}I^{\rm sur}_{\mu\eta}(\omega)\label{MMInfinityS_omega},
\end{equation}
with
\begin{align}
I^{\rm sur}_{\mu\eta}(\omega)=&\frac{1}{16\pi^2}\int_0^{2\pi}{\rm d\varphi}\int_{0}^{\pi}{\rm d}\vartheta \sin(\vartheta) \times
\nonumber \\
\ \ &\tilde{\mathbf{Z}}_\mu(\varphi,\vartheta,\omega)\cdot\tilde{\mathbf{Z}}_\eta^*(\varphi,\vartheta,\omega),
\end{align}
where the function $\tilde{\mathbf{Z}}_\mu(\varphi,\vartheta,\omega)$ is given as
\begin{align}
\begin{split}
&\tilde{\mathbf{Z}}_\mu(\varphi,\vartheta,\omega)=i\omega\mu_0\oint_{S'} {\rm d}S' e^{-in_{\rm B}\omega\hat{\mathbf{R}}\cdot\mathbf{r}_s'/c}\\
& \ \ \ \ \bigg[\tilde{\mathbf{J}}_\mu(\mathbf{r}_s') - \left(\tilde{\mathbf{J}}_\mu(\mathbf{r}_s')\cdot\hat{\mathbf{R}}\right)\hat{\mathbf{R}}
-n_{\rm B}c\epsilon_0\hat{\mathbf{R}}\times\tilde{\mathbf{M}}_\mu(\mathbf{r}_s')\bigg],
\end{split}
\end{align}
and $\hat{\mathbf{R}}=\hat{\mathbf{R}}(\varphi,\vartheta)$ is the radial basis vector in spherical coordinates, namely:
\begin{equation}
\hat{\mathbf{R}} =
 \Big(\sin(\vartheta)\cos(\varphi), ~\sin(\vartheta)\sin(\varphi),~\cos(\vartheta)\Big)\label{eq: SpherCoord}.
\end{equation}
Notably, the above form in Eq.~\eqref{MMInfinityS_omega} is independent of the radius of the sphere $S_\infty$, as long as it is chosen in the very far field, which  significantly simplifies the numerical evaluation of the radiative part.

 Equations (\ref{eq:S-definition_QNMs}-\ref{eq: SpherCoord}) show how to use the NF2FF transform to model a quantum
 emitter coupled to the quantized QNMs. For example,
 placing an emitter with dipole moment $\mathbf{d}$ at $\mathbf{r}_{0}$,
and assuming the bad cavity limit (i.e., a weakly coupled emitter), then
the quantum SE rate, i.e., the SE rate obtained from the QNM quantization model,  is~\cite{franke_quantization_2018}
\begin{equation}
     \Gamma_{\rm quan}=\Gamma_{\rm quan}^{\rm diag}+\Gamma_{\rm quan}^{\rm ndiag}, \label{GammaQM}
 \end{equation}
 where the diagonal contribution is
 \begin{equation}
     \Gamma^{\rm diag}_{\rm quan}=\sum_{\mu}S_{\mu\mu}\frac{\big|\tilde{g}_{\mu}\big|^{2}\gamma_{\mu}}{\Delta_{\mu\mu}^{2}+\gamma^{2}_{\mu}},
 \end{equation}
 and non-diagonal contribution is
 \begin{equation}
     \Gamma^{\rm ndiag}_{\rm quan}=\sum_{_{\mu,\eta\neq\mu}}\tilde{g}_{\mu}S_{\mu\eta}\tilde{g}_{\eta}^{\ast}K_{\mu\eta},
 \end{equation}
 with
 \begin{equation}
   K_{\mu\eta}=\frac{\big[i(\omega_{\mu}-\omega_{\eta})+\gamma_{\mu}+\gamma_{\eta}\big]}{\big[2(\Delta_{\mu{\rm e}}-i\gamma_{\mu})(\Delta_{\eta{\rm e}}+i\gamma_{\eta})\big]}.
 \end{equation}
 Here, $\Delta_{\mu{\rm e}}=\omega_{\mu}-\omega_{\rm e}$ is the frequency detuning between the emitter and QNMs, and $\tilde{g}_{\mu}=\sqrt{\omega_{\mu}/(2\epsilon_{0}\hbar)}\mathbf{d}\cdot\tilde{\mathbf{f}}_{\mu}(\mathbf{r}_{0})$ is the emitter-QNM coupling. The total quantum Purcell factor is
\begin{equation}\label{quantumpurcell}
F_{\rm P}^{\rm quan}=\frac{\Gamma_{\rm quan}}{\Gamma^{0}},
\end{equation}
where $\Gamma^{0}$ is the spontaneous emission rate in a homogeneous medium.

Note we refer to Eq.~\eqref{GammaQM}
as the ``quantum SE rate'' in the sense that
it is derived using a system-level quantized
mode theory for the photons. In the limit of
a single photon subspace, as appropriate for a SE
description, we naturally expect agreement with
the semiclassical theory for SE. Nevertheless, for
effects beyond the single quantum regime, the
quantum approach is required, so we use this label merely to label the rate that is computed using the quantized QNM
theory.

For the single mode case, $S_{\mu\eta}$ becomes a simple photon normalization factor $S$, and takes the simplified form 
\begin{align}\label{eq:S-definition}
\begin{split}
S &=\frac{2}{\pi\omega_{\rm c}}\int_0^{\infty}\!{\rm d}\omega |A_{\rm c}(\omega)|^2\left[S^{\rm nrad}(\omega){+}S^{\rm rad}(\omega)\right] \\
&\equiv S^{\rm nrad} + S^{\rm rad},
\end{split}
\end{align}
where
\begin{equation}
S^{\rm nrad}(\omega)=\int_{V} {\rm d}\mathbf{r}\,\epsilon_I(\mathbf{r},\omega)\,
|\tilde {\mathbf{f}}({\bf r})|^2,
\end{equation}
%
and
\begin{equation}
S^{\rm rad}(\omega)= \frac{1}{\epsilon_0\omega}\int_{S_{V}} {\rm d}A_{\mathbf{s}}\mathbf{n}_{\mathbf{s}}\cdot {\rm Re}(\tilde{\mathbf{F}}(\mathbf{s},\omega)\times\tilde{\mathbf{H}}^*(\mathbf{s},\omega)).
\end{equation}
Thus the full expressions of $S^{\rm nrad}$ and $S^{\rm rad}$ are
\begin{align}\label{Snrad_full}
\begin{split}
S^{\rm nrad} & =\frac{2}{\pi\omega_{\rm c}}\int_{0}^{\infty}d\omega\big|A_{\rm c}(\omega)\big|^{2}\int_{V}d\mathbf{r}\epsilon_{\rm I}(\mathbf{r},\omega)\big|\tilde{\mathbf{f}}_{\rm c}(\mathbf{r})\big|^{2},\\
& =\frac{2}{\pi\omega_{\rm c}}\int_{0}^{\infty}d\omega|A_{\rm c}(\omega)|^{2}\epsilon_{\rm I}(\omega)\int_{V}d\mathbf{r}|\tilde{\mathbf{f}}_{\rm c}(\mathbf{r})|^{2}.
\end{split}
\end{align}
and
\begin{align}\label{Sradfull}
\begin{split}
S^{\rm rad}  =\frac{2}{\pi\omega_{\rm c}}&\int_{0}^{\infty}d\omega\big|A_{\rm c}(\omega)\big|^{2}\\
&\frac{1}{\epsilon_{0}\omega}\int_{S_{V}}dA_{\rm s}\mathbf{n_{\rm s}}\cdot {\rm Re}\big(\mathbf{\tilde{F}}(\mathbf{s},\omega)\times\mathbf{\tilde{H}}^{*}(\mathbf{s},\omega)\big).
\end{split}
\end{align}
Once again, choosing  $S_V$ as sphere in the very far field at $S_{\rm \infty}$ (and applying the same approximations as in the multi-mode case) leads to
\begin{equation}
S^{\rm rad}\approx \frac{2}{\pi\omega_{\rm c}}\int_{0}^{\infty}d\omega\big|A_{\rm c}(\omega)\big|^{2}\frac{n_{\rm B}c}{\omega}I^{\rm sur}(\omega),
\end{equation}
with
\begin{equation}
I^{\rm sur}(\omega)=\frac{1}{16\pi^2}\int_0^{2\pi}{\rm d\varphi}\int_{0}^{\pi}{\rm d}\vartheta \sin(\vartheta)|\tilde{\mathbf{Z}}_{\rm c}(\varphi,\vartheta,\omega)|^2.
\end{equation}

Placing an emitter with dipole moment $\mathbf{d}$ at $\mathbf{r}_{0}$,
and assuming the bad cavity limit,
the quantum SE rate is
 \begin{equation}
     \Gamma_{\rm quan,c}=S\frac{\big|\tilde{g}_{\rm c}\big|^{2}\gamma_{\rm c}}{\Delta_{\rm ce}^{2}+\gamma^{2}_{\rm c}},
     \label{GammaPF}
 \end{equation}
where $\Delta_{\rm ce}=\omega_{\rm c}-\omega_{\rm e}$ is the frequency detuning between the emitter and single QNM, and $\tilde{g}_{\rm c}=\sqrt{\omega_{\rm c}/(2\epsilon_{0}\hbar)}\mathbf{d}\cdot\tilde{\mathbf{f}}_{\rm c}(\mathbf{r}_{0})$ is the emitter-QNM coupling. Then the quantum Purcell factor is
\begin{equation}\label{quantumpurcellsingle}
F_{\rm P}^{\rm quan,c}=\frac{\Gamma_{\rm quan,c}}{\Gamma^{0}}.
\end{equation}
In the limit that $S\rightarrow 1$,
Eqs.~\eqref{GammaPF}-\eqref{quantumpurcellsingle} recover the well-known
decay rate and  Purcell factor from the
dissipative Jaynes-Cummings model
~\cite{PhysRevA.46.4354}.

Furthermore, the radiative and non-radiative contributions are associated with the beta factors, defined also classically in Eq.~\eqref{betanradQNMsingle} and Eq.~\eqref{betaradQNMsingle}.
In the quantized QNM theory, the beta factors are defined from
\begin{align}
 \beta_{\rm quan}^{\rm rad}& =\frac{S^{\rm rad}}{S}, \label{beta_rad_quan}
 \\
 \beta_{\rm quan}^{\rm nrad}&=\frac{S^{\rm nrad}}{S} \label{beta_nrad_quan}.
\end{align}
 These quantum-derived  $S$ factors are unitless quantities, and for well isolated single QNMs, we have found that~\cite{franke_quantization_2018}  $S{\approx}1$ (see also calculations below for gold dimers).

Although the Purcell factors obtained from the quantized QNM theory
have been shown to be in excellent agreement with the semiclassical results (also using the QNM approximation)~\cite{franke_quantization_2018}, and therefore with the full Maxwell solution, there can be generally a discrepancy between both approaches;
the reason for this is because
different approximations to the full Green function are imposed on different stages of the derivation. Whereas in the semiclassical case, the QNM approximation is done at the emitter position $\mathbf{r}_0$ only, in the quantum case, the approximations is applied to all positions within the resonator region. This is deeply connected to the relation
 \begin{align} \label{GXE}
     {\rm Im}\mathbf{G}(\mathbf{r}_0,\mathbf{r}_0)=&\int_V d\mathbf{r}\epsilon_I(\mathbf{r})\mathbf{G}(\mathbf{r}_0,\mathbf{r})\cdot\mathbf{G}(\mathbf{r},\mathbf{r}_0) \nonumber \\
     &+i\frac{c^2}{2\omega^2}\int_{S}dA_{s}\mathbf{C}(\mathbf{s},\mathbf{r}_0)-\mathbf{C}^{\dagger}(\mathbf{s}, \mathbf{r}_0)
 \end{align}
 with
 \begin{equation}
     \mathbf{C}(\mathbf{s},\mathbf{r}_0)=\left[\nabla\times\mathbf{G}(\mathbf{s}, \mathbf{r}_0)\right]^t\cdot\left[\mathbf{n}_s\times\mathbf{G}^*(\mathbf{s}, \mathbf{r}_0)\right].
 \end{equation}
 In the semiclassical case, the QNM approximation is done on the lhs (left hand side) of \eqref{GXE};
 however, in the quantized QNM theory, the approximation must be done on the rhs in order to construct Fock states, that are independent on the emitter positions and in order to formulate the electric field operator in the cavity with few mode operators, i.e.,  instead of an infinite set of position-dependent operators.


So summarize this subsection, the quantum $S$ factors for quantized QNM theory are greatly desired in the process of the quantization of the open cavities, and the calculation of the related quantum quantities, such as coupling coefficient, spontaneous emission rate, Purcell factors, and single photon source figures of merit \cite{Hughes_SPS_2019}. Thus it is important
to have accurate efficient numerical techniques to obtain the required fields
and integrals.

 \subsection{Practical evaluation of the frequency integrals in the
 quantized QNM model}
 \label{subSec2.5}

Due to the introduction of a rectangular function in $A_{\rm c}(\omega)$ (Eq.~\eqref{A1}), the frequency integral in $S$ is restricted on a finite frequency band around the QNM center frequency $\omega_{\rm c}$. If the non-Lorentzian contributions are nearly constant in this effective frequency regime, we can approximate $S$ as
\begin{align}
    \int_0^{\infty} d\omega S(\omega) \approx S_{\rm p}.
\end{align}

In the multi-mode case, 
the frequency integrals are performed, in the same approximation, as
\begin{equation}
    \int_0^{\infty} d\omega S_{\mu\eta}(\omega)\approx S_{{\rm p},\mu\eta},  
\end{equation}
where $S_{{\rm p},\mu\eta}$ is an average of the non-modal contributions calculated at $\omega_\mu, \omega_\eta$.

Within these approximations, the pole terms of the photon coupling matrices and photon normalization factors take the form
\begin{align}
S_{{\rm p},\mu\eta}^{\rm nrad}= \frac{\sqrt{\omega_\mu\omega_\eta}}{i(\tilde{\omega}_\mu - \tilde{\omega}_\eta^*)}\int_{V}& {\rm d}\mathbf{r}\,\sqrt{\epsilon_I(\mathbf{r},\omega_{\rm \mu})\epsilon_I(\mathbf{r},\omega_{\eta})}\\
&\times\tilde {\mathbf{f}}_\mu({\bf r})\cdot\tilde {\mathbf{f}}_\eta({\bf r}),\label{eq:SnradapproxMulti}
\end{align}
and
\begin{align}
S_{{\rm p1},\mu\eta}^{\rm rad}  = & \frac{1}{i(\tilde{\omega}_\mu -\tilde{\omega}_\eta)2\epsilon_0\omega }\\
&\int_{S_{\rm V}}{\rm d}A_{\rm s}\hat{\mathbf{n}}_{\rm s}\!\cdot\!\big(\tilde{\mathbf{F}}_{\mu}(\mathbf{s},\omega_\mu)\!\times\!\tilde{\mathbf{H}}_{\eta}^{\ast}(\mathbf{s},\omega_\eta) \!+\!  \underset{(\mu\leftrightarrow \eta)}{{\rm H.c.}}\big).\label{Sradpole1Multi}
\end{align}
For the latter case, we can also derive
an alternative pole approximation  if one is interested
in the integrated value over a far field surface:
\begin{align}
S_{{\rm p2},\mu\eta}^{\rm rad} = \frac{n_{\rm B}c}{i(\tilde{\omega}_\mu - \tilde{\omega}_\eta)}I_{\rm \mu\eta}^{\rm sur}, \label{Sradpole2Multi}
\end{align} where
\begin{align}
    I_{\rm \mu\eta}^{\rm sur}& = \frac{1}{16\pi^2}\int_0^{2\pi}{\rm d\varphi}\int_{0}^{\pi}{\rm d}\vartheta \sin(\vartheta) \times
    \nonumber \\
    &\tilde{\mathbf{Z}}_\mu(\varphi,\vartheta,\omega_\mu)\cdot\tilde{\mathbf{Z}}_\eta^*(\varphi,\vartheta,\omega_\eta)\label{I_sur_mueta}.
\end{align}

For the single QNM case, then
%
%
\begin{equation}
S^{\rm nrad}_{\rm p}= Q\int_{V} {\rm d}\mathbf{r}\,\epsilon_I(\mathbf{r},\omega_{\rm c})\,|\tilde {\mathbf{f}}({\bf r})|^2,\label{eq:Snradapprox}
\end{equation}
\begin{equation}
S_{\rm p1}^{\rm rad}  =  \frac{1}{2\epsilon_{0}\gamma_{\rm c}}\int_{S_{V}}dA_{\rm s}\mathbf{n_{\rm s}}\cdot {\rm Re}(\mathbf{\tilde{F}}(\mathbf{s},\omega_{\rm c})\times\mathbf{\tilde{H}}^{*}(\mathbf{s},\omega_{\rm c}))\label{Sradpole1},
\end{equation}
\begin{equation}
S^{\rm rad}_{\rm p2} = \frac{n_{\rm B}c}{2\gamma_{\rm c}}I_{\rm c}^{\rm sur}, \label{Sradpole2}
\end{equation}
and
\begin{equation}
    I_{\rm c}^{\rm sur} = \frac{1}{16\pi^2}\int_0^{2\pi}{\rm d\varphi}\int_{0}^{\pi}{\rm d}\vartheta \sin(\vartheta)|\tilde{\mathbf{Z}}_{\rm c}(\varphi,\vartheta,\omega_{\rm c})|^2\label{I_sur_c}.
\end{equation}

\section{Numerical results for single quasinormal modes
of metal nanorod dimers}\label{Sec3}

\subsection{Single quasinormal modes for metal dimers: role of
material losses}\label{subSec3.1}

\begin{figure}[th]
  \centering
  \includegraphics[width=0.99\columnwidth]{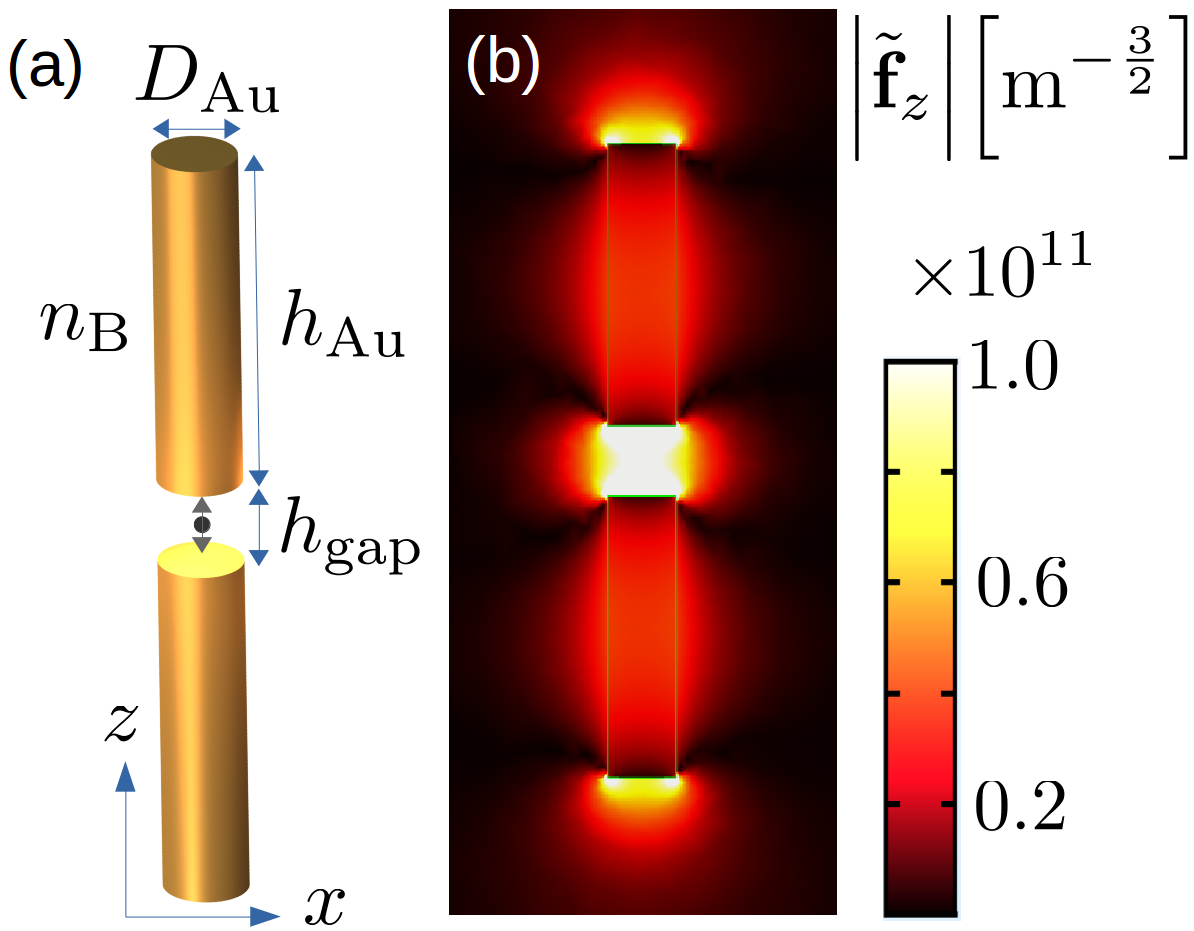}
  \caption{(a) Schematic diagram of Au dimer in free space. (b) Normalized QNM field $\big|\tilde{\bf f}_{z}\big|$ (dominant component) distribution at plane $y=0$ for $\gamma_{\rm p}=\gamma_{\rm p0}$. Here the absolute value means that both the real and the imaginary parts are taken into account. The origin of the coordinate system is at the gap center of the dimer. The other QNMs for different loss values look similar.
  }\label{schematic}
  \centering
  \includegraphics[width=0.99\columnwidth]{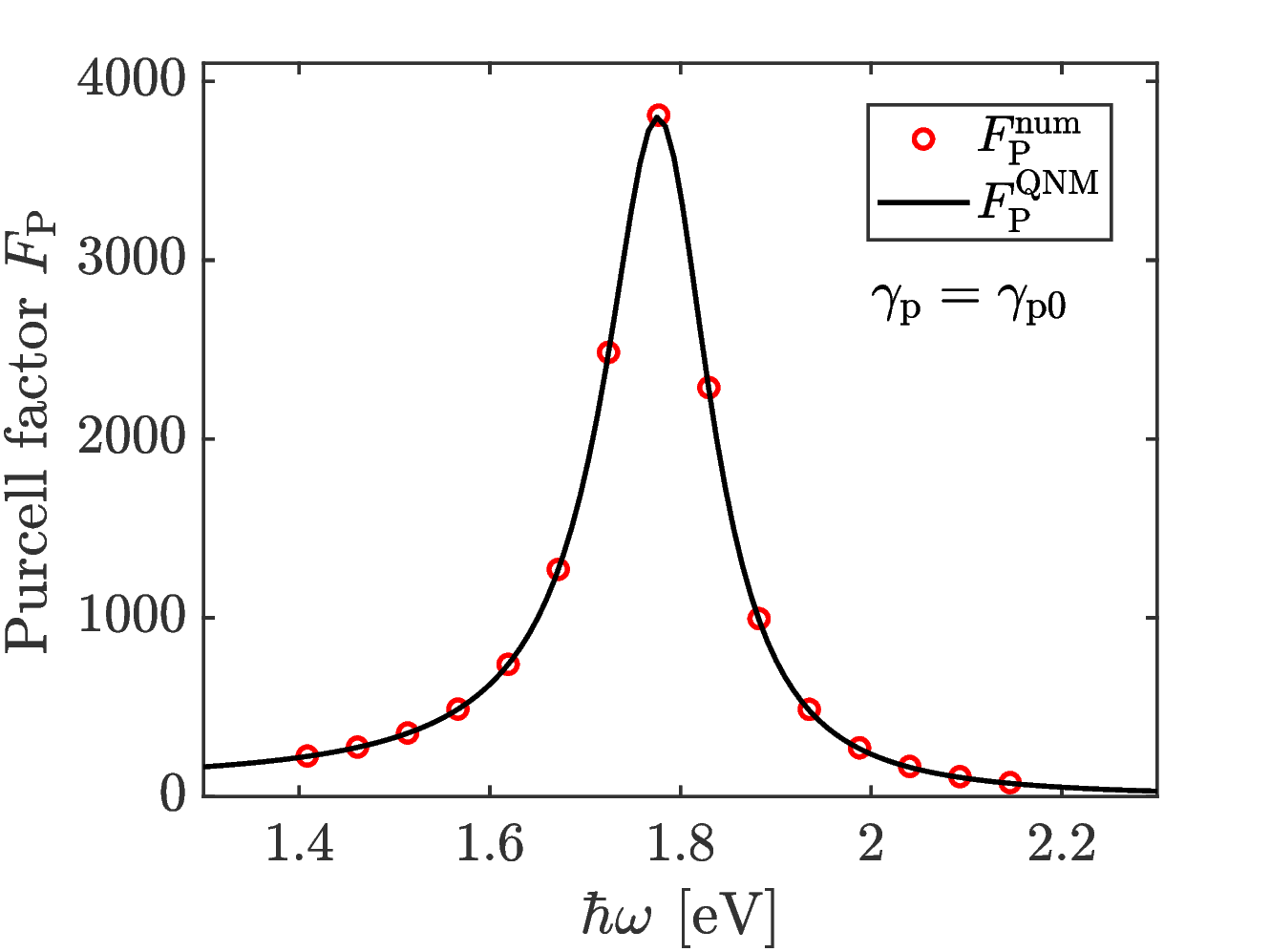}
  \caption{Classical Purcell factor calculations for a $z$-polarized  dipole at dimer center, with $\gamma_{\rm p}=\gamma_{\rm p0}$ (gold), using the analytical QNM formula (Eq.~\eqref{QNMpurcell}) and the full dipole formula (Eq.~\eqref{Purcellfulldipole}).
 }\label{agree1}
\end{figure}

As shown in Fig.~\ref{schematic} (a),
we first consider a gold (Au) rod dimer (with diameter of $D_{\rm Au}=20$ nm, length of $h_{\rm Au}=80$ nm and gap of $h_{\rm gap}=20$ nm ) in free space ($\epsilon_{\rm B}=n_{\rm B}^{2}=1.0$), with the same parameters as used in Ref.~\onlinecite{franke_quantization_2018}.
The local dielectric function of Au is described by the Drude model,
\begin{equation}\label{Drude}
    \epsilon_{\rm Au}=1-\frac{\omega_{\rm p}^{2}}{\omega^{2}+i\omega\gamma_{\rm p}},
\end{equation}
where $\hbar\omega_{\rm p}=8.2934$ eV ($\omega_{\rm p}=1.26\times 10^{16}$ rad/s) and $\hbar\gamma_{\rm p}=0.0928$ eV ($\gamma_{\rm p}=\gamma_{\rm p0}=1.41\times 10^{14}$ rad/s).

In order to understand the effect of material losses on the dimer QNMs properties, we will also artificially change the loss term $\gamma_{p}$ in the metal Drude model to $3\gamma_{\rm p0}$, $2\gamma_{\rm p0}$, $(2/3)\gamma_{\rm p0}$ and $(1/3)\gamma_{\rm p0}$.
The corresponding complex QNM frequency $\tilde{\omega}_{\rm c}=\omega_c-i\gamma_c$, quality factor $Q_{\rm c}$ (=$\omega_{\rm c}/2\gamma_{\rm c}$), QNM field $\tilde{\mathbf{f}}_{z}(\mathbf{r}_{0})$ ($z$-component) at the center of the Au dimer gap, and effective mode volume $V_{\rm c}^{\rm eff}({\bf r}_0)$ ($V_{\rm c}^{\rm eff}={1}/{{\rm Re}\big[\tilde{\mathbf{f}}^{2}_{z}(\mathbf{r}_{0})\big]}$ and $\lambda_{\rm c}=2\pi c/{\rm Re}(\tilde{\omega}_{\rm c})$) are shown in Table~\ref{Q}.
The real part of the $\tilde{\omega}_{\rm c}$ and $\tilde{\mathbf{f}}_{z}(\mathbf{r}_{0})$ are noted to be very similar in all cases, while the
quality factors $Q_{\rm c}$ increases with the decrease of $\gamma_{\rm p}$ as expected.

Using the method that Bai {\em et al.} proposed~\cite{bai_efficient_2013-1}, we can obtain the normalized QNM fields from a simple
dipole excitation. Specifically,
the scattered electric field of a point dipole at position $\mathbf{r}_{0}$ is related to the Green's function, and given by
\begin{equation}
\mathbf{E}^{\rm s}(\mathbf{r},\omega)=\frac{1}{\epsilon_{0}}\mathbf{G}(\mathbf{r},\mathbf{r}_{0},\omega)\cdot \mathbf{d},
\end{equation}
where $\mathbf{d}$ is the dipole moment of the emitter.
If only a single mode is dominant, we can expand the Green's function with one QNM, so that
\begin{equation}\label{E_scatter}
\mathbf{E}^{\rm s}(\mathbf{r},\omega)=\frac{1}{\epsilon_{0}}A(\omega)\tilde{\mathbf{f}}_{\rm c}(\mathbf{r})\tilde{\mathbf{f}}_{\rm c}(\mathbf{r}_{0})\cdot \mathbf{d}.
\end{equation}
Multiplying Eq. \eqref{E_scatter} with $\mathbf{d}$ and first using
$\mathbf{r}=\mathbf{r}_{0}$, then
\begin{equation}
\mathbf{d}\cdot\tilde{\mathbf{f}}_{\rm c}(\mathbf{r}_{0})=\sqrt{\frac{\epsilon_{0}\mathbf{d}\cdot\mathbf{E}^{\rm s}(\mathbf{r}_{0},\omega)}{A(\omega)}}.
\end{equation}
Substituting this back to Eq.~\eqref{E_scatter}, we obtain the normalized field as
a function of space
\begin{align}
\begin{split}
\tilde{\mathbf{f}}_{\rm c}(\mathbf{r})&=\sqrt{\frac{\epsilon_{0}}{A(\omega)\mathbf{d}\cdot\mathbf{E}^{\rm s}(\mathbf{r}_{0},\omega)}}\mathbf{E}^{\rm s}(\mathbf{r},\omega),\\
&=\sqrt{\frac{2\epsilon_{0}(\tilde{\omega}_{\rm c}-\omega)}{\omega\mathbf{d}\cdot\mathbf{E}^{\rm s}(\mathbf{r}_{0},\omega)}}\mathbf{E}^{\rm s}(\mathbf{r},\omega),
\end{split}
\end{align}
and corresponding effective mode volume is simply~\cite{kristensen_generalized_2012}:
\begin{equation}
V_{\rm c}^{\rm eff}({\bf r}_0)
=\frac{1}{\epsilon({\bf r}_0){\rm Re}[{\bf f}^2_{\rm c}({\bf r}_0)]}.
\end{equation}

\begin{table}[b]
\caption {Single QNM resonance frequency $\tilde{\omega}_{c}$, quality factor $Q_{\rm c} $, normalized QNM field $\tilde{\bf f}_{z}$ at the center of the dimer gap, and corresponding effective mode volume for various material losses. All parameters are calculated using the
classical QNM theory.} \label{Q}
    \centering
    \begin{tabular}{|c|c|c|c|c|}
 \hline
 $\gamma_{\rm p}$ & $\hbar\tilde{\omega}_{c}~[{\rm eV}]$ &  $Q_{\rm c} $ & $\tilde{\bf f}_{z}(\mathbf{r}_{0})~[10^{9}\cdot$ m$^{-\frac{3}{2}}]$ & $V_{\rm c}^{\rm eff}({\bf r}_0)/\lambda_{\rm c}^{3}$\\
 \hline
 $3\gamma_{\rm p0}$                  & $(1.773 - 0.147i)$         & $6.0$  & $(106.2+i3.287) $ & $2.594\times10^{-4}$        \\
 \hline
  $2\gamma_{\rm p0}$                 & $(1.777 - 0.107i)$         & $8.3$ & $(106.1+i2.367) $   & $2.615\times10^{-4}$      \\
 \hline
 $\gamma_{\rm p0}$                  & $(1.780 - 0.068i)$         & $13.1$ & $(106.0+i1.451) $   & $2.631\times10^{-4}$        \\
 \hline
  $\frac{2}{3}\gamma_{\rm p0}$             & $(1.781 - 0.055i)$         & $16.3$ & $(106.0+i1.146) $ & $2.636\times10^{-4}$        \\
  \hline
  $\frac{1}{3}\gamma_{\rm p0}$             & $(1.781 - 0.041i)$         & $21.5$ & $(106.0+i0.842) $  & $2.640\times10^{-4}$      \\
 \hline
    \end{tabular}
\end{table}
\begin{figure*}
  \includegraphics[width=0.499\columnwidth]{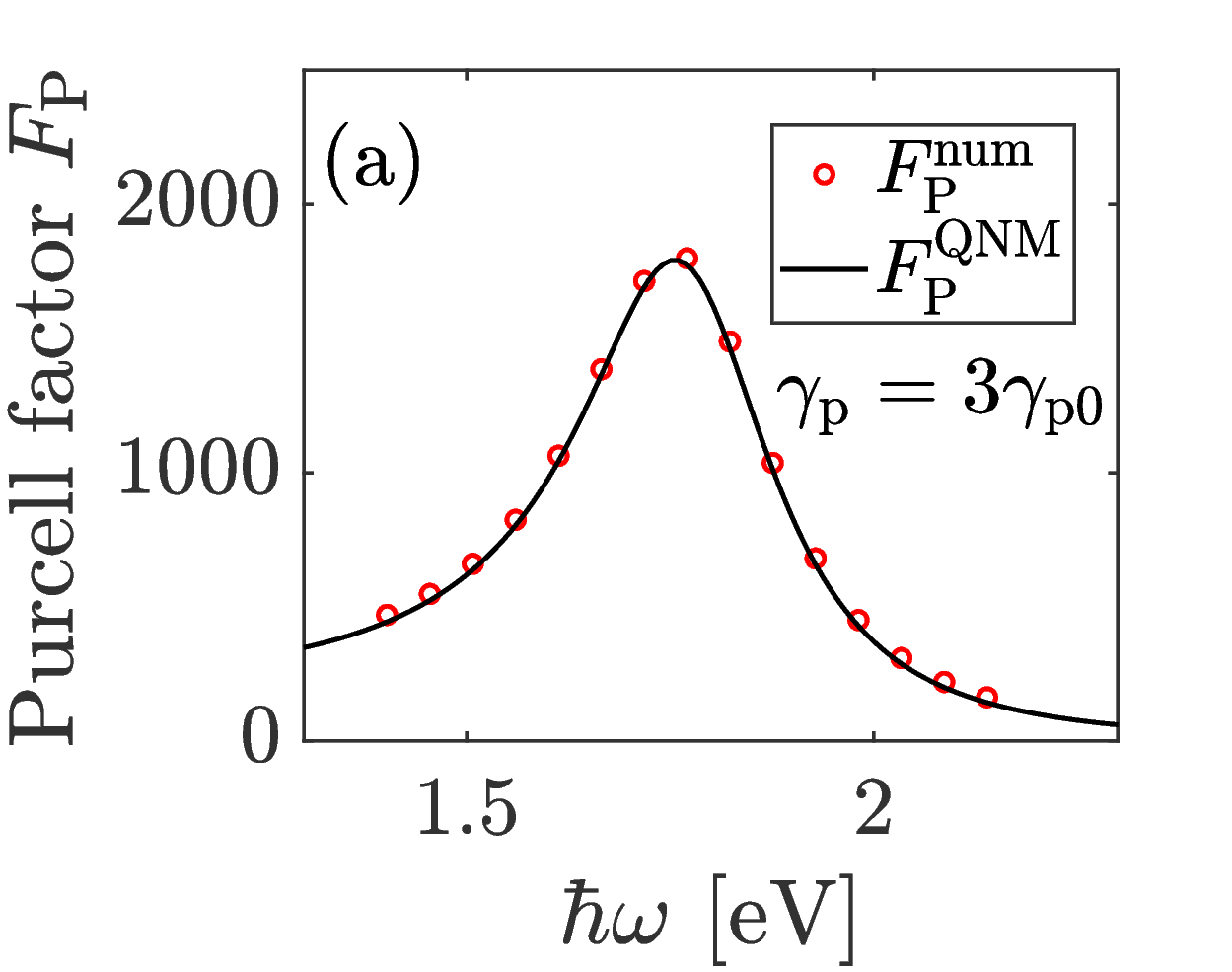}
   \includegraphics[width=0.499\columnwidth]{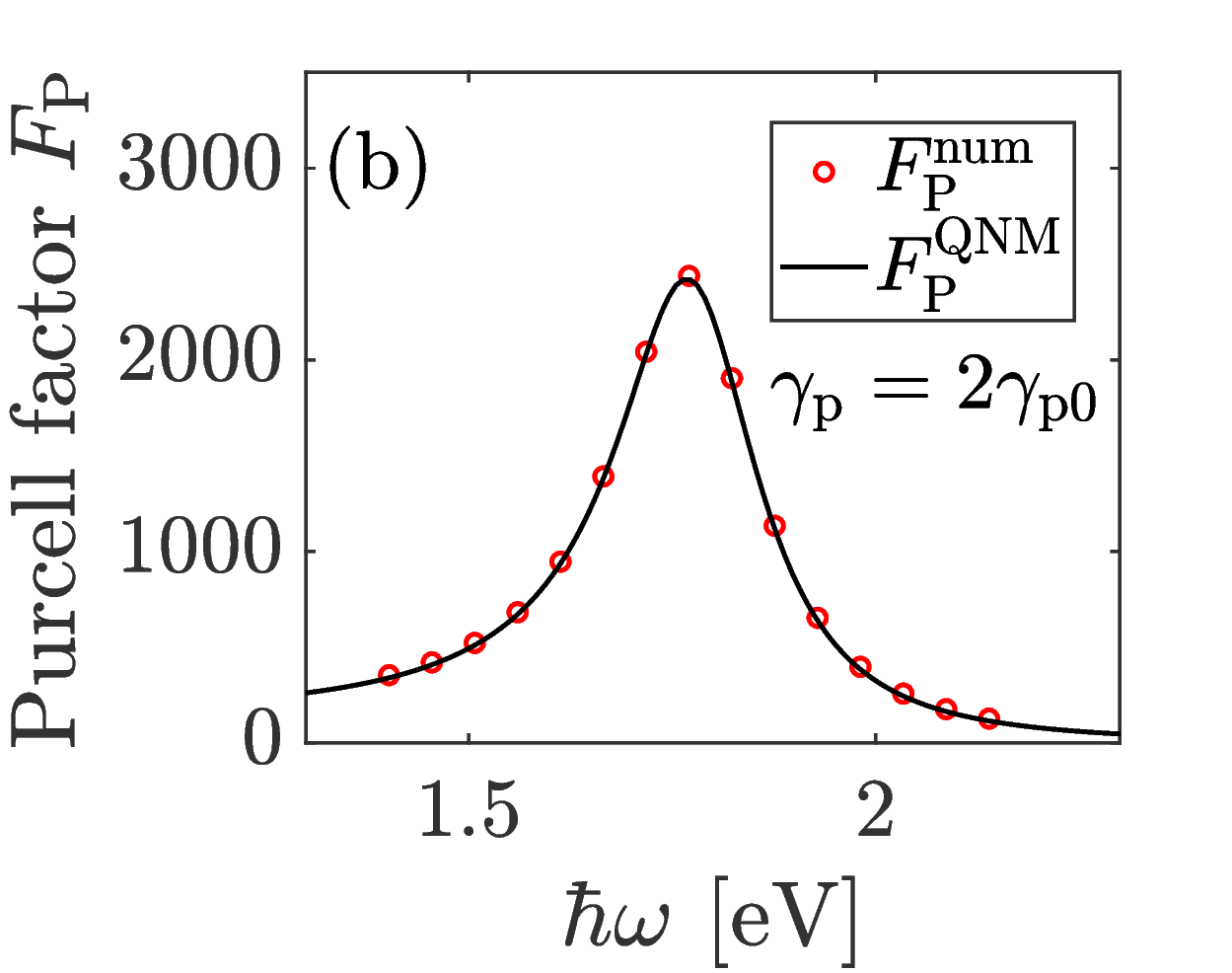}
   \includegraphics[width=0.499\columnwidth]{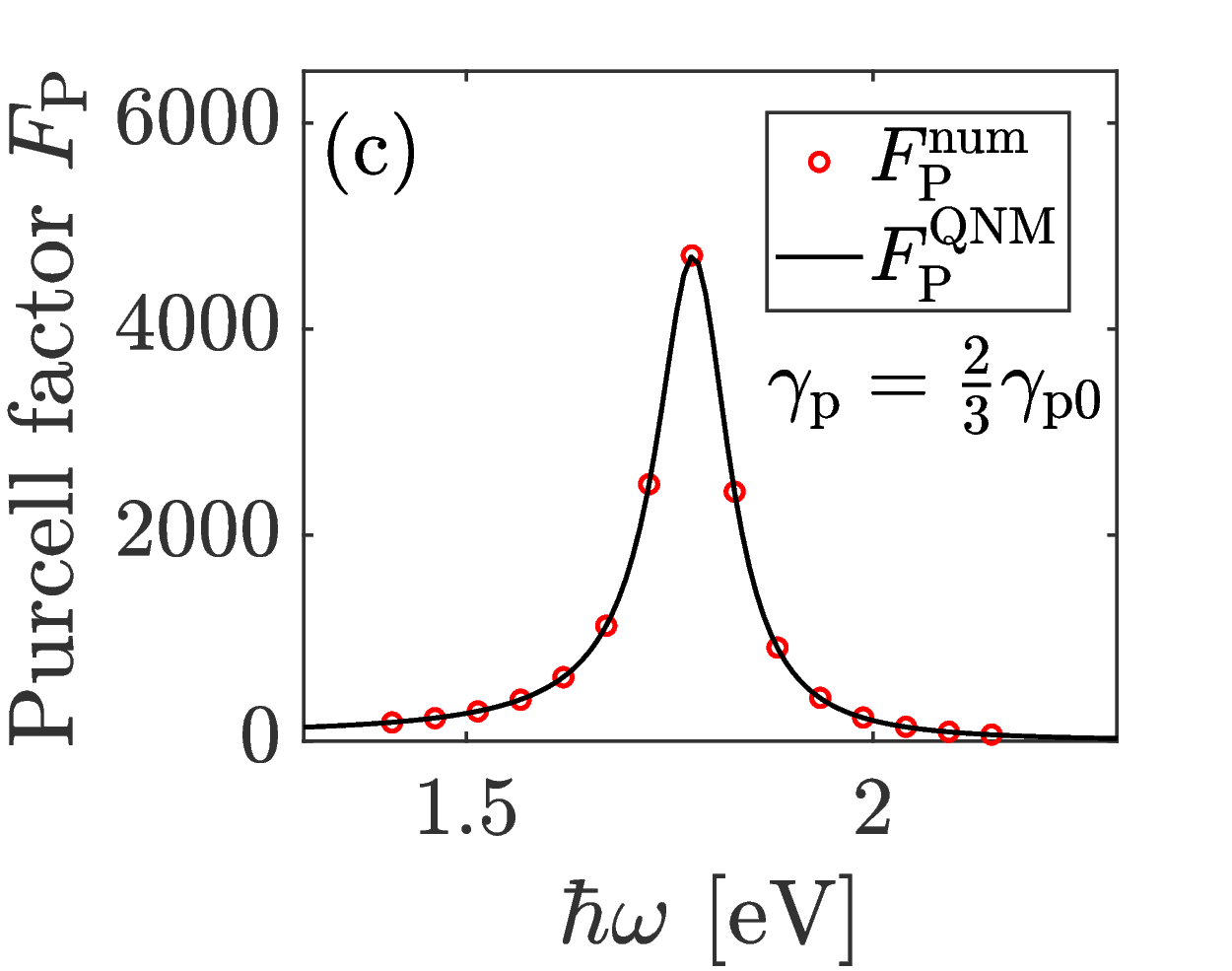}
   \includegraphics[width=0.499\columnwidth]{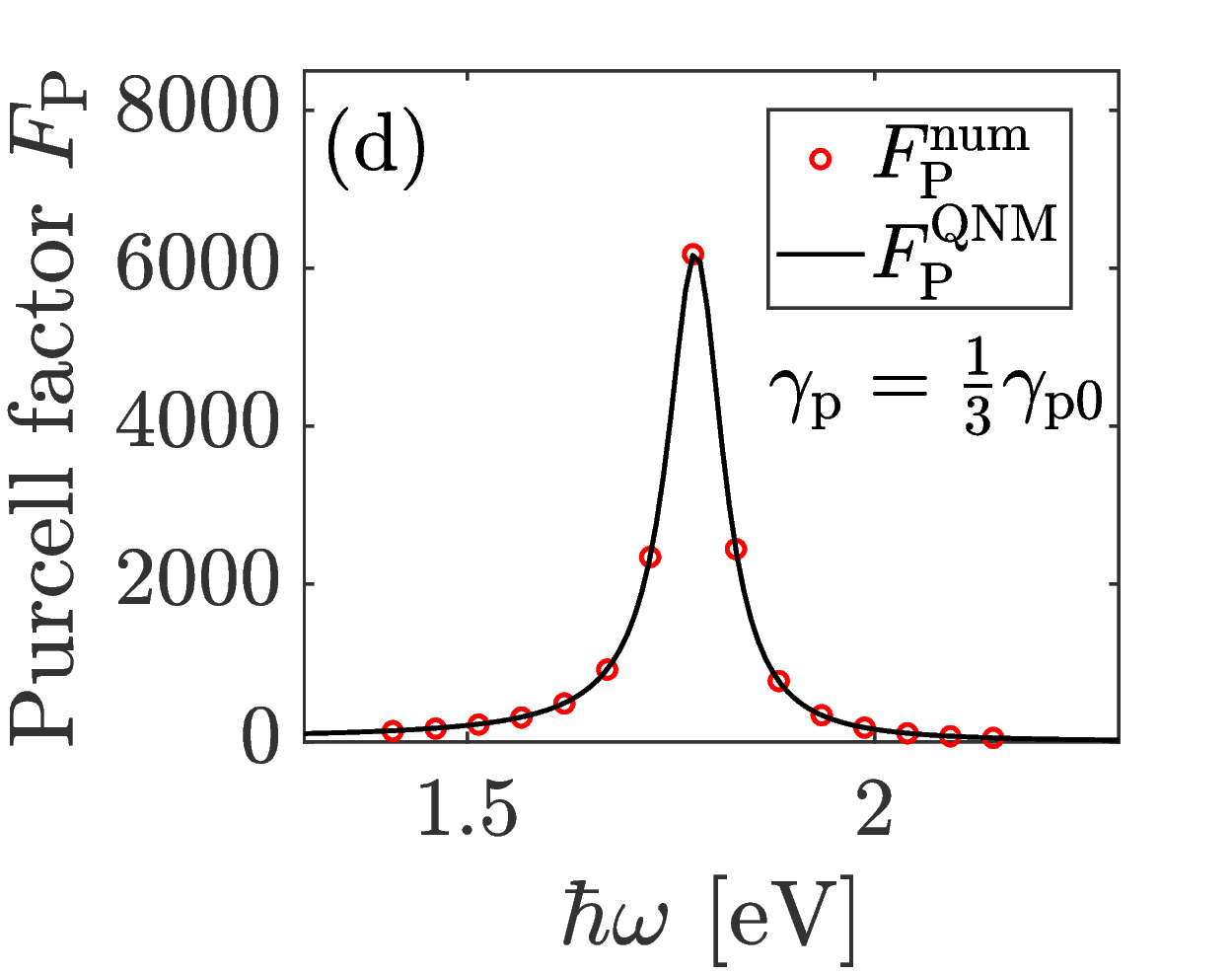}
   \caption{Classical Purcell factor
   calculations for a $z$-polarized point dipole placed at dimer center (full dipole result $F^{\rm num}_{\rm P}$ from Eq. \eqref{Purcellfulldipole} versus $F^{\rm QNM}_{\rm P}$ from single QNM with Eq. \eqref{QNMpurcell}) for (a) $\gamma_{\rm p}=3\gamma_{\rm p0}$, (b) $2\gamma_{\rm p0}$, (c) $(2/3)\gamma_{\rm p0}$, and (d) $(1/3)\gamma_{\rm p0}$. }\label{agree2}
\end{figure*}

We performed the the QNM simulations in a commercial COMSOL software~\cite{comsol}, where $\omega=(1-10^{-5})\times\tilde{\omega}_{\rm c}$, very close to the pole frequency.
For the single mode case, we define
$\tilde{\bf f} = \tilde{\bf f}_{\rm c}$.
%
The computational domain (including perfectly matched layers (PMLs)) is
around $3$ $\mu$m$^{3}$, where the maximum mesh element sizes are $0.1$ nm, $2$ nm and $80$ nm at the dipole point (center of the gap), inside and outside the metal. To minimize boundary reflections,
we used
$5$ perfectly matched layers (PMLs) with a total thickness of $300$ nm,
which was found to be well converged numerically.

The numerical Purcell factors with full dipole method is defined as follows
(the analytic QNM one is  defined through
Eq.~\eqref{QNMpurcell} in Sec. \ref{subSec2.1}):
\begin{equation}\label{Purcellfulldipole}
    F_{\rm P}^{\rm num}(\mathbf{r}_{0},\omega)=\frac{\int_{\rm S}\hat{\mathbf{n}}\cdot {\bf S}_{\rm dipole,total}(\mathbf{r},\omega)d{\rm A} }{\int_{\rm S}\hat{\mathbf{n}}\cdot {\bf S}_{\rm dipole,background}(\mathbf{r},\omega)d{\rm A} },
\end{equation}
where $\rm S$ is a small spherical surface (with radius $1$ nm) surrounding dipole point  and $\hat{\mathbf{n}}$ is a unit vector normal to $\rm S$, pointing outward.
The vector ${\bf S}(\mathbf{r},\omega)$ is the Poynting vector at this small surface and the subscript `total' and `background' represent the case with and without resonator.
The excellent agreement with the Purcell factors using the QNM method
(Eq.~\eqref{QNMpurcell}) and full dipole method (Eq.~\eqref{Purcellfulldipole}) indicate the validity of the QNM results (see Figs. \ref{agree1}, \ref{agree2}).

In addition to the full-dipole numerical Purcell factors, the numerical radiative beta factor (assuming single QNM behaviour) is defined as
\begin{equation}\label{betaradfull}
    \beta_{\rm num}^{\rm rad}(\mathbf{r}_{0},\omega)=\frac{\int_{\rm S'}\hat{\mathbf{n}}\cdot {\bf S}_{\rm PML,total}(\mathbf{r}',\omega)d{\rm A'} }{\int_{\rm S}\hat{\mathbf{n}}\cdot {\bf S}_{\rm dipole,total}(\mathbf{r},\omega)d{\rm A} },
\end{equation}
where $\rm S'$ is the interface of PML and internal module, and ${\bf S}_{\rm PML,total}(\mathbf{r}',\omega)$ is the Poynting vector at this interface.
Similarly, the numerical nonradiative beta factor is
\begin{equation}\label{betanradfull}
\beta_{\rm num}^{\rm nrad}(\mathbf{r}_{0},\omega)=1-\beta_{\rm num}^{\rm rad}(\mathbf{r}_{0},\omega).
\end{equation}
Note that in contrast to the quantum
beta factors (Eqs.~\eqref{beta_rad_quan}-\eqref{beta_nrad_quan}), the classical beta factors
are frequency dependent, but are most important
near ${\omega_{\rm c}}$.

\subsection{Calculation of the regularized QNM  fields
using  a near-field to far-field transformation}\label{subSec3.2}

\begin{figure}[hbt]
  \centering
  \includegraphics[width=0.99\columnwidth]{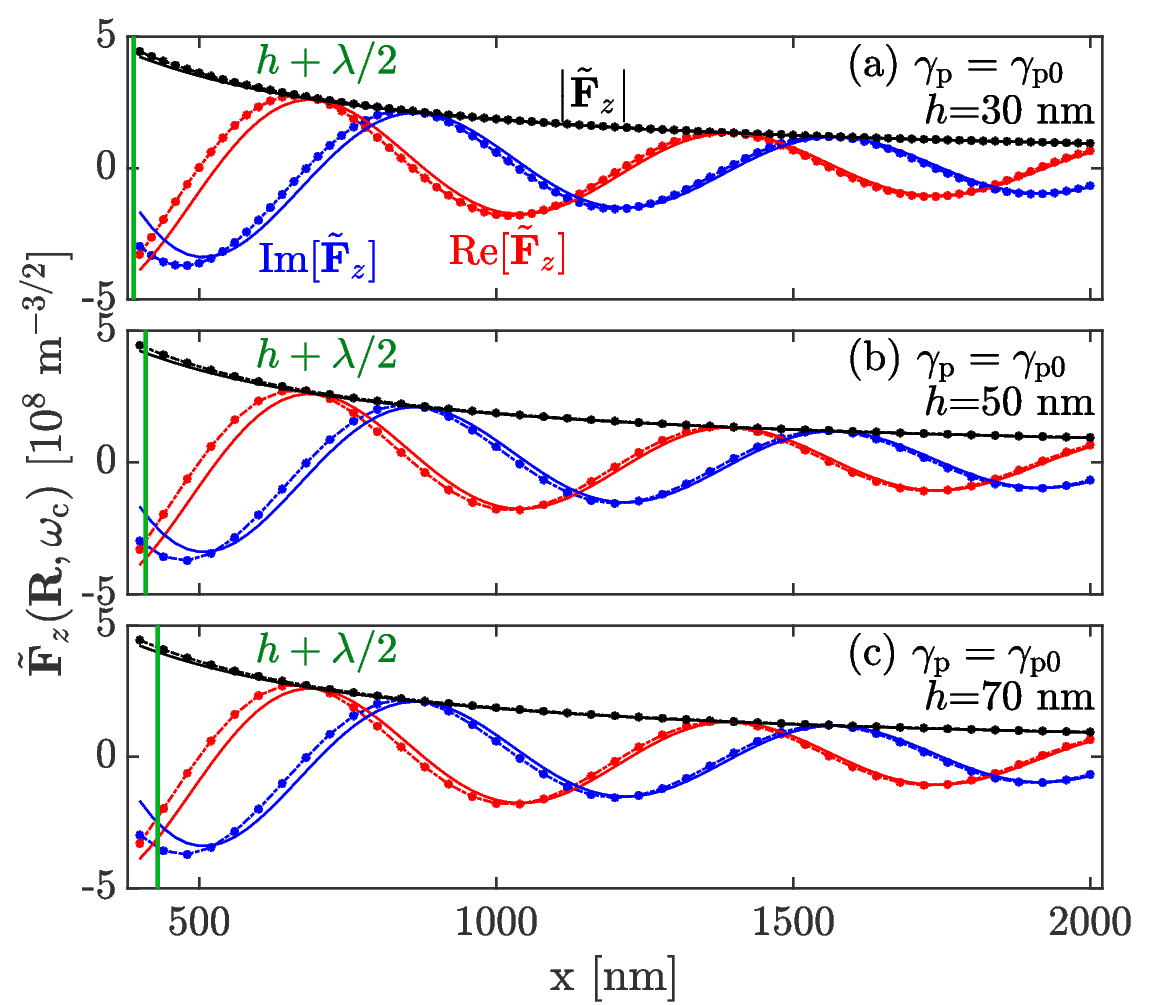}
  \caption{Comparison of $\tilde{\mathbf{F}}_{z}(\mathbf{R},\omega_{\rm c})$ (the real part, the imaginary part, and the absolute value) obtained from Dyson equation (solid line, Eq. \eqref{Dysoneq}) and NF2FF transformation (dotted line, Eq.~\eqref{bigFnear2far}) for $\gamma_{\rm p}=\gamma_{\rm p0}$. Here $y=z=0$, thus $\mathbf{R}=(x,0,0)$. The two approaches are seen to agree very well
  after about one wavelength outside the resonator ($\lambda$).
  }\label{F2}
\end{figure}

In order to verify the accuracy of the NF2FF transformation, and to confirm that it works correctly,  we compare the $\tilde{\mathbf{F}}(\mathbf{R},\omega_{\rm c})$ obtained from the NF2FF transformation with
the Dyson equation as selected spatial points outside the resonator, as shown in Fig.~\ref{F2}. As expected, in the far field zone, these
fields compare extremely well.

\begin{figure}[htp]
  \centering
  \includegraphics[width=0.6\columnwidth]{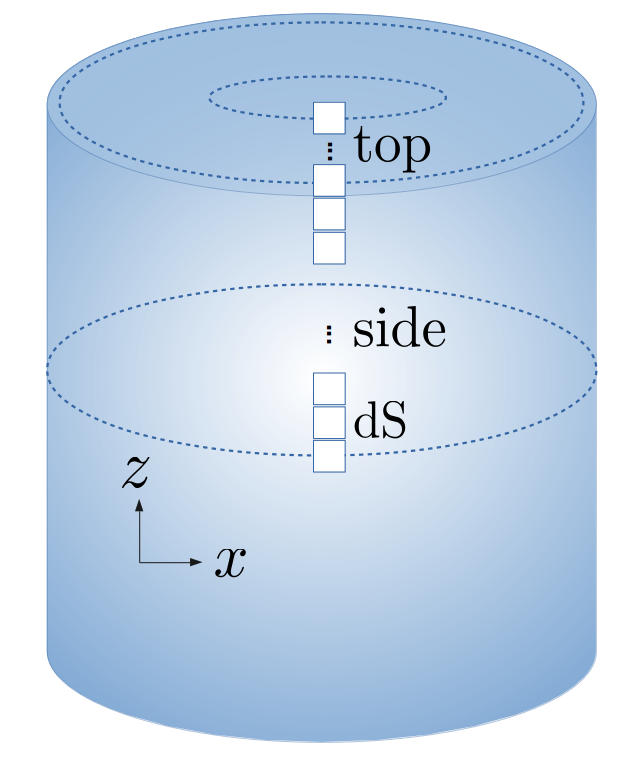}
  \caption{Schematic of the simplification  that can be used for calculating $S^{\rm rad}$ (Eq. \eqref{Snrad_full}) and $S^{\rm rad}_{\rm p1}$ (Eq. \eqref{Sradpole1}) using symmetry of QNM field (dimer).
  }\label{simple}
\end{figure}

The computational  run time of the NF2FF transformation, for a spatial single point, is $0.44$ minutes from the surface $h=50$ nm (grid size $0.5$ nm), and $0.24$ minutes using $h=30$ nm (grid size $0.5$ nm). However, since some of the vector potential points needed for the transformation are also used at other points, the scaling to more points is much faster than linear. For example, using an average over 37 points,  it take about $0.22$ minutes per point from the surface $h=50$ nm.
Using spatial points inside the resonator
with a grid size of 0.2\,nm, then the Dyson equation
takes about 10 minutes for a single point.
In the following calculation of $S^{\rm rad}$, we need to calculate many points on a surface, as shown in Fig.~\ref{simple}, and we give the total run time needed for calculations using
Matlab on a single computer workstation. More detailed computational run times are given later.

\begin{figure}[htp]
\centering
\includegraphics[width=0.99\columnwidth]{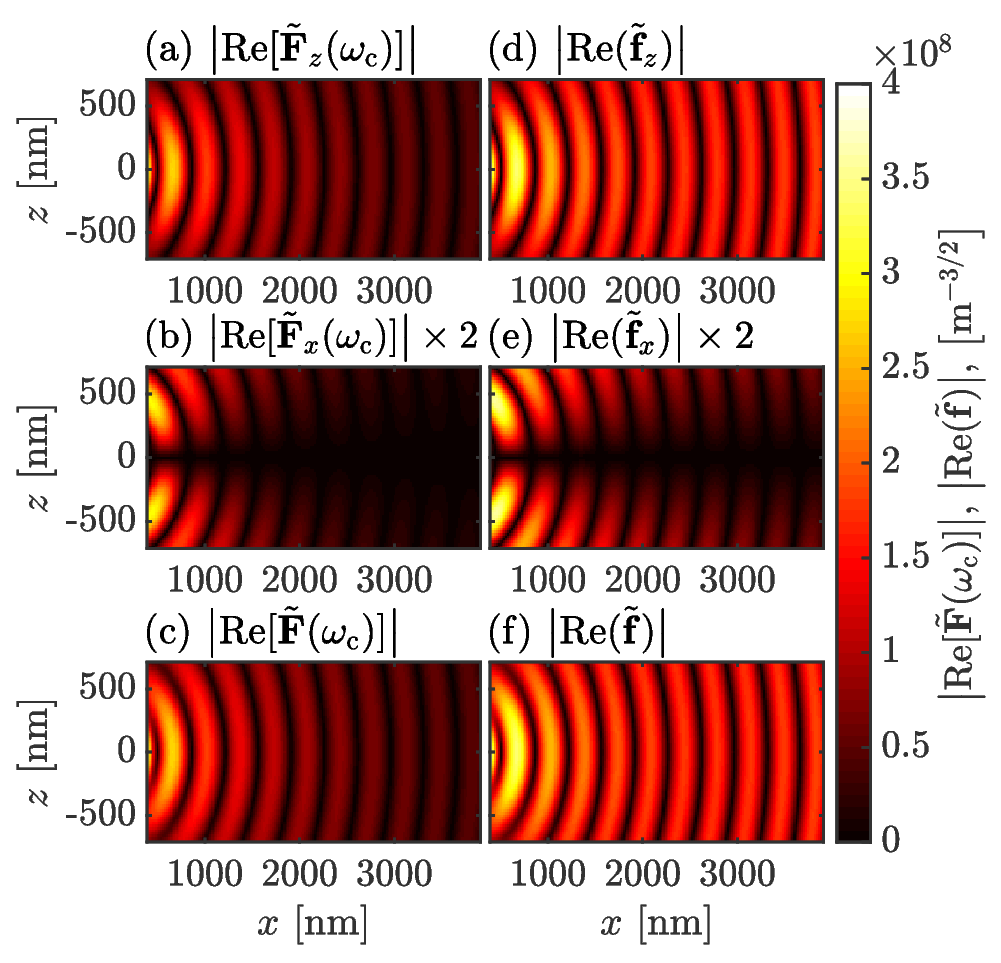}
\caption{(a)~$\big|{\rm Re}[\tilde{\mathbf{F}}_{z}(\omega_{\rm c})]\big|$, (b)~$2\big|{\rm Re}[\tilde{\mathbf{F}}_{x}(\omega_{\rm c})]\big|$, (c)~$\big|{\rm Re}[\tilde{\mathbf{F}}(\omega_{\rm c})]\big|$,
(d)~$\big|{\rm Re}(\tilde{\mathbf{f}}_{z})\big|$,
(e)~$2\big|{\rm Re}(\tilde{\mathbf{f}}_{x})\big|$, and (f)~$\big|{\rm Re}(\tilde{\mathbf{f}})\big|$
at plane $y=0$ nm for $\gamma_{\rm p}=\gamma_{\rm p0}$, using the NF2FF transformation (Eq. \eqref{bigFnear2far}) from near field surface $h=50$ nm. Ranges of $z$ and $x$ are $(-\lambda,\lambda)$ and $(h+0.5\lambda,h+5.5\lambda)$, where $\lambda\approx700$~nm. Note the  $z$-component is the dominant one, and the $y$-component can be ignored.}\label{bigF_xz}
\end{figure}

\begin{figure}[htp]
  \centering
  \includegraphics[width=0.99\columnwidth]{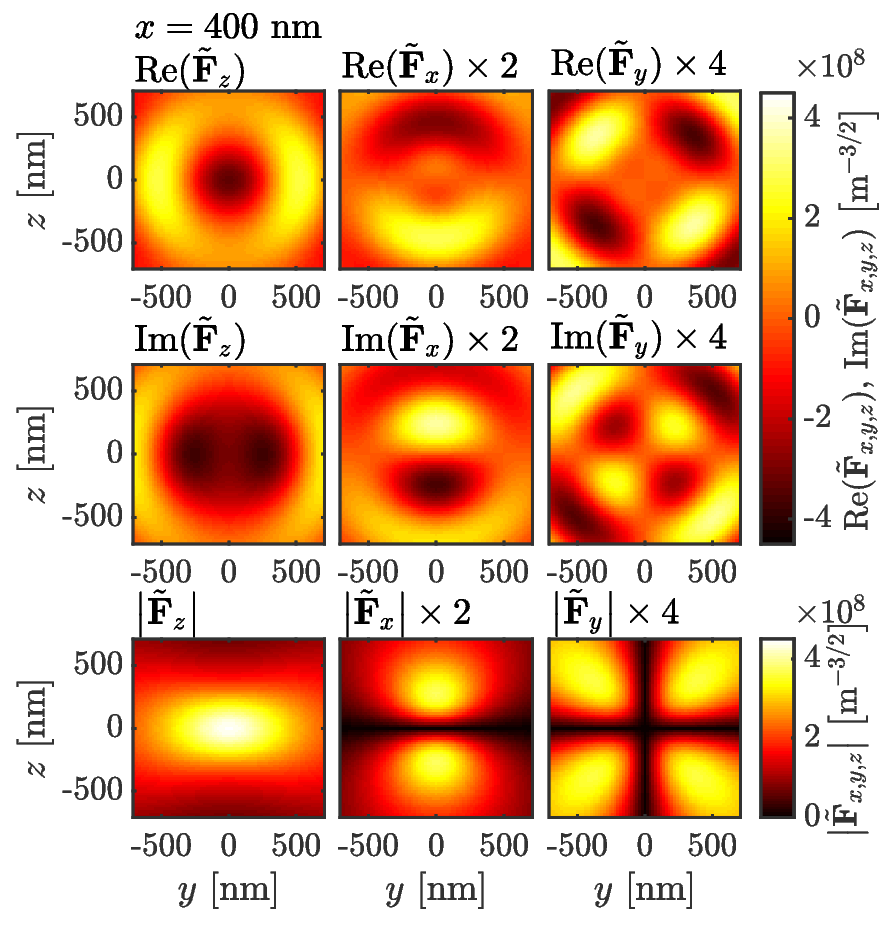}
  \caption{$\tilde{\mathbf{F}}_{z}$, $\tilde{\mathbf{F}}_{x}$ and $\tilde{\mathbf{F}}_{y}$ at plane $x=400$ nm for $\gamma_{\rm p}=\gamma_{\rm p0}$ using NF2FF transformation (Eqs.~\eqref{bigFnear2far}) from near field surface $h=50$ nm. Ranges of $z$ and $x$ are $(-\lambda,\lambda)$, where $\lambda\approx700$~nm. $z$-component is the dominant one.
  The $x$ and $y$ components are multiplied by a factor of $2$ and $4$ to better display the field distribution.
  }\label{bigF_yz_x400}
\end{figure}

\begin{figure}[htp]
  \centering
  \includegraphics[width=0.99\columnwidth]{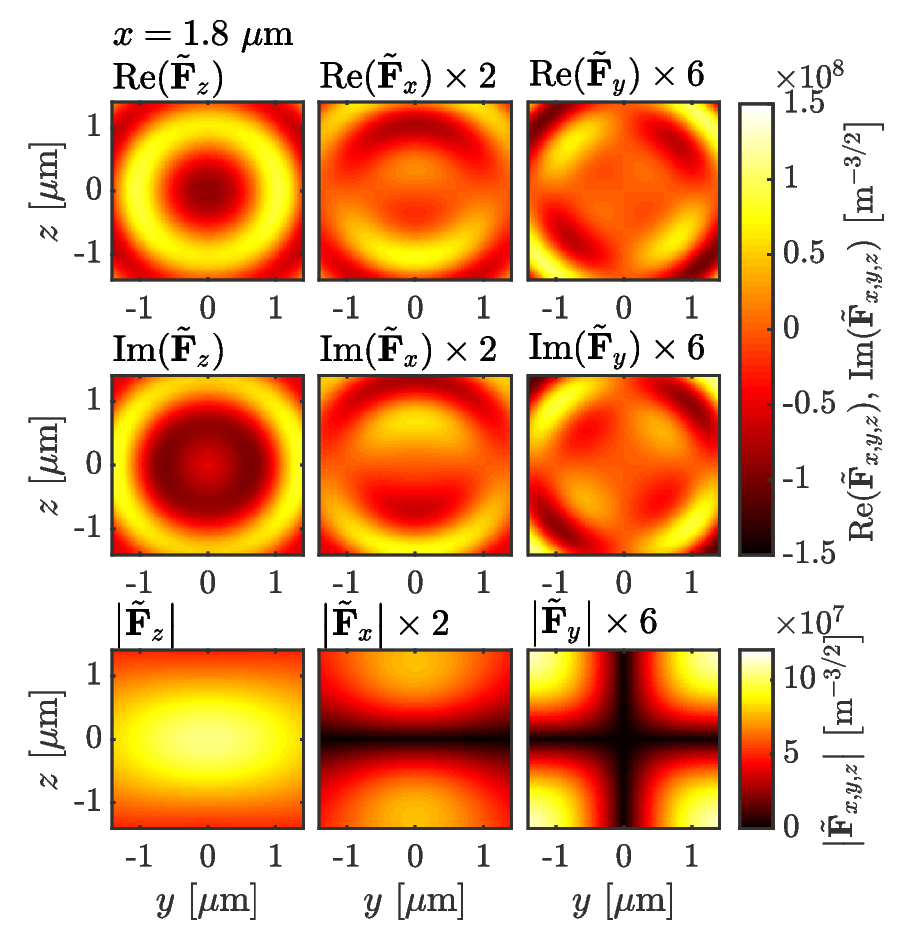}
  \caption{$\tilde{\mathbf{F}}_{z}$, $\tilde{\mathbf{F}}_{x}$ and $\tilde{\mathbf{F}}_{y}$ at plane $x=1800$ nm for $\gamma_{\rm p}=\gamma_{\rm p0}$ using NF2FF transformation (Eqs.~\eqref{bigFnear2far}) from the near field surface $h=50$ nm. The ranges of $z$ and $x$ are $(-2\lambda,2\lambda)$, where $\lambda\approx700$~nm. $z$-component is the dominant one.
  The $x$ and $y$ components are multiplied by a factor of $2$ and $6$ to better display the field distribution. }
  \label{bigF_yz_x1800}
\end{figure}

Using a NF2FF transformation from a near field surface at $h=50$ nm,
we also display surface plots of $\big|{\rm Re}(\tilde{\mathbf{F}}_i(\omega_{\rm c}))\big|$, $\big|{\rm Re}(\tilde{\mathbf{f}}_i)\big|$ in Fig. \ref{bigF_xz},
where we show the  $z$ and $x$ components  at the $xz$ plane ($y=0$ nm); the $y$-component can be ignored since it is much smaller than other two components. To better display $z$ and $x$ components with the same scale,
the $x$ component is multiplied by a factor of $2$.
The ranges of $z$ and $x$ are $(-\lambda,\lambda)$ and $(h+0.5\lambda,h+5.5\lambda)$, where $\lambda\approx700$~nm;
$\big|{\rm Re}(\tilde{\mathbf{f}})\big|$ show increasing behavior (eventually divergent) while $\big|{\rm Re}(\tilde{\mathbf{F}}(\omega_{\rm c}))\big|$ is convergent. Also, these fields show periodic distribution along $x$ direction with a period.
%
Moreover, we also show $\tilde{\mathbf{F}}$ at two $yz$ surfaces with $x=400$ nm$=h+0.5\lambda$ (Fig. \ref{bigF_yz_x400}) and $x=1800$ nm$=h+2.5\lambda$ (Fig. \ref{bigF_yz_x1800}) for $\gamma_{\rm p}=\gamma_{\rm p0}$ using NF2FF transformation (Eqs.~\eqref{bigFnear2far}) from near field surface $h=50$ nm.
These  far field regions are directly related to the output fields that experiments can detect. However, importantly,
these fields are obtained directly from the QNMs, and they also remain orthogonal to each other, which
is precisely why we call them QNM regularized fields or QNM reservoir fields.



\subsection{Calculation of
\texorpdfstring{$S^{\rm nrad}$}{Lg}
using the QNM fields}\label{subSec3.3}

To calculate $S^{\rm nrad}$ (Eq. \eqref{Snrad_full}) and $S^{\rm nrad}_{\rm p}$ (Eq. \eqref{eq:Snradapprox}), a spatial volume integration is needed, within the metal.
Below, we performed two approaches to obtain the numerical space integration:
(1) the normalized QNM fields were extracted from COMSOL with some extrapolated mesh size, then the calculation was performed in Matlab; or (2)
the spatial integration was performed directly in COMSOL with its own grid selection, which will be more accurate because there is no need to artificially choose the grid,
minimizing interpolation errors.

\begin{figure}[tp]
  \centering
  \includegraphics[width=8cm]{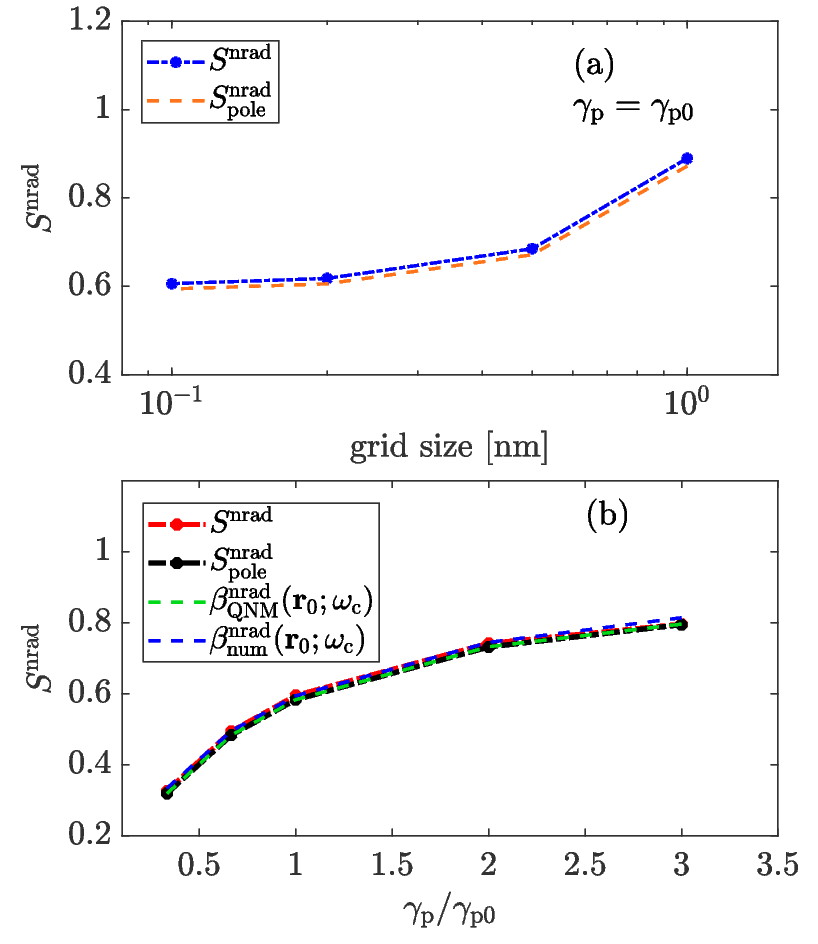}
  \caption{ (a) Numerical grid dependence of $S^{\rm nrad}$ (Eq. \eqref{Snrad_full}) and $S^{\rm nrad}_{\rm p}$ (Eq. \eqref{eq:Snradapprox}) for $\gamma_{\rm p}=\gamma_{\rm p0}$. If the spatial volume integration is directly performed in COMSOL (i.e., no need to artificially choose the grid), the results are $S^{\rm nrad}=0.5947$ and $S^{\rm nrad}_{\rm p}=0.5830$, which is very close to the results with grid size of $0.1$ nm. (b) $S^{\rm nrad}$ (Eq. \eqref{Snrad_full}), $S^{\rm nrad}_{\rm p}$ (Eq. \eqref{eq:Snradapprox}), $\beta_{\rm QNM}^{\rm nrad}$ (Eq. \eqref{betanradQNMsingle}) from single QNM, and $\beta_{\rm num}^{\rm nrad}$ (Eq. \eqref{betanradfull}) from full dipole method for various losses, where only the results directly from COMSOL are presented. Once gaain, the
  full dipole simulations are performed at gap center.
}\label{Snrad_plot}
\end{figure}

As shown in Fig.~\ref{Snrad_plot} (a), we tested four different grid sizes;
as expected, smaller grids lead to more accurate calculations
and eventual convergence of the integral.
If we use the second approach, we obtain $S^{\rm nrad}=0.595$ and $S^{\rm nrad}_{\rm p}=0.583$, which is very close to the results from first approach with
a grid size of $0.1$ nm, and is also very close to the result of $0.58$, reported  in Ref.~\onlinecite{franke_quantization_2018}.

Moreover, as shown in Fig.~\ref{Snrad_plot} (b), $S^{\rm nrad}$ and $S^{\rm nrad}_{\rm p}$ increase with larger material losses, where only the results directly from COMSOL are presented.
Meanwhile, $\beta_{\rm QNM}^{\rm nrad}$ (Eq. \eqref{betanradQNMsingle}) from single QNM, and $\beta_{\rm num}^{\rm nrad}$ (Eq. \eqref{betanradfull}) from the full dipole method are also shown in Fig. \ref{Snrad_plot} (b), which are very close to each other, indicating that single mode approximation is an excellent one for these resonators (and also confirms the accuracy of our numerical calculations).
Furthermore, these two classical nonradiative beta factors are very close to the quantum $S^{\rm nrad}$ and $S^{\rm nrad}_{\rm p}$.

\begin{table}[b]
\caption {Spatial integration $S^{\rm nrad}|_{\rm space}$ (Eq.~\eqref{Space}), quoted to the third decimal place } \label{space_table}
    \centering
    \begin{tabular}{|c|c|c|c|c|c|}
 \hline
 spatial integral & $3\gamma_{\rm p0}$ & $2\gamma_{\rm p0}$ & $\gamma_{\rm p0}$ & $(2/3)\gamma_{\rm p0}$& $(1/3)\gamma_{\rm p0}$  \\
 \hline
  grid $1$ nm & $0.059$ & $0.059$ & $0.059$ & $0.059$ & $0.059$ \\
 \hline
 grid $0.5$ nm & $0.045$& $0.045$ & $0.045$ & $0.045$ & $0.045$ \\
 \hline
 grid $0.2$ nm & $0.041$ & $0.041$ & $0.041$ &$0.041$ & $0.041$ \\
 \hline
 grid $0.1$ nm & $0.040$ & $0.040$ &$0.040$ &$0.040$  & $0.040$ \\
 \hline
 COMSOL & $0.039$   & $0.039$ & $0.039$  & $0.039$  & $0.039$ \\
\hline
    \end{tabular}
\end{table}

The spatial integration
contribution to $S^{\rm nrad}$, is defined through:
\begin{equation}\label{Space}
S^{\rm nrad}|_{\rm space}  =\int_{V}d\mathbf{r}|\tilde{\mathbf{f}}_{\rm c}(\mathbf{r})|^{2},
\end{equation}
which was found to be very similar for all the
five cases studies, as shown in Table \ref{space_table}.
Thus, according to Eq. \eqref{eq:Snradapprox}, the corresponding $S_{\rm p}^{\rm nrad}$ will be proportional to $Q \epsilon_{\rm I}(\omega_{\rm c})$.
If $\gamma_{\rm p}$ ($0.0928$ eV) is much smaller than $\omega_{\rm c}$ ($1.7798$ eV), then
\begin{equation}
\epsilon_{\rm I}(\omega_{\rm c}) =\frac{\omega_{\rm p}^2\gamma_{\rm p}}{\omega_{\rm c}^3+\omega_{\rm c}\gamma_{\rm p}^2}\approx\frac{\omega_{\rm p}^2\gamma_{\rm p}}{\omega_{\rm c}^3}.
\end{equation}
Since $\omega_{\rm c}$ for these five cases are also very close,
we expect that $S_{\rm p}^{\rm nrad}$ will be proportional to $Q\gamma_{\rm p}$.
We calculate $S_{\rm p}^{\rm nrad}=0.5830$ for $\gamma_{\rm p}=\gamma_{\rm p0}$, and can estimate $S_{\rm p}^{\rm nrad} \approx 0.805, 0.736, 0.483, 0.319$ for the other four cases, which are very close to the
full computed values $S_{\rm p}^{\rm nrad}$ ($0.790, 0.731, 0.483, 0.319$) shown in
Fig.~\ref{Snrad_plot} (b); as expected,
more accurate agreements with this simpler scaling
argument  is obtained for smaller $\gamma_{\rm p}$ i.e., larger $Q_{\rm c}$.

\subsection{Calculation of \texorpdfstring{$S^{\rm rad}$}{Lg}
using the regularized QNMs}\label{subSec3.4}

\subsubsection{Pole approximation \texorpdfstring{$S_{\rm p1}^{\rm rad}$}{Lg} (\texorpdfstring{Eq. \eqref{Sradpole1}}{Lg}) }

Next we use the  NF2FF approach to calculate $S^{\rm rad}$, which is a
much more  involved numerical calculation.
We begin by considering on the first pole result $S_{\rm p1}^{\rm rad}$ (Eq. \eqref{Sradpole1}).
The grid size in the near field surface $h$ is set as $0.5$ nm;
the far field surface is fixed at $h_{\rm far}=630$ nm, and the grid on that surface is $20$ nm (in both transverse directions).
The near field surface dependence of pole $S_{\rm p1}^{\rm rad}$ for various material loss cases are shown in Table \ref{NF_p1}.
$S_{\rm p1}^{\rm rad}$ from $h=20$ nm, $h=30$ nm, and $h=50$ nm are very close, which appears to be more robust as expected. Deviations then
start to occur for distances
greater than 70\,nm or so, as the QNM is no longer
a good approximation to use for the near field
currents.
In principle, the near field surface should be as close as possible to the metal surface.
However, numerically,
the fields close to the metal surface have a large gradient (and convergence problems at the metal surface), so it will (at least) need smaller grid size (smaller than $0.5$ nm now used) at near field surfaces to guarantee the accuracy of numerical results.
So in following calculations, we mainly choose $h=50$ nm as near field surface.
The run time is around $16$ minutes.

\begin{table}[h]
\caption {Near field surface dependence of $S_{\rm p1}^{\rm rad}$ (Eq. \eqref{Sradpole1}), quoted to the second decimal place for various material loss cases.} \label{NF_p1}
    \centering
    \begin{tabular}{|c|c|c|c|c|c|}
 \hline
 NF surface & $3\gamma_{\rm p0}$ & $2\gamma_{\rm p0}$ & $\gamma_{\rm p0}$ & $(2/3)\gamma_{\rm p0}$& $(1/3)\gamma_{\rm p0}$ \\
 \hline
 $h=10$ nm & $0.19$ & $0.26$ & $0.41$ & $0.51$ & $0.68$ \\
 \hline
 $h=20$ nm & $0.19$ & $0.26$ & $0.42$ & $0.52$ & $0.68$ \\
 \hline
 $h=30$ nm & $0.19$ & $0.26$ & $0.42$ & $0.52$ & $0.68$ \\
 \hline
 $h=50$ nm & $0.20$ & $0.27$ & $0.42$ & $0.52$ & $0.69$ \\
 \hline
 $h=70$ nm & $0.21$ & $0.28$ & $0.44$ & $0.54$ & $0.70$ \\
\hline
    \end{tabular}
\caption {Influence of far field location on
the pole calculation for  $S_{\rm p1}^{\rm rad}$ (Eq. \eqref{Sradpole1}),  using $\gamma_{\rm p}=\gamma_{\rm p0}$. Here the near field surface is fixed at $h=50$ nm and resonance wavelength is about $\lambda\sim700$ nm ($1.78$ eV). When the propagation distance $h_{\rm far}-h$ is larger than $0.5\lambda$, $S_{\rm p1}^{\rm rad}$ is well converged. The values are quoted to the second decimal place} \label{soutpolel1_far}
    \centering
    \begin{tabular}{|c|c|}
 \hline
  $h_{\rm far}=410$ nm &  $0.42$  \\
 \hline
  $h_{\rm far}=590$ nm &  $0.42$  \\
 \hline
$h_{\rm far}=630$ nm &  $0.42$  \\
 \hline
 $h_{\rm far}=750$ nm &  $0.42$ \\
 \hline
 $h_{\rm far}=990$ nm &  $0.42$  \\
\hline
    \end{tabular}
\end{table}

In Table \ref{soutpolel1_far},  we
summarize
the impact of
 the far field surface $h_{\rm far}$ selection on the  $S_{\rm p1}^{\rm rad}$ (Eq. \eqref{Sradpole1}), with the near field surface fixed at $h=50$ nm.
The resonance wavelength here is about $\lambda\sim700$ nm ($1.78$ eV).
With $\gamma_{\rm p}=\gamma_{\rm p0}$, choosing far field surfaces at $h_{\rm far}=410,~590,~630,~750,~990$ nm gave the same pole result
of $S_{\rm p1}^{\rm rad}=0.42$.

\subsubsection{The second pole approximation \texorpdfstring{$S_{\rm p2}^{\rm rad}$}{Lg} (\texorpdfstring{Eq. \eqref{Sradpole2}}{Lg}) with field equivalence}

Next we considerthe second pole result  $S_{\rm p2}^{\rm rad}$ from Eq.~\eqref{Sradpole2}.
There are several influencing factors when performing the
numerical integrals in $I_c$ (Eq.~\eqref{I_sur_c}), including the selection of near field surface ($S'$) and angle grid size for angle integral.
Here we fix the grid size in near field surface with a spacing of $0.5$ nm,
and  we use the same angle grid for integration over both $\vartheta$ and $\varphi$.

\begin{table}[hb]
\caption {Pole calculation for $S_{\rm p2}^{\rm rad}$ from Eq. \eqref{Sradpole2} with $h=50$ nm and $\gamma_{\rm p}=\gamma_{\rm p0}$, quoted to the second decimal place. } \label{sout_eq44}
    \centering
    \begin{tabular}{|c|c|c|}
    \hline
near field surface  $h=50$ nm &    &   \\
 \hline
 angle grid for $\vartheta$ and $\varphi$ & run time   & pole $S_{\rm p2}^{\rm rad}$  \\
 \hline
 $\pi/2$    & $3.2$ secs   & $0.51$  \\
 \hline
 $\pi/3$    & $4.8$ secs  & $0.42$  \\
 \hline
 $\pi/5$    & $8.7$ secs   & $0.42$  \\
 \hline
 $\pi/10$   & $25.3$ secs   & $0.42$  \\
 \hline
    \end{tabular}
%
\caption {Pole calculation for  $S_{\rm p2}^{\rm rad}$ from Eq. \eqref{Sradpole2} with $h=30$ nm and $\gamma_{\rm p}=\gamma_{\rm p0}$, quoted to the second decimal place.  } \label{sout_eq44_1}
    \centering
    \begin{tabular}{|c|c|c|}
    \hline
near field surface  $h=30$ nm &    &   \\
 \hline
 angle grid for $\vartheta$ and $\varphi$ & run time    & pole $S_{\rm p2}^{\rm rad}$  \\
 \hline
 $\pi/2$    & $1.9$ secs   & $0.51$  \\
 \hline
 $\pi/3$    & $2.8$ secs   & $0.42$  \\
 \hline
 $\pi/5$    & $4.9$ secs   & $0.42$  \\
 \hline
 $\pi/10$   & $13.9$ secs   & $0.42$  \\
 \hline
    \end{tabular}
\end{table}

The main numerical results
for obtaining the pole calculation $S_{\rm p2}^{\rm rad}$ (Eq.~\eqref{Sradpole2})
are shown in Tables \ref{sout_eq44} and \ref{sout_eq44_1},
showing a convergent solutions of $0.42$ (quoted to the second decimal place) with near field surface $h=$50 nm and 30 nm for $\gamma_{\rm p}=\gamma_{\rm p0}$. These also
 agree with the first pole result $S_{\rm p1}^{\rm rad}$ of Eq.~\eqref{Sradpole1} (Table \ref{NF_p1}), but with a significantly faster run time (around $100-200$ times faster).

Moreover, the results of $S_{\rm p2}^{\rm rad}$ (Eq.~\eqref{Sradpole2}) for various material losses and with fixed near field surface $h=50$ nm (grid size is $0.5$ nm) are shown in Table \ref{compare}, which increase with the decrease of the loss, and they are very close to the corresponding $S_{\rm p1}^{\rm rad}$ (Eq.~\eqref{Sradpole1}).
 We have also found that pole $S_{\rm p2}^{\rm rad}$ always converges at a relatively large angle grid size, and thus the calculation is extremely fast, only a few seconds to several tens of seconds.

\subsubsection{\texorpdfstring{ Computing $S^{\rm rad}$}{Lg} (\texorpdfstring{Eq. \eqref{Sradfull}}{Lg}) with a numerical frequency integration versus the two pole approximations \texorpdfstring{$S_{\rm p1}^{\rm rad}$ (Eq.~\eqref{Sradpole1})}{Lg} and \texorpdfstring{$S_{\rm p2}^{\rm rad}$ (Eq.~\eqref{Sradpole2})}{Lg}}

\begin{figure}[htp]
  \centering
  \includegraphics[width=0.99\columnwidth]{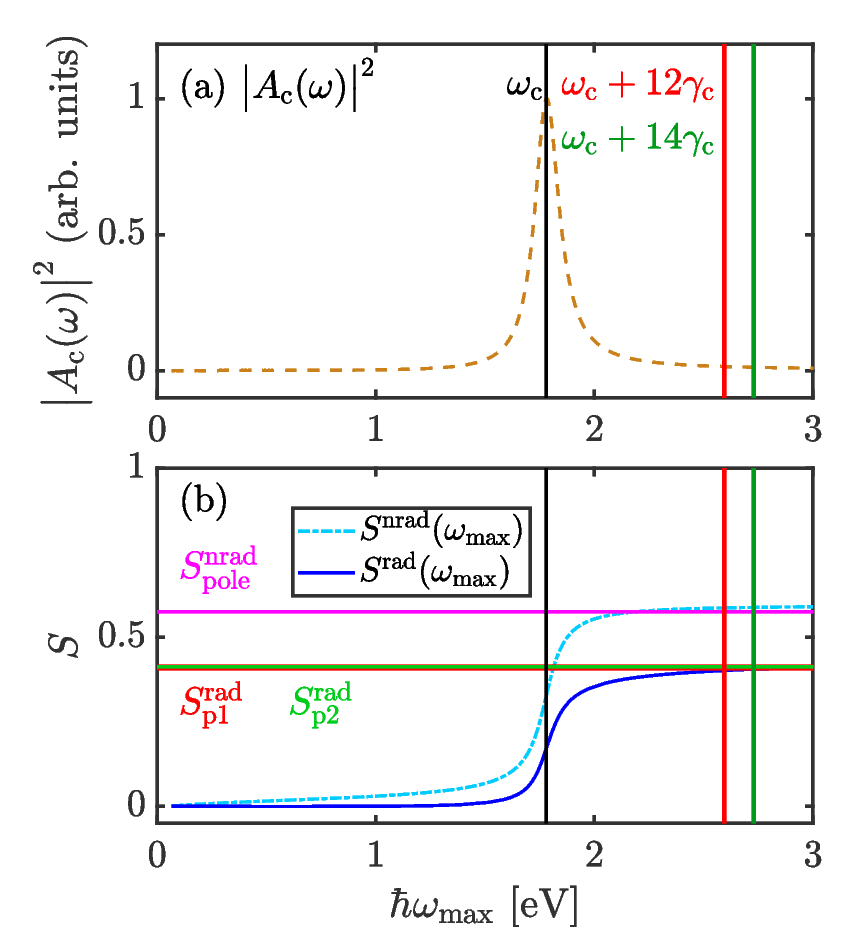}
  \caption{(a) Plot of $|A_{\rm c}(\omega)|^2$
  versus frequency (normalized to $1$). The black, red, and green perpendicular lines indicate the frequency position at $\omega_{\rm c}$, $\omega_{\rm c}+12\gamma_{\rm c}$, $\omega_{\rm c}+14\gamma_{\rm c}$. (b) Full frequency integral $S^{\rm nrad}$ (Eq. \eqref{Snrad_full}), pole result $S^{\rm nrad}_{\rm p}$ (Eq. \eqref{eq:Snradapprox}), full frequency result $S^{\rm rad}$ (Eq. \eqref{Sradfull}), pole result $S_{\rm p1}^{\rm rad}$ (Eq. \eqref{Sradpole1}) and pole results $S_{\rm p2}^{\rm rad}$ (Eq. \eqref{Sradpole2}) for $h$=50 nm and $h_{\rm far}=630$ nm with $\gamma_{\rm p}=\gamma_{\rm p0}$. In detail, $S^{\rm nrad}_{\rm p}=0.583$. $S^{\rm nrad}(\omega_{\rm max}=\omega_{\rm c}+12\gamma_{\rm c})=0.587$. $S^{\rm nrad}(\omega_{\rm max}=\omega_{\rm c}+14\gamma_{\rm c})=0.589$. $S_{\rm p1}^{\rm rad}=0.423$. $S_{\rm p2}^{\rm rad}=0.418$. $S^{\rm rad}(\omega_{\rm max}=\omega_{\rm c}+12\gamma_{\rm c})=0.402$. $S^{\rm rad}(\omega_{\rm max}=\omega_{\rm c}+14\gamma_{\rm c})=0.406$.}
   \label{sout_full_l1_f}
\end{figure}

Next, we carry out the full frequency integration results $S^{\rm rad}$ (Eq.~\eqref{Sradfull}), again using the NF2FF transformation.
Figure \ref{sout_full_l1_f} (a) show the normalized function $|A_{\rm c}(\omega)|^2$
  versus frequency.
The black, red, and green perpendicular lines indicate the frequency position at $\omega_{\rm c}$, $\omega_{\rm c}+12\gamma_{\rm c}$, $\omega_{\rm c}+14\gamma_{\rm c}$.
The integrated frequency region $(0,~\omega_{\rm c}+12\gamma_{\rm c})$ and $(0,~\omega_{\rm c}+14\gamma_{\rm c})$ cover the vast majority of
the QNM lineshape, as discussed earlier.
Figure \ref{sout_full_l1_f} (b) show the full frequency integration results vs the pole results with $\gamma_{\rm p}=\gamma_{\rm p0}$, $h=50$ nm and $h_{\rm far}=630$ nm.

With regards to the pole results,
for the nonraditive contribution,
 $S^{\rm nrad}_{\rm p}=0.583$ (Eq. \eqref{Snrad_full}) (magenta horizontal line). Cyan line show $S^{\rm nrad}$ as a function of the maximum integral frequency;
integrating over frequency, then
we obtain $S^{\rm nrad}(\omega_{\rm max}=\omega_{\rm c}+12\gamma_{\rm c})=0.587$. $S^{\rm nrad}(\omega_{\rm max}=\omega_{\rm c}+14\gamma_{\rm c})=0.589$.
These values are very close to the pole result, which confirms the accuracy of the pole approximation $S^{\rm nrad}_{\rm p}$.
For the radiative contribution,
$S_{\rm p1}^{\rm rad}=0.423$ (Eq. \eqref{Sradpole1}) (red horizontal line), $S_{\rm p2}^{\rm rad}=0.418$ (Eq. \eqref{Sradpole2}) (green horizontal line), while the blue curve shows $S^{\rm rad}$ (Eq. \eqref{Sradfull}) as a function of the maximum integral frequency (56 frequency points are used).
These results confirm the accuracy of both two pole approximations $S^{\rm rad}_{\rm p1}$ and $S^{\rm rad}_{\rm p2}$, though the
latter is considerably more efficient.

\subsubsection{Summary of quantum \texorpdfstring{$S$}{Lg} parameter for metal dimers}

Here we summarize the quantum pole $S$ parameters, $S^{\rm nrad}_{\rm p}$ (Eq. \eqref{eq:Snradapprox}) and $S_{\rm p}^{\rm rad}=S_{\rm p2}^{\rm rad}$ (Eq. \eqref{Sradpole2}), for gold dimer
with different materials loss $\gamma_{\rm p}$ in Table \ref{compare}.
As the material loss $\gamma_{\rm p}$ decreases ($Q_{\rm c}$ increases), $S^{\rm nrad}_{\rm p}$ decreases and $S^{\rm rad}_{\rm p}$ increases.
Somewhat remarkably though, the total $S_{\rm p}=S^{\rm nrad}_{\rm p}+S^{\rm rad}_{\rm p}$ for these five cases are all close to $1.0$.
While this may be expected for a single QNM, a general proof is not known,
and the complexity of the numerical calculations
also confirm the general accuracy of the numerical implementation.


For completeness, we have also listed the quantum radiative beta factor $\beta^{\rm rad}_{\rm quan}=S^{\rm rad}_{\rm p}/S_{\rm p}$ (Eq. \eqref{beta_rad_quan}), classical  $\beta_{\rm QNM}^{\rm rad}=\beta_{\rm QNM}^{\rm rad}(\mathbf{r}_{0},\omega_{\rm c})$ (Eq. \eqref{betanradQNMsingle}) with single QNM, and classical $\beta_{\rm num}^{\rm rad}=\beta_{\rm num}^{\rm rad}(\mathbf{r}_{0},\omega_{\rm c})$ (Eq. \eqref{betaradfull}) with full dipole method in Table \ref{compare}.
The last two classical beta factor are evaluated at the pole frequency $\omega_{\rm c}$ for a emitter placed at dimer center $\mathbf{r}_{0}$.
We found that the quantum and classical radiative beta factor are relatively close to each other and they increase with $Q_{\rm c}$ increase. However, note that the classical estimation (with full dipole method) is really a total beta calculation as opposed to a single mode, but they are likely very close in this regime.

\begin{table}[hb]
\caption {Pole result for $S^{\rm nrad}_{\rm p}$ (Eq. \eqref{eq:Snradapprox}) and $S_{\rm p}^{\rm rad}=S_{\rm p2}^{\rm rad}$ (Eq. \eqref{Sradpole2}) with different materials loss $\gamma_{\rm p}$. The total $S_{\rm p}=S^{\rm nrad}_{\rm p}+S^{\rm rad}_{\rm p}$. The quantum radiative beta factor $\beta^{\rm rad}_{\rm quan}=S^{\rm rad}_{\rm p}/S_{\rm p}$. Here, classical beta factor $\beta_{\rm QNM}^{\rm rad}=\beta_{\rm QNM}^{\rm rad}(\mathbf{r}_{0},\omega_{\rm c})$ (Eq. \eqref{betanradQNMsingle}) and $\beta_{\rm num}^{\rm rad}=\beta_{\rm num}^{\rm rad}(\mathbf{r}_{0},\omega_{\rm c})$ (Eq. \eqref{betaradfull}), which are evaluated at the pole frequency $\omega_{\rm c}$ for a emitter placed at dimer center $\mathbf{r}_{0}$. All $S$ and beta factors are quoted to the second decimal place.} \label{compare}
    \centering
    \begin{tabular}{|c|c|c|c|c|c|c|c|c|}
 \hline
  $\gamma_{\rm p}$ & $Q_{\rm c}$ & $S^{\rm nrad}_{\rm p}$      & $S^{\rm rad}_{\rm p}$ & $S_{\rm p}$       & $\beta_{\rm quan}^{\rm rad}$ & $\beta_{\rm QNM}^{\rm rad}$& $\beta_{\rm num}^{\rm rad}$  \\
 \hline
 $3\gamma_{\rm p0}$  & $6.0$  & $0.79$ & $0.19$ & $0.99$   & $0.20$ & $0.20$ & $0.19$ \\
  \hline
 $2\gamma_{\rm p0}$  &$8.3$  &  $0.73$ & $0.26$ & $1.00$   & $0.27$ & $0.27$ & $0.26$ \\
 \hline
  $\gamma_{\rm p0}$ &$13.1$ & $0.58$ & $0.42$ & $1.00$   & $0.42$ &$0.42$  &$0.41$   \\
 \hline
  $\frac{2}{3}\gamma_{\rm p0}$ & $16.3$ &$0.48$ & $0.52$ & $1.00$&   $0.52$ & $0.52$ & $0.50$\\
 \hline
  $\frac{1}{3}\gamma_{\rm p0}$ &$21.5$ & $0.32$ & $0.68$ & $1.00$  & $0.68$ & $0.68$ & $0.67$ \\
 \hline
    \end{tabular}
\end{table}

\subsubsection{Quantum Purcell factor for the metal gold dimer}

\begin{figure}[!tp]
  \centering
  \includegraphics[width=0.99\columnwidth]{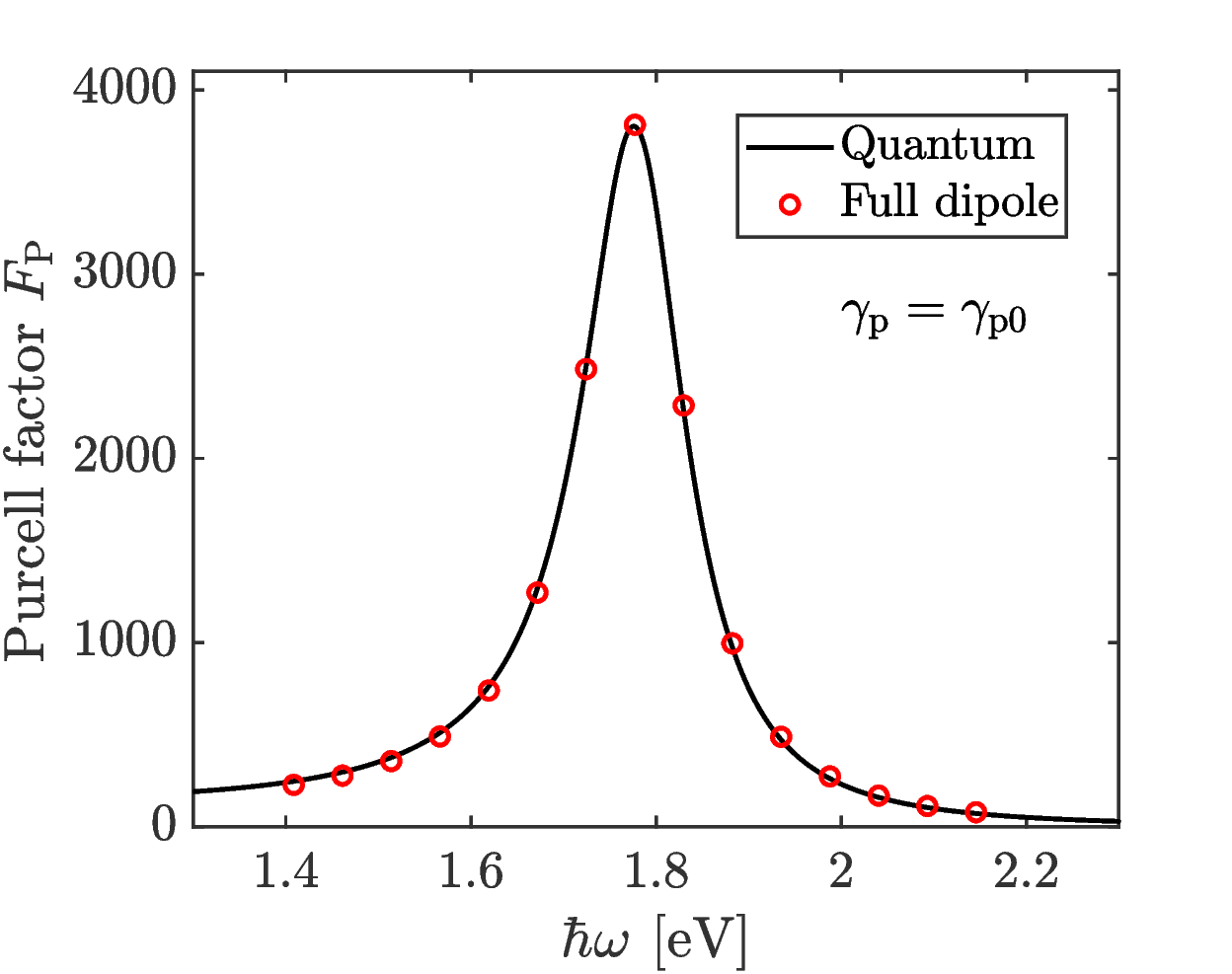}
  \caption{Quantum Purcell factor calculations for a $z$-polarized  dipole at dimer gap center, with $\gamma_{\rm p}=\gamma_{\rm p0}$ (gold), using the quantum formula (Eq.~\eqref{quantumpurcellsingle}), where pole $S_{\rm p}=1.0$ (shown in Table \ref{compare}). The quantum Purcell factors (black curve) show excellent agreement with the result from full dipole formula (red circle, Eq.~\eqref{Purcellfulldipole}).
 }\label{quansingle}
\end{figure}

For completeness, we now show the
quantum Purcell factor in the bad cavity limit,
using Eq.~\eqref{quantumpurcellsingle}.
Considering again a $z$-polarized dipole placed at gold cylindrical dimer gap center with $\gamma_{\rm p}=\gamma_{\rm p0}$,
then Fig.~\ref{quansingle}
shows the excellent agreement with the  full dipole formula Eq.~\eqref{Purcellfulldipole} (Fig. \ref{quansingle}).
Here, as shown in in Table \ref{compare},the calculated pole $S_{\rm p}=1.00$ is used.

\subsection{Run time comparison between the Dyson approach and the NF2FF approach to compute the quantum \texorpdfstring{$S$}{Lg} parameters}\label{subSec3.5}

Before showing the run times, we briefly clarify the simplification used to calculate $S^{\rm rad}$ (Eq. \eqref{Sradfull}) and the pole
result for $S^{\rm rad}_{\rm p1}$ (Eq. \eqref{Sradpole1}).
We need to perform an integral  involving $\tilde{\mathbf{F}}$ and $\tilde{\mathbf{H}}$ over a far field surface ($h_{\rm far}$), and
we also know that the radiation of the QNM we are working with is axial-symmetric to the long axis of the Au dimer, and is also symmetrical to the center plane perpendicular to the long axis.  Thus we can simplify this surface integral using symmetry (for this
specific problem at least).

As shown in Fig. \ref{simple}, we select a cylindrical surface as the far field surface.
The surface integral unit is labeled as $\rm dS$.
If we select the grid size of $20$ nm, then $\rm dS=$($20$ nm)$^{2}$.
On the one hand, because it is symmetrical to the center plane, we can just calculate the integral over upper part of the cylindrical surfacel, and simply multiply the result by the factor of $2$.
On the other hand, since it is axial-symmetric, we only need to
perform the integral over a line (both in side surface and top surface), and then multiply the results by the length of the circumference (it is actually circumference$/$grid size). Note that this simplification may not work for other resonator structures if they do not have a similar symmetry, such as the coupled QNM case shown in Section~\ref{Sec4}.

\subsubsection{Dyson equation run times for computing the quantum \texorpdfstring{$S$}{Lg} parameters}

Here we estimate typical computational
run times to obtain $S^{\rm rad}$ (Eq. \eqref{Sradfull}), $S^{\rm rad}_{\rm p1}$ (Eq. \eqref{Sradpole1}) and $S^{\rm rad}_{\rm p2}$ (Eq. \eqref{Sradpole2}) at Table \ref{runtime}.
First, for the Dyson approach, we selected the grid size as $0.2$ ($0.5$) nm for the spatial integration over
the metal volume. As mentioned in Sec.~\ref{subSec2.2}, the run time for a single $\tilde{\mathbf{F}}$ point at single frequency is about $\approx8.5-10$ ($0.6-0.7$) minutes.
We choose the far field cylindrical surface at $h_{\rm far}=630$ nm,
and the grid at this surface is $20$ nm. Exploiting the symmetry mentioned above, then $33$ and $37$ points (both $\tilde{\mathbf{F}}$ and $\tilde{\mathbf{H}}$) are needed for the top surface   (same for the bottom surface) and (half) side surface.
Also note that $\tilde{\mathbf{H}}$ need to be calculated from curl of $\tilde{\mathbf{F}}$.
Then $167$ and $187$ $\tilde{\mathbf{F}}$ points are needed for the top (same for down) surface and (half) side surface of the far field.
Using the $S^{\rm rad}$ (Eq.~\eqref{Sradfull}) results with a numnerical frequency integration, we use $56$ frequency points (this is a typical number, and $51$ points used in Ref.~\onlinecite{franke_quantization_2018}.)
Thus, with inside grid of $0.2$ ($0.5$) nm, it will take about $8.5\times354\times56$ minutes $\approx117$ days ($0.6\times354\times56$ minutes $\approx8.3$ days) for $S^{\rm rad}$ (Eq.~\eqref{Sradfull}), and $8.5\times354$ minutes $\approx2.1$ days ($0.6\times354$ minutes $\approx3.5$ hours) for the
pole result $S^{\rm rad}_{\rm p1}$ (Eq.~\eqref{Sradpole1}).

\subsubsection{Near-field to far-field run times for computing the quantum \texorpdfstring{$S$}{Lg} parameters}

Next, if we employ the NF2FF transformation, we show how the run time will be greatly reduced. For example,
considering a grid size of $0.5$ nm at the near field surface (3D grid),
and, similar to the above Dyson approach, the far field surface is selected at $h_{\rm far}=630$ nm with a grid size of $20$ nm;
the averaged time per spatial point at a single frequency is about $0.224$ minutes (also shown in Sec. \ref{subSec3.2}) and $0.123$ minutes from the near field surface $h=50$ nm and $h=30$ nm.
Also note that, from the NF2FF theory, $\tilde{\mathbf{H}}$ does not need to be calculated from the curl of $\tilde{\mathbf{F}}$; these two are obtained at the same time from the vector potential  (Eq. \eqref{bigFnear2far} and Eq. \eqref{bigHnear2far}).
Thus for $h=50$ nm, it will take about $0.224\times70\times56$ minutes $\approx14.6$ hours for $S^{\rm rad}$ (Eq. \eqref{Sradfull}), and $0.2244\times70$ minutes $\approx15.7$ minutes for pole result $S^{\rm rad}_{\rm p1}$ (Eq. \eqref{Sradpole1}).
Similarly, for $h=30$ nm, it will take about $0.123\times70\times56$ minutes $\approx8$ hours for $S^{\rm rad}$ (Eq. \eqref{Sradfull}), and $0.123\times70$ minutes $\approx8.6$ minutes for pole result $S^{\rm rad}_{\rm p1}$ (Eq. \eqref{Sradpole1}).

The run times for the pole result $S^{\rm rad}_{\rm p2}$ (Eq. \eqref{Sradpole2}) comes from Table \ref{sout_eq44} and Table \ref{sout_eq44_1}, using the smallest time it take to get the convergent value. Clearly this
method is extremely efficient and full calculations are completed
in a few seconds.

We summarize the above run times using the Dyson and NF2FF approaches in Table \ref{runtime}.
For $S^{\rm rad}$ (Eq. \eqref{Sradfull}) and $S_{\rm p1}^{\rm rad}$ (Eq. \eqref{Sradpole1}), the run times with the Dyson equation using grid $0.2$ ($0.5$) nm will be about 192 (13.5) times) and 348 (24.6) times longer than the NF2FF transformation with the near field surface $h=50$ nm and $h=30$ nm.
Moreover,  the run time for $S_{\rm p2}^{\rm rad}$ (Eq. \eqref{Sradpole2}) is only several seconds, which is promising to use
with more complicated geometries such as the example
below with coupled plasmon PC modes.

\begin{table}
\caption {Comparison of example run times for $S^{\rm rad}$ (Eq.~\eqref{Sradfull}), $S_{\rm p1}^{\rm rad}$ (Eq.~\eqref{Sradpole1}), and $S_{\rm p2}^{\rm rad}$ (Eq.~\eqref{Sradpole2}) between the Dyson approach and the NF2FF approach. For the Dyson approach, the grid size for volume integration inside metal is $0.2$ or $0.5$ nm. For NF2FF approach, the grid size used for near field surface (here $h=30$ nm and $h=50$ nm are shown) integration is selected as $0.5$ nm. Far field surface for both approaches is at $h_{\rm far}=630$ nm, and the grid at this surface is $20$ nm.} \label{runtime}
\centering
\begin{tabular}{|p{0.8cm}|p{3.0cm}|p{3.0cm}|}
\hline
 time  & Dyson, grid$=0.2$ nm &  NF2FF, $h=50$ nm \\
\hline
$S^{\rm rad}$  & $117$ days   &  $14.6$ hours \\
\hline
$S_{\rm p1}^{\rm rad}$  &$2.1$ days & $15.7$ mins \\
\hline
$S_{\rm p2}^{\rm rad}$  &   & $4.8$ secs \\
\hline
   & Dyson, grid$=0.5$ nm    & NF2FF, $h=30$ nm \\
\hline
 $S^{\rm rad}$ & $8.3$ days  & $8$ hours \\
 \hline
$S_{\rm p1}^{\rm rad}$  &  $3.5$ hours  & $8.6$ mins \\
\hline
 $S_{\rm p2}^{\rm rad}$  &   & $2.8$ secs \\
\hline
\end{tabular}
\end{table}

\section{Numerical results for coupled quasinormal modes and hybrid metal-dielectric systems}\label{Sec4}

In this section, we focus on a much more complex example, which uses coupled modes formed by a metal-dielectric system, with peak Purcell factors
in excess of 1 million.
These systems can exhibit rich interference effects
and exploit some of the advantages of both cavity parts.
For example, the plasmonic structure posses extreme localized fields enhancement (small mode volumes), but with relative low quality factor due to metallic losses \cite{maier_plasmonics:_2007, novotny_antennas_2011, andersen_strongly_2011}.
In contrast, a PC cavity generally has a very high quality factor, but with a smaller mode volume that is limited by diffraction.
Combining these systems together will get a range of cavity mode properties and line shapes (including Fano-like lineshapes \cite{doi:10.1143/JPSJ.80.104707,thakkar_sculpting_2017}), which offer new possibilities \cite{barth_nanoassembled_2010,doeleman_antennacavity_2016,2017PRA_hybrid,palstra_hybrid_2019,dezfouli_molecular_2019} that can benefit from the high quality factor of the dielectric structure and the significant field enhancements of the plasmonic structure. The challenge for obtaining the
regularized modes of such a system is that the simple symmetry of the dimer cannot
be exploited, and the spatial size of the system region is much larger
in the case for the PC-like mode.

\begin{figure}[thp]
  \centering
  \includegraphics[width=0.99\columnwidth]{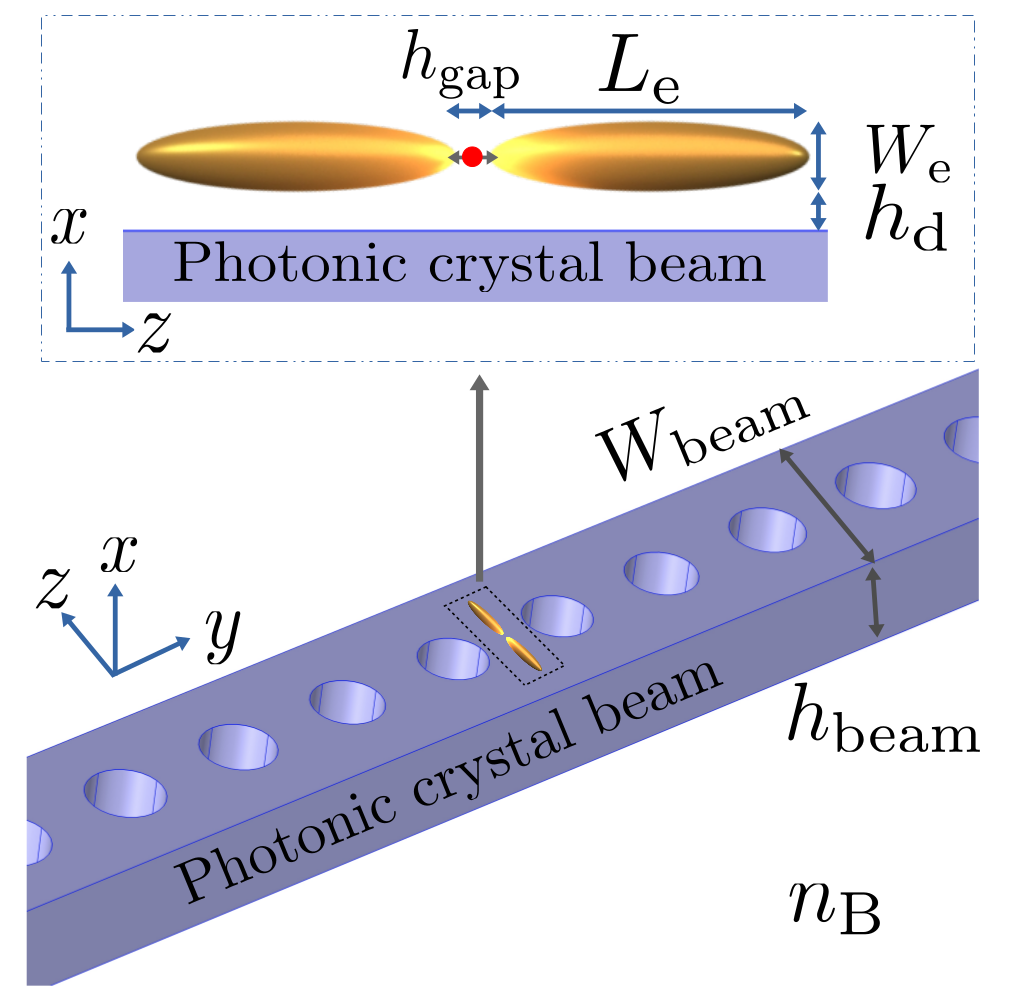}
  \caption{Schematic diagram of metal-dielectric hybrid structure, where a gold ellipsoid dimer is placed close to a PC cavity. A point dipole (red dot) polarized along $z$-direction is put at the gap center of the dimer. The origin of Cartesian coordinate system is located at dimer gap center. The length and width of the single ellipsoid are $L_{\rm e}=50$ nm and $W_{\rm e}=10$ nm. The gap of dimer is $h_{\rm gap}=2$ nm,  and the minimum distance between dimer surface and photonic crystal beam surface is $h_{\rm d}=5$ nm. The width and height of the beam are $W_{\rm beam}=376$ nm and $h_{\rm beam}=200$ nm. The refractive index of beam is $n_{\rm pc}=2.04$, and the  background medium is free space ($n_{\rm B}=1$).
  }\label{sche}
\end{figure}

The hybrid system we model is shown in Fig.~\ref{sche}, which uses a gold (Au) ellipsoid dimer (with center width of $W_{\rm e}=10$ nm, center length of $L_{\rm e}=50$ nm and gap of $h_{\rm gap}=2$ nm ),  put above a silicon-nitride photonic crystal beam (index $n_{\rm PC}=2.04$), similar to the coupled mode structures  used in Refs.~\onlinecite{dezfouli_molecular_2019, franke_quantization_2018,palstra_hybrid_2019}.
Specifically, the width and height of the beam is $W_{\rm beam}=376$ nm
and $L_{\rm beam}=200$ nm; the length of the finite beam is $8.5~\mu$m.
The nearest distance between the dimer surface and the beam surface is $h_{\rm d}=5$ nm.
The background medium is free space with refractive index $n_{\rm B}=1.0$, and again we
use a Drude model with
the same parameters as above for gold.

We use the same approach as shown in Sec. \ref{subSec3.1}, to compute the QNMs:
 The scattered electric field of a point dipole at position $\mathbf{r}_{0}$ is related to the Green's function, and given by
\begin{equation}
\mathbf{E}^{\rm s}(\mathbf{r},\omega)=\frac{1}{\epsilon_{0}}\mathbf{G}(\mathbf{r},\mathbf{r}_{0},\omega)\cdot \mathbf{d},
\end{equation}
where $\mathbf{d}$ is the dipole moment of the emitter.
If several modes are dominant in the regime of interest, then we can expand the Green's function with several QNMs,
\begin{equation}\label{E_scatter_multi}
\mathbf{E}^{\rm s}(\mathbf{r},\omega)=\sum_{\mu}\frac{1}{\epsilon_{0}}A(\omega)\tilde{\mathbf{f}}_{\mu}(\mathbf{r})\tilde{\mathbf{f}}_{\mu}(\mathbf{r}_{0})\cdot \mathbf{d}.
\end{equation}
We assume these modes  are orthogonal with each other, so that
\begin{equation}\label{E_scatter_multimu}
\mathbf{E}^{\rm s}_{\mu}(\mathbf{r},\omega)=\frac{1}{\epsilon_{0}}A(\omega)\tilde{\mathbf{f}}_{\mu}(\mathbf{r})\tilde{\mathbf{f}}_{\mu}(\mathbf{r}_{0})\cdot \mathbf{d},
\end{equation}
and $\mathbf{E}^{\rm s}(\mathbf{r},\omega)=\sum_{\mu}\mathbf{E}^{\rm s}_{\mu}(\mathbf{r},\omega)$.
Subsequently, using a similar method shown in Sec. \ref{subSec3.1}, we
obtain the normalized QNMs field as
\begin{align}
\begin{split}
\tilde{\mathbf{f}}_{\mu}(\mathbf{r})=\sqrt{\frac{2\epsilon_{0}(\tilde{\omega}_{\mu}-\omega)}{\omega\mathbf{d}\cdot\mathbf{E}_{\mu}^{\rm s}(\mathbf{r}_{0},\omega)}}\mathbf{E}_{\mu}^{\rm s}(\mathbf{r},\omega),
\end{split}
\end{align}
where $\omega=(1-10^{-5})\times\tilde{\omega}_{\mu}$, very close to the pole frequency.

\begin{figure*}[htp]
  \centering
  \includegraphics[width=0.95\columnwidth]{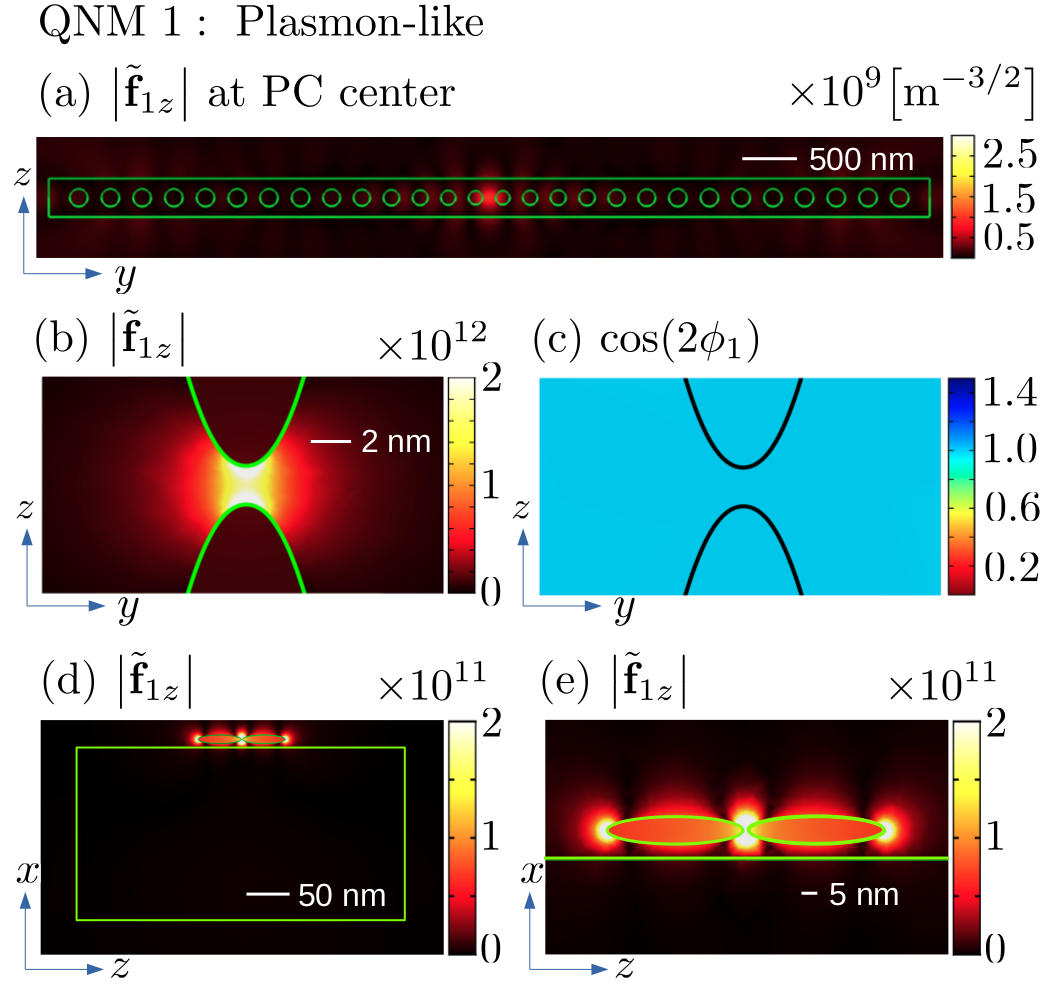}
  \includegraphics[width=0.95\columnwidth]{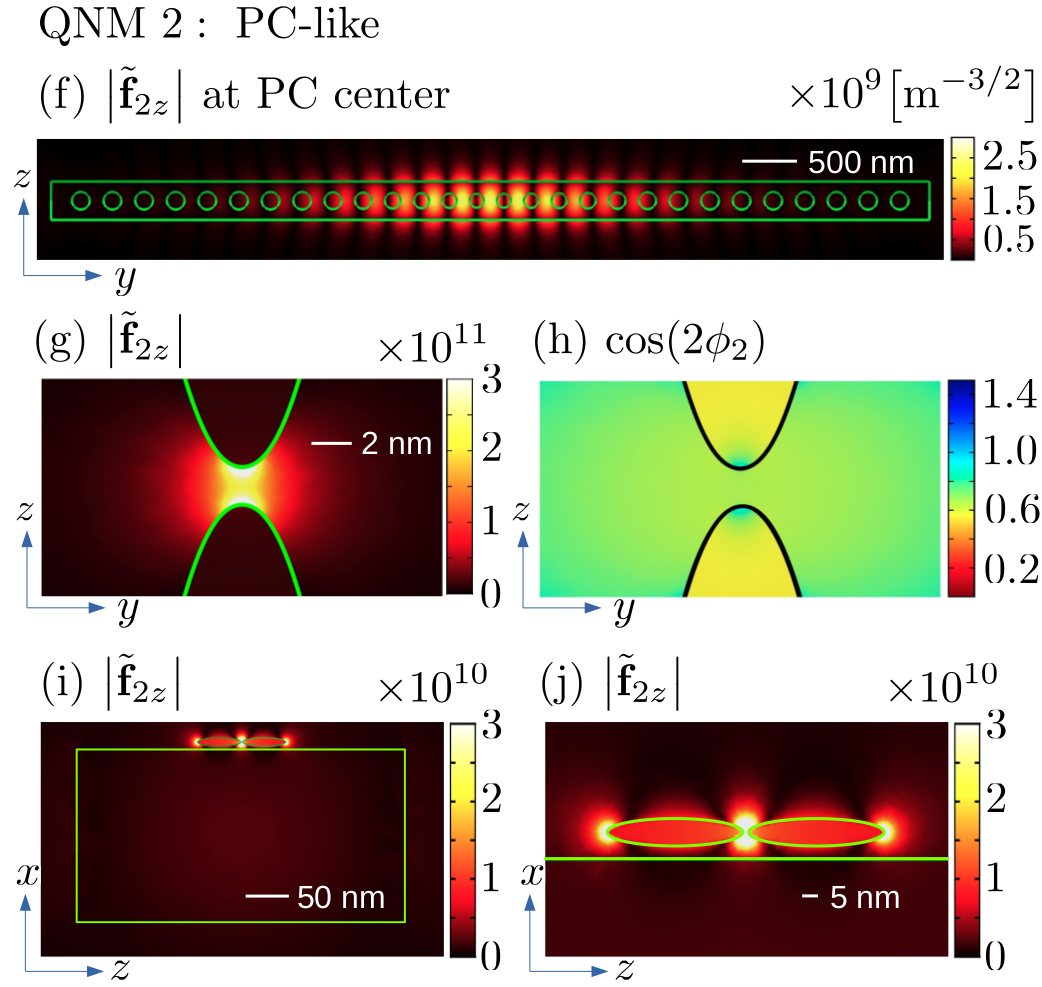}
  \caption{ QNM spatial fields and phase distribution for the two QNMs of the hybrid metal-dielectric system. (a)  $\big|\tilde{\mathbf{f}}_{1z}\big|$  of QNM 1 (plasmon-like mode, low $Q$) at beam center surface ($x=-W_{\rm e}/2-h_{\rm d}-h_{\rm beam}/2$). Note that the origin of Cartesian coordinate system is located at dimer gap center point. (b) $\big|\tilde{\mathbf{f}}_{1z}\big|$  of QNM 1 at surface $x=0$ nm (dimer center surface). (c) QNM phase $\cos(2\phi_1)$ at $x=0$ nm, where phase is defined by $\tilde{\mathbf{f}}_{1z}=\big|\tilde{\mathbf{f}}_{1z}\big|e^{i\phi_{1}}$. At gap center point, $\cos(2\phi_1(\mathbf{r}_{0}=[0,0,0]))=0.999$. (d) $\big|\tilde{\mathbf{f}}_{1z}\big|$ at $y=0$ nm. (e) Zoom in of (d). (f), (g), (h), (i), (j) Corresponding $\big|\tilde{\mathbf{f}}_{2z}\big|$ and QNM phase $\cos(2\phi_2)$ for QNM 2 (PC-like mode, high $Q$). At the gap center point, $\cos(2\phi_2(\mathbf{r}_{0}=[0,0,0]))=0.655$.
  }\label{M1field}
\end{figure*}

\begin{figure*}[htp]
  \centering
  \includegraphics[width=0.99\columnwidth]{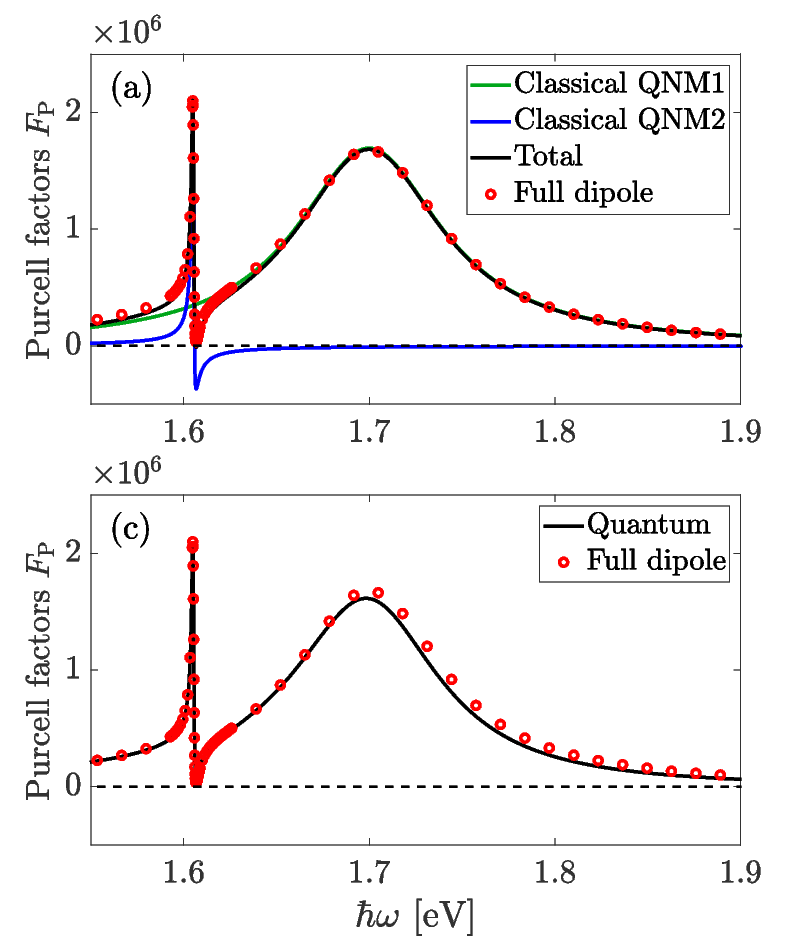}
  \includegraphics[width=0.99\columnwidth]{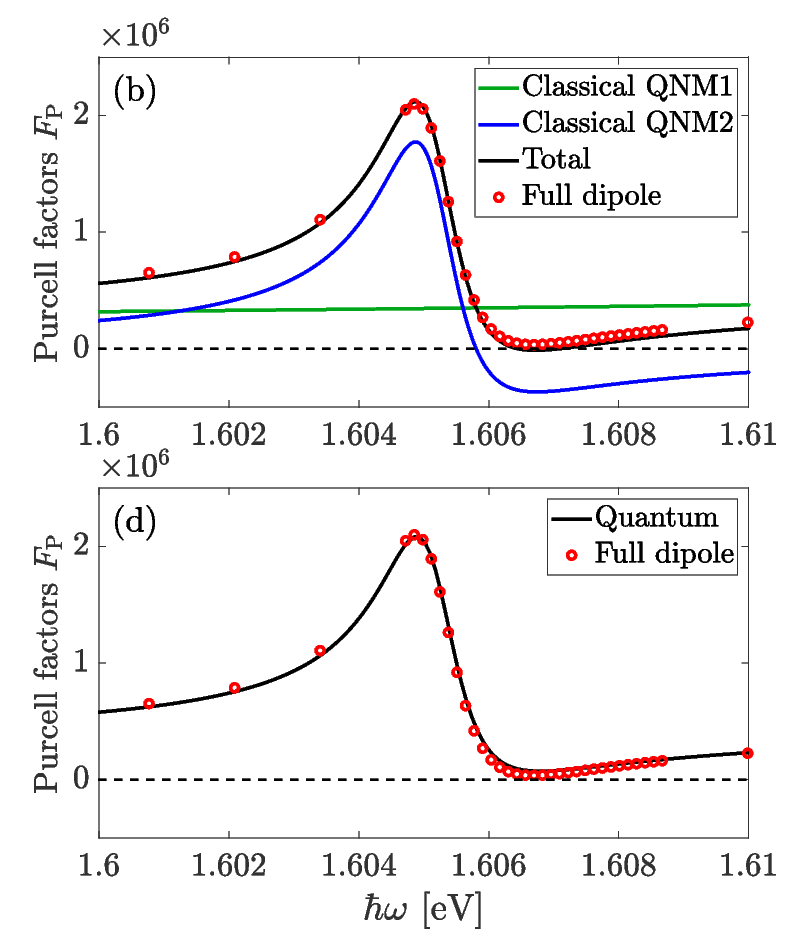}
  \vspace{-0.5cm}
  \caption{ Purcell factors for a $z$-polarized point dipole  at the dimer gap center point (as shown Fig .\ref{sche}) from  (a) classical QNMs theory  and (c) quantum theory. Panels (b) and (d) show a zoom in of (a) and (d) near the Fano resonance. The green curve is from QNMs theory (Eq. \eqref{QNMpurcelltotal}), which show excellent qualitative agreement with full dipole calculation (red circles). There are two dominated modes with $\tilde{\omega}_{1}=1.6999 - 0.0479i$ eV, $Q_{1}=17.8$ and $\tilde{\omega}_{1}=1.6052 - 0.0007i$ eV, $Q_{2}=1139.3$. Black line and blue line present their respective contribution to Purcell factors.
  (c) Quantum Purcell factors (Eq. \eqref{quantumpurcell}) for coupled structures, which show nice agreement with full dipole results (red circles). The corresponding 4 $S$ parameter are as follows, $S_{11}=0.894+0.068$, $S_{22}=0.904+0.134$, $S_{12}=(-0.0042-0.0967i)+(-0.0021-0.0024i)$, $S_{21}=S_{12}^{\ast}$.
  }\label{PFcoupled}
\end{figure*}

The simulation volume of the cylindrical module is about $85$ $\mu$m$^{3}$ (including PMLs), where the maximum mesh sizes  are $0.1$ nm, $3$ nm, $50$ nm and $120$ nm at the dipole point (center of the gap), inside the ellipsoid dimer, PC beam and elsewhere. The hybrid structure is significantly larger than gold dimer on its own, and we use $10$ perfectly matched layers (PMLs) to minimize boundary reflections (the number of layers and the total thickness of PMLs should be adjusted according to the size of the inside simulation domain and the inside mesh settings to show the better performance).

Over a broad bandwidth of several eV, there
 are two dominant modes of interest in this coupled structure.
The first one we term `QNM 1' is a plasmon-like mode, with resonance frequency $\tilde{\omega}_{1}=1.6999 - 0.0479i$ eV and relatively low quality factor $Q_{1}=17.8$. The QNM field at the gap center of the ellipsoid dimer is $\tilde{\mathbf{f}}_{1z}(\mathbf{r}_{0})=1.800\cdot10^{12}-4.692i\cdot10^{10}$~[m$^{-3/2}$] ($z$-component).
The corresponding effective mode volume is $V_1^{\rm eff}/\lambda_1^3=7.9575\times10^{-7}$, where $V_1^{\rm eff}={1}/{{\rm Re}\big[\tilde{\mathbf{f}}^{2}_{1z}(\mathbf{r}_{0})\big]}$ and $\lambda_1=2\pi c/{\rm Re}(\tilde{\omega}_{1})$.
The second QNM,  `QNM 2', is a PC-like mode, with resonance frequency $\tilde{\omega}_{1}=1.6052 - 0.0007i$ eV and a relatively high quality factor $Q_{2}=1139.3$.
The QNM field at the center of the ellipsoid dimer gap is $\tilde{\mathbf{f}}_{2z}(\mathbf{r}_{0})=2.108\cdot10^{11}+9.623i\cdot10^{10}$~[m$^{-3/2}$];
the effective mode volume is $V_2^{\rm eff}/\lambda_2^3=6.1697\times10^{-5}$, where $V_2^{\rm eff}={1}/{{\rm Re}\big[\tilde{\mathbf{f}}^{2}_{2z}(\mathbf{r}_{0})\big]}$ and $\lambda_2=2\pi c/{\rm Re}(\tilde{\omega}_{2})$.
Note that these effective mode volumes can also be negative~\cite{2017PRA_hybrid}.

To show the differences of the two QNMs more clearly, the QNM spatial fields and phase distribution are shown in Fig. \ref{M1field}, where the phase is defined by $\tilde{\mathbf{f}}_{1z}=\big|\tilde{\mathbf{f}}_{1z}\big|e^{i\phi_{1}}$ and $\tilde{\mathbf{f}}_{2z}=\big|\tilde{\mathbf{f}}_{2z}\big|e^{i\phi_{2}}$.
At the PC beam center surface ($x=-W_{\rm e}/2-h_{\rm d}-h_{\rm beam}/2$), two QNMs show very different fields distributions (Fig. \ref{M1field} (a) and (f)).
While at dimer center surface ($x=0$), the two modes show similar fields distribution (Fig. \ref{M1field} (b) and (g)), except $\big|\tilde{\mathbf{f}}_{1z}\big|$ is an order of magnitude larger than $\big|\tilde{\mathbf{f}}_{2z}\big|$.
Also note that the fields at dimer center surface ($x=0$) are three (two) orders of magnitude larger than those at PC beam center surface for plasmon-like mode (PC-like mode), which means the two modes mainly live around the dimer region.
In addition, as shown in Fig. \ref{M1field} (c) and (h), the
QNM phases for the two modes are very different at the dimer center surface.
At the dimer gap center point ($x=y=z=0$ nm), $\cos(2\phi_1)=0.999$ and $\cos(2\phi_2)=0.655$ for two modes.
The phase difference between them will result in the Fano-like lineshape \cite{doi:10.1143/JPSJ.80.104707} in total Purcell factors for a dipole placed at dimer gap center (Fig. \ref{PFcoupled}).
 Figures \ref{M1field} (d), (e), (i), and (j) show the
 QNM fields distribution at the surface $y=0$ nm.
Close to dimer region, they look similar, except $\big|\tilde{\mathbf{f}}_{1z}\big|$ is an order of magnitude larger than $\big|\tilde{\mathbf{f}}_{2z}\big|$. Clearly,  most of the QNM fields live in near dimer region, especially for plasmon-like mode.

We also stress that we are testing an extreme example here, as the
gap size is only 2 nm. To test the accuracy of a two QNM
description,
the generalized classical Purcell factors for a point dipole placing at the dimer gap center are shown in Fig. \ref{PFcoupled} (a).
The results from QNMs theory (black curve) (Eq. \eqref{QNMpurcelltotal}) show excellent qualitative agreement with full dipole calculation (red circles, Eq. \eqref{Purcellfulldipole}).
This clearly indicates the validity of the QNMs results.
The green and blue lines present their respective contribution to Purcell factors, with QNM 2 contributing negatively (in a certain frequency regime).
Note also that,  if we focus on dip region, the results from QNMs theory (black curve) net negative (below the black dashed line) (lowest point $-1.2\times10^{4}$). This is most probably caused by the onset of quasi-static contributions,
whose contribution would naturally result in a net positive
total Purcell factor; such contributions can be added
into the theory using a quasi-static Green function theory
 \cite{ge_quasinormal_2014}. If quasi-static contributions are considered, then the total Purcell factor from QNMs theory will be net-positive.
For larger dimer gaps,
the contribution of quasi-static modes
is negligible~\cite{2017PRA_hybrid}.

Next, we discuss the calculation of the quantized QNM
parameters.
First,
it is quite simple to calculate $S_{\mu\eta}^{\rm nrad}$ for coupled modes, since we only need the QNM fields inside the metal, which is the
same level of difficulty as the dimer calculation on its own.
Using Eq. \eqref{SnradfullwMulti}, with a numerical frequency integral, we obtain $S^{\rm nrad}_{11}=0.905$, $S^{\rm nrad}_{22}=0.904$, $S^{\rm nrad}_{12}=-0.0014-0.0975i$, and $S^{\rm nrad}_{21}=S^{{\rm nrad} \ast}_{12}$.
Moreover, using Eq. \eqref{eq:SnradapproxMulti}, with an
accurate pole approximation, we obtain $S^{\rm nrad}_{\rm p,11}=0.894$, 
$S^{\rm nrad}_{\rm p,22}=0.904$, 
$S^{\rm nrad}_{\rm p,12}=-0.0042-0.0967i$, 
and $S^{\rm nrad}_{\rm p,21}=S^{{\rm nrad}{\ast}}_{\rm p,12}$.
Note that for $S^{\rm nrad}_{11}$, $S^{\rm nrad}_{22}$, $S^{\rm nrad}_{\rm p,11}$, and $S^{\rm nrad}_{\rm p,22}$, the spatial integrals over metal volume are directly done in COMSOL; and for $S^{\rm nrad}_{12}$ and $S^{\rm nrad}_{\rm p,12}$, the grid size is $0.1$ nm.

Numerically, the main challenge is to obtain the radiative part
of the quantum $S$ factors, $S_{\mu\eta}^{\rm rad}$, which generally require the  regularized QNM fields ($\tilde{\mathbf{F}}$ and $\tilde{\mathbf{H}}$) at a surface surrounding the entire hybrid structures.
To compute $\tilde{\mathbf{F}}$ at a single spatial point with the Dyson Equation requires a spatial integration over the entire hybrid structure (PC beam and dimer).
If the grid size $0.2$ ($0.5$) nm is used, then it will take about $48$ days ($3.4$ days) to get $\tilde{\mathbf{F}}$ at a single spatial point for a single frequency.
The run time for Dyson equation is proportional to the integral volume if the same grid size is used.
The integral volume for this structure, with  just the PC beam
and the ellipsoid dimer, is around 8122 times larger than that for gold cylindrical dimer.
Then correspondingly, the run times will increase
in the same way.
Moreover, since the size of the hybrid structure is on the order of several micrometers, then the area of the outside surface is very large.
For instance, the outside surface is chosen as a cuboid surface at $h_{\rm far}=630$ nm (the smallest vertical distance to the dimer surface or PC beam surface); and
the grid size at this surface is selected as $20$ nm.
Although there is no similar symmetry as dimer only scenario, the fields are symmetric with respect to the  $x$-$z$  and  $x$-$y$ planes. Then we only need to do the integral over $1/4$ of this cuboid surface.
Thus, in sumamry, it will take about $16114$ ($1141$) years (!) with an inside grid size $0.2$ ($0.5$) nm for even pole $S^{\rm rad}_{\rm p1}$ (Eq. \eqref{Sradpole1}) with Dyson equation.
So clearly it is impractical to employ the Dyson equation to calculate $S_{\mu\eta}^{\rm rad}$ for such coupled modes.

To address this significant problem,
we  use the pole approximation (Eq. \eqref{Sradpole2Multi}) with a NF2FF transformation to calculate $S_{\mu\eta}^{\rm rad}$.
For the plasmon-like mode, the near field surface is chosen as a cuboid surface just surrounding the dimer, since most of fields are located around the dimer (Fig. \ref{M1field}). 
The smallest vertical distance from the dimer to the five surfaces of this cuboid  (not including the surface below) is $50$ nm; and in the $x$- direction, the sixth surface is $4$ nm below the lowest part of the dimer surface.
For the PC-like mode, the near field surface is set as a cuboid surface surrounding the entire coupled structures, where the smallest vertical distance from the hybrid structure to every surface of this cuboid is $50$ nm ($h=50$ nm). Note that, as shown in Appendix \ref{Model1_Sradp2},  for the PC-like mode, the results from near field surface just surrounding dimer alone yield a very good approximation as well, since one can see that most of the PC-like mode fields are also located in the dimer region (see Fig.~\ref{M1field}).

\begin{table}[b]
\caption {Pole $S_{\rm p2,11}^{\rm rad}$ from Eq. \eqref{Sradpole2Multi}, quoted to the third decimal place. For the calculations, the
plasmon-like mode uses a cuboid surface (with $h=50$ nm) that surrounds the dimer.} \label{sout_11}
    \centering
    \begin{tabular}{|c|c|c|}
    \hline
near field surface  $h=50$ nm &    &   \\
 \hline
 angle grid for $\vartheta$ and $\varphi$ & run time   & pole $S_{\rm p2,11}^{\rm rad}$  \\
 \hline
 $\pi/2$    & $2.0$ secs   & $0.081$  \\
 \hline
 $\pi/3$    & $2.8$ secs  & $0.069$  \\
 \hline
 $\pi/5$    & $5.0$ secs   & $0.068$  \\
  \hline
 $\pi/8$    & $9.4$ secs   & $0.068$  \\
 \hline
 $\pi/10$    & $12.4$ secs   & $0.068$  \\
  \hline
    \end{tabular}
\end{table}

The grid sizes at these near field surfaces are $0.5$ nm (the same as that for dimer only).
The convergence tests for angle resolutions are shown in Table \ref{sout_11}, \ref{sout_22} and \ref{sout_12}.
We got $S^{\rm rad}_{\rm p2,11}=0.068$, $S^{\rm rad}_{\rm p2,22}=0.134$, $S^{\rm rad}_{\rm p2,12}=-0.0021-0.0024i$, and $S^{\rm rad}_{\rm p2,21}=S^{{\rm rad}{\ast}}_{\rm p2,12}$.
As  also shown in these table, the run times are several seconds to $1.7$ hours, which is acceptable and significantly faster than those with Dyson approach (which are untractable for this geometry).

\begin{table}
\caption {Pole $S_{\rm p2,22}^{\rm rad}$ from Eq. \eqref{Sradpole2Multi} with $h=50$ nm (surrounding entire coupled structures), quoted to the third decimal place. For the calculations, the PC-like mode
uses a cuboid surface surrounding the entire structure. } \label{sout_22}
    \centering
    \begin{tabular}{|c|c|c|}
    \hline
near field surface  $h=50$ nm &    &   \\
 \hline
 angle grid for $\vartheta$ and $\varphi$ & run time   & pole $S_{\rm p2,22}^{\rm rad}$  \\
 \hline
 $\pi/2$    & $4.6$ mins   & $0.194$  \\
  \hline
 $\pi/5$    & $16.8$ mins  & $0.150$  \\
 \hline
 $\pi/10$    & $43.3$ mins  & $0.140$  \\
 \hline
  $\pi/15$    & $56.3$ mins  & $0.137$  \\
 \hline
 $\pi/20$    & $1.26$ hours   & $0.134$  \\
 \hline
  $\pi/30$    & $1.7$ hours   & $0.134$  \\
 \hline
    \end{tabular}
\end{table}

\begin{table}
\caption {Pole results for $S_{\rm p2,12}^{\rm rad}$ from Eq. \eqref{Sradpole2Multi}, quoted to the 4th decimal place. For the calculations, the PC-like mode
uses a cuboid surface surrounding the entire structure, while the
plasmon-like mode uses a cuboid surface that surrounds the dimer.} \label{sout_12}
    \centering
    \begin{tabular}{|c|c|c|}
    \hline
near field surface  $h=50$ nm &    &   \\
 \hline
 angle grid for $\vartheta$ and $\varphi$ & run time   & pole $S_{\rm p2,12}^{\rm rad}$  \\
 \hline
 $\pi/2$    & $4.5$ mins   & $-0.0062+0.0006i$  \\
  \hline
 $\pi/3$    & $7.0$ mins   & $-0.0040-0.0051i$  \\
 \hline
 $\pi/5$    & $13.3$ mins   & $-0.0023+0.0003i$  \\
  \hline
 $\pi/10$    & $41.9$ mins   & $-0.0019-0.0013i$  \\
  \hline
 $\pi/20$    & $53.8$ mins   & $-0.0021-0.0024i$  \\
 \hline
 $\pi/30$    & $1.4$ hours   & $-0.0021-0.0024i$  \\
  \hline
    \end{tabular}
\end{table}

In summary,
for the total $S$ values
of the hybrid modes:
$S_{\rm p,11}=0.894+0.068$, $S_{\rm p,22}=0.904+0.134$, $S_{\rm p,12}=(-0.0042-0.0967i)+(-0.0021-0.0024i)$, $S_{\rm p,21}=S_{\rm p,12}^{\ast}$, where the first is nonradiative part and the second is radiative part.
Then quantum Purcell factors
(Eq.~\eqref{quantumpurcell}) for coupled structures are shown in Fig. \ref{PFcoupled}(c) and (d) (black curve), which show nice agreement with full dipole results (red circles, Eq.~\eqref{Purcellfulldipole}).
Interestingly, the quantum result is net positive, since the quantum
basis states constitute a different representation
of the modes that are needed from quantum mechanical (namely, the symmetrized
QNMs).

Given the extreme localization of a 2-nm gap, and the complex hybrid structure, the agreement with full dipole calculations is rather striking and shows the power of the quantum model, which can then be used to explore a wide range of nonlinear quantum optical effects.

\section{Conclusions}\label{Sec5}

In summary, we have introduced
a timely and efficient NF2FF approach for QNMs, which allows one to quickly and accurately
obtain the regularized QNMs in the far field.
We also showed how to compliment this NF2FF transformation with
an efficient pole approximation and a very far field
extrapolation for obtaining the quantum optical parameters for quantized QNM theory~\cite{franke_quantization_2018}: For example,
the quantized QNM theories require the regularized fields flowing radiatively through a closed surface.
In this regard, the NF2FF approach is shown to be several orders
of magnitude faster to that of a direct Dyson equation
approach, which has been used in the literature to obtain the
regularized fields. In a practical calculation of a 3D metal dimer QNM,
we  obtain the desired
quantization parameters in under 1 minute run time on a standard computer
workstation, which previously takes weeks to one month of computational time. We also
reported detailed numerical convergence checks on the QNM quantization parameters, both for the radiative and nonradiative contributions. The former requires the QNM field within or close to the metal (lossy structure), while the
latter required the regularized QNM field (non-divergent), far outside the resonator. Together these contributions form the input-output channel contributions in quantum optics and form the basis for Fock space quantization for these
dissipative modes. More generally, the regularized QNMs are the physical fields that connect to experiments for detection of the modes
outside the resonator.

The  general
formalism presented here solves a major
computational problem
in quantized open-cavity mode theories
by efficiently returning regularized QNMs for a wide range of structures,
yielding
QNM parameters for immediate use in quantum optics
and quantum plasmonics with the underlying quantized open-cavity modes.
We applied the theory to compute far-field
(regularized) mode profiles and quantum optical
parameters for quantized QNMs theory,
for both metal dimer structures and a complex
hybrid structure that formed coupled QNMs
between a PC cavity and a dimer. In all cases,
we show excellent agreement with full dipole classical calculations in the bad cavity limit, using both semiclassical and fully quantum simulations. This approach can thus easily be used as input to explore system level quantum optics with these modes, where unique few quanta effects beyond the single Fock state and coherent state can be probed
and explored beyond the semiclassical limit. Such effects
will be reported elsewhere.


\acknowledgements
We acknowledge Queen's University and the Natural Sciences and Engineering Research Council of Canada for financial support, and CMC Microsystems for the provision of COMSOL Multiphysics to facilitate this research.
We also acknowledge support from the Deutsche Forschungsgemeinschaft (DFG) through SFB 951 Project B12 (Project number 182087777), Project BR1528/8-2 (Project number 177864488) and the
Alexander von Humboldt Foundation through a Humboldt Research Award.
We thank Mohsen  Kamandar  Dezfouli
for useful discussions.
This project has also received
funding  from  the  European  Unions  Horizon  2020
research and innovation program under Grant Agreement
No. 734690 (SONAR).



\vspace{0.2cm}

\appendix

\section{Simpler calculation for \texorpdfstring{$S^{\rm rad}_{\rm p2,22}$}{Lg} and \texorpdfstring{$S^{\rm rad}_{\rm p2,12}$}{Lg}, using
only fields surrounding the dimer}\label{Model1_Sradp2}

In the main text, for the PC-like mode of the hybrid device we uses a cuboid surrounding the entire coupled structures to calculate the corresponding $S^{\rm rad}_{\rm p2,22}$ and $S^{\rm rad}_{\rm p2,22}$ (Eq. \eqref{Sradpole2Multi}).
Here we show it is also a good approximation if the near field surface is replaced by a cuboid just surrounding the dimer.
The grid sizes at this near field surface are set as $0.5$ nm (the same as that used in main text).
The convergence check over angle integral are shown in Table \ref{sout_22_model1} and \ref{sout_12_model1}.
We got $S^{\rm rad}_{\rm p2,22}=0.061$, $S^{\rm rad}_{\rm p2,12}=-0.0008-0.0063i$, and $S^{\rm rad}_{\rm p2,21}=S^{{\rm rad}{\ast}}_{\rm p2,12}$.
 As  also shown in these tables, the run times
 are only several seconds.

\begin{figure}[htp]
  \centering
  \includegraphics[width=0.95\columnwidth]{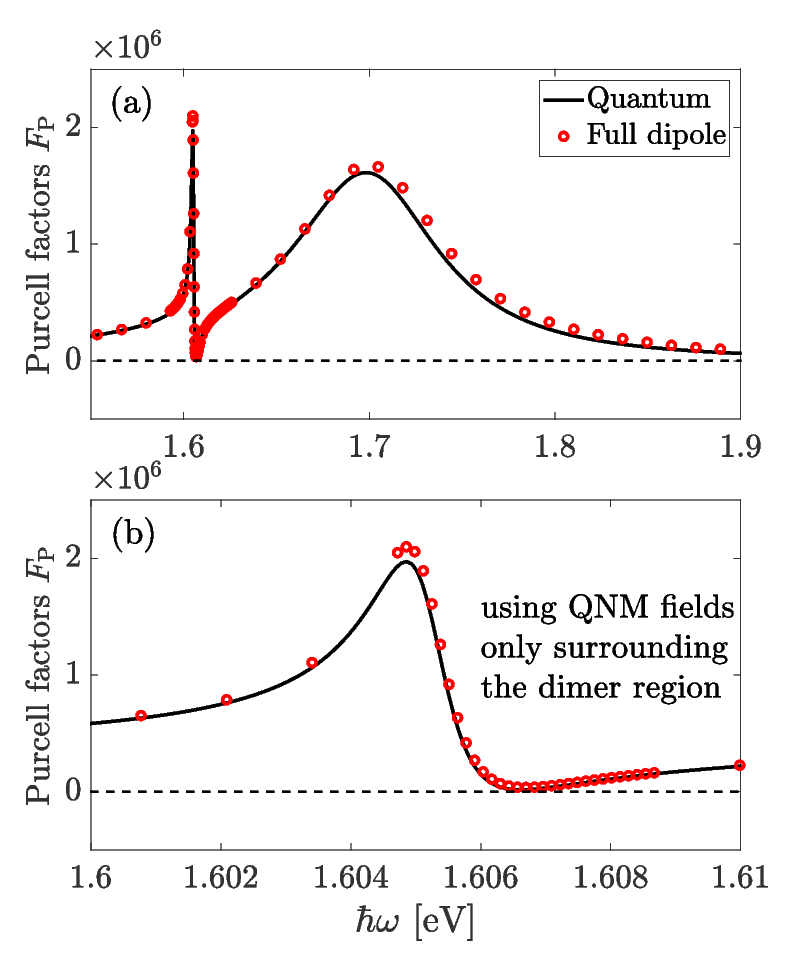}
  \vspace{-0.5cm}
  \caption{Quantum Purcell factors (Eq. \eqref{quantumpurcell}) for coupled structures, which again show good agreement with full dipole results (red circles). In contrast to the results  shown in
  Figs.~\ref{PFcoupled}(c)-(d),  when calculating
  $S_{{\rm p2},\mu\eta}^{\rm rad}$ (Eq. \eqref{Sradpole2Multi}),
  the near field surfaces are now chosen as a small cuboid ($h=50$ nm) just surrounding dimer for both plasmon-like mode and PC-like mode.
  The corresponding 4 $S$ parameter are as follows, $S_{11}=0.894+0.068$, $S_{22}=0.904+0.061$, $S_{12}=(-0.0042-0.0967i)+(-0.0008-0.0063i)$, $S_{21}=S_{12}^{\ast}$.
  In this way the
  run times to obtain the radiative $S$ parameters are reduced from a few hours to a few seconds, while still maintaining a very good level of accuracy with no fitting parameters.
 }\label{PFcoupledModel1}
\end{figure}

\begin{table}[th]
\caption {Pole $S_{\rm p2,22}^{\rm rad}$ from Eq. \eqref{Sradpole2Multi} with $h=50$ nm ({with small cuboid around the dimer}), quoted to the third decimal place. } \label{sout_22_model1}
    \centering
    \begin{tabular}{|c|c|c|}
    \hline
near field surface  $h=50$ nm &    &   \\
 \hline
 angle grid for $\vartheta$ and $\varphi$ & run time   & pole $S_{\rm p2,22}^{\rm rad}$  \\
 \hline
 $\pi/2$    & $1.5$ secs   & $0.073$  \\
  \hline
 $\pi/3$    & $2.8$ secs  & $0.062$  \\
 \hline
 $\pi/5$    & $4.9$ secs  & $0.061$  \\
  \hline
  $\pi/8$    & $9.4$ secs  & $0.061$  \\
 \hline
  $\pi/10$    & $13.4$ secs  & $0.061$  \\
 \hline
    \end{tabular}
\end{table}
\begin{table}
\caption {Pole $S_{\rm p2,12}^{\rm rad}$ from Eq. \eqref{Sradpole2Multi} ({both with small cuboid surface}), quoted to the 4th decimal place.  } \label{sout_12_model1}
    \centering
    \begin{tabular}{|c|c|c|}
    \hline
near field surface  $h=50$ nm &    &   \\
 \hline
 angle grid for $\vartheta$ and $\varphi$ & run time   & pole $S_{\rm p2,12}^{\rm rad}$  \\
 \hline
 $\pi/2$    & $4.0$ secs   & $-0.0010-0.0075i$  \\
  \hline
 $\pi/3$    & $5.6$ secs   & $-0.0008-0.0064i$  \\
 \hline
 $\pi/5$    & $10.2$ secs   & $-0.0008-0.0063i$  \\
  \hline
 $\pi/8$    & $19.1$ secs   & $-0.0008-0.0063i$  \\
 \hline
    \end{tabular}
\end{table}

A summary of the quantum parameters are now as follows:
we use the  previous $S^{\rm rad}_{\rm p2,11}=0.068$ and just replace $S^{\rm rad}_{\rm p2,22}$ and $S^{\rm rad}_{\rm p2,12}$ with the new ones, $S_{\rm p,11}=0.894+0.068$, $S_{\rm p,22}=0.904+0.061$, $S_{\rm p,12}=(-0.0042-0.0967i)+(-0.0008-0.0063i)$, $S_{\rm p,21}=S_{\rm p,12}^{\ast}$, where the first is nonradiative part and the second is radiative part.
Then quantum Purcell factors (Eq. \eqref{quantumpurcell}) for the coupled structure
is shown in Fig. \ref{PFcoupledModel1}, which show good agreement with full dipole results (red circles, Eq. \eqref{Purcellfulldipole}).


\bibliography{refs}

\begin{thebibliography}{53}%
\makeatletter
\providecommand \@ifxundefined [1]{%
 \@ifx{#1\undefined}
}%
\providecommand \@ifnum [1]{%
 \ifnum #1\expandafter \@firstoftwo
 \else \expandafter \@secondoftwo
 \fi
}%
\providecommand \@ifx [1]{%
 \ifx #1\expandafter \@firstoftwo
 \else \expandafter \@secondoftwo
 \fi
}%
\providecommand \natexlab [1]{#1}%
\providecommand \enquote  [1]{``#1''}%
\providecommand \bibnamefont  [1]{#1}%
\providecommand \bibfnamefont [1]{#1}%
\providecommand \citenamefont [1]{#1}%
\providecommand \href@noop [0]{\@secondoftwo}%
\providecommand \href [0]{\begingroup \@sanitize@url \@href}%
\providecommand \@href[1]{\@@startlink{#1}\@@href}%
\providecommand \@@href[1]{\endgroup#1\@@endlink}%
\providecommand \@sanitize@url [0]{\catcode `\\12\catcode `\$12\catcode
  `\&12\catcode `\#12\catcode `\^12\catcode `\_12\catcode `\%12\relax}%
\providecommand \@@startlink[1]{}%
\providecommand \@@endlink[0]{}%
\providecommand \url  [0]{\begingroup\@sanitize@url \@url }%
\providecommand \@url [1]{\endgroup\@href {#1}{\urlprefix }}%
\providecommand \urlprefix  [0]{URL }%
\providecommand \Eprint [0]{\href }%
\providecommand \doibase [0]{http://dx.doi.org/}%
\providecommand \selectlanguage [0]{\@gobble}%
\providecommand \bibinfo  [0]{\@secondoftwo}%
\providecommand \bibfield  [0]{\@secondoftwo}%
\providecommand \translation [1]{[#1]}%
\providecommand \BibitemOpen [0]{}%
\providecommand \bibitemStop [0]{}%
\providecommand \bibitemNoStop [0]{.\EOS\space}%
\providecommand \EOS [0]{\spacefactor3000\relax}%
\providecommand \BibitemShut  [1]{\csname bibitem#1\endcsname}%
\let\auto@bib@innerbib\@empty
\bibitem [{\citenamefont {Vahala}(2013)}]{vahala_optical_2003}%
  \BibitemOpen
  \bibfield  {author} {\bibinfo {author} {\bibfnamefont {{K. J.}}\ \bibnamefont
  {Vahala}},\ }\bibfield  {title} {\enquote {\bibinfo {title} {Optical
  microcavities},}\ }\href {\doibase 10.1038/nature01939} {\bibfield  {journal}
  {\bibinfo  {journal} {Nature}\ }\textbf {\bibinfo {volume} {424}},\ \bibinfo
  {pages} {839--846} (\bibinfo {year} {2013})}\BibitemShut {NoStop}%
\bibitem [{\citenamefont {Chang}\ and\ \citenamefont
  {Campillo}(1996)}]{chang1996optical}%
  \BibitemOpen
  \bibfield  {author} {\bibinfo {author} {\bibfnamefont {Richard~Kounai}\
  \bibnamefont {Chang}}\ and\ \bibinfo {author} {\bibfnamefont {Anthony~J}\
  \bibnamefont {Campillo}},\ }\href@noop {} {\emph {\bibinfo {title} {Optical
  processes in microcavities}}},\ Vol.~\bibinfo {volume} {3}\ (\bibinfo
  {publisher} {World scientific},\ \bibinfo {year} {1996})\BibitemShut
  {NoStop}%
\bibitem [{\citenamefont {Bergman}\ and\ \citenamefont
  {Stockman}(2003)}]{bergman_surface_2003}%
  \BibitemOpen
  \bibfield  {author} {\bibinfo {author} {\bibfnamefont {D.~J.}\ \bibnamefont
  {Bergman}}\ and\ \bibinfo {author} {\bibfnamefont {M.~I.}\ \bibnamefont
  {Stockman}},\ }\bibfield  {title} {\enquote {\bibinfo {title} {Surface
  plasmon amplification by stimulated emission of radiation: Quantum generation
  of coherent surface plasmons in nanosystems},}\ }\href {\doibase
  10.1103/PhysRevLett.90.027402} {\bibfield  {journal} {\bibinfo  {journal}
  {Phys. Rev. Lett.}\ }\textbf {\bibinfo {volume} {90}},\ \bibinfo {pages}
  {027402} (\bibinfo {year} {2003})}\BibitemShut {NoStop}%
\bibitem [{\citenamefont {Maier}(2007)}]{maier_plasmonics:_2007}%
  \BibitemOpen
  \bibfield  {author} {\bibinfo {author} {\bibfnamefont {Stefan~Alexander}\
  \bibnamefont {Maier}},\ }\href@noop {} {\emph {\bibinfo {title} {Plasmonics:
  fundamentals and applications}}}\ (\bibinfo  {publisher} {Springer Science \&
  Business Media},\ \bibinfo {year} {2007})\BibitemShut {NoStop}%
\bibitem [{\citenamefont {Noginov}\ \emph {et~al.}(2009)\citenamefont
  {Noginov}, \citenamefont {Zhu}, \citenamefont {Belgrave}, \citenamefont
  {Bakker}, \citenamefont {Shalaev}, \citenamefont {Narimanov}, \citenamefont
  {Stout}, \citenamefont {Herz}, \citenamefont {Suteewong},\ and\ \citenamefont
  {Wiesner}}]{noginov_demonstration_2009}%
  \BibitemOpen
  \bibfield  {author} {\bibinfo {author} {\bibfnamefont {M.~A.}\ \bibnamefont
  {Noginov}}, \bibinfo {author} {\bibfnamefont {G.}~\bibnamefont {Zhu}},
  \bibinfo {author} {\bibfnamefont {A.~M.}\ \bibnamefont {Belgrave}}, \bibinfo
  {author} {\bibfnamefont {R.}~\bibnamefont {Bakker}}, \bibinfo {author}
  {\bibfnamefont {V.~M.}\ \bibnamefont {Shalaev}}, \bibinfo {author}
  {\bibfnamefont {E.~E.}\ \bibnamefont {Narimanov}}, \bibinfo {author}
  {\bibfnamefont {S.}~\bibnamefont {Stout}}, \bibinfo {author} {\bibfnamefont
  {E.}~\bibnamefont {Herz}}, \bibinfo {author} {\bibfnamefont {T.}~\bibnamefont
  {Suteewong}}, \ and\ \bibinfo {author} {\bibfnamefont {U.}~\bibnamefont
  {Wiesner}},\ }\bibfield  {title} {\enquote {\bibinfo {title} {Demonstration
  of a spaser-based nanolaser},}\ }\href {\doibase 10.1038/nature08318}
  {\bibfield  {journal} {\bibinfo  {journal} {Nature}\ }\textbf {\bibinfo
  {volume} {460}},\ \bibinfo {pages} {1110--1112} (\bibinfo {year}
  {2009})}\BibitemShut {NoStop}%
\bibitem [{\citenamefont {Novotny}\ and\ \citenamefont {van
  Hulst}(2011)}]{novotny_antennas_2011}%
  \BibitemOpen
  \bibfield  {author} {\bibinfo {author} {\bibfnamefont {Lukas}\ \bibnamefont
  {Novotny}}\ and\ \bibinfo {author} {\bibfnamefont {Niek}\ \bibnamefont {van
  Hulst}},\ }\bibfield  {title} {\enquote {\bibinfo {title} {Antennas for
  light},}\ }\href {\doibase 10.1038/nphoton.2010.237} {\bibfield  {journal}
  {\bibinfo  {journal} {Nature Photonics}\ }\textbf {\bibinfo {volume} {5}},\
  \bibinfo {pages} {83--90} (\bibinfo {year} {2011})}\BibitemShut {NoStop}%
\bibitem [{\citenamefont {Chang}\ \emph {et~al.}(2007)\citenamefont {Chang},
  \citenamefont {Sørensen}, \citenamefont {Demler},\ and\ \citenamefont
  {Lukin}}]{chang_single-photon_2007}%
  \BibitemOpen
  \bibfield  {author} {\bibinfo {author} {\bibfnamefont {Darrick~E.}\
  \bibnamefont {Chang}}, \bibinfo {author} {\bibfnamefont {Anders~S.}\
  \bibnamefont {Sørensen}}, \bibinfo {author} {\bibfnamefont {Eugene~A.}\
  \bibnamefont {Demler}}, \ and\ \bibinfo {author} {\bibfnamefont {Mikhail~D.}\
  \bibnamefont {Lukin}},\ }\bibfield  {title} {\enquote {\bibinfo {title} {A
  single-photon transistor using nanoscale surface plasmons},}\ }\href
  {https://doi.org/10.1038/nphys708} {\bibfield  {journal} {\bibinfo  {journal}
  {Nature Physics}\ }\textbf {\bibinfo {volume} {3}},\ \bibinfo {pages} {807}
  (\bibinfo {year} {2007})}\BibitemShut {NoStop}%
\bibitem [{\citenamefont {Andersen}\ \emph {et~al.}(2011)\citenamefont
  {Andersen}, \citenamefont {Stobbe}, \citenamefont {Sørensen},\ and\
  \citenamefont {Lodahl}}]{andersen_strongly_2011}%
  \BibitemOpen
  \bibfield  {author} {\bibinfo {author} {\bibfnamefont {Mads~Lykke}\
  \bibnamefont {Andersen}}, \bibinfo {author} {\bibfnamefont {Søren}\
  \bibnamefont {Stobbe}}, \bibinfo {author} {\bibfnamefont {Anders~Søndberg}\
  \bibnamefont {Sørensen}}, \ and\ \bibinfo {author} {\bibfnamefont {Peter}\
  \bibnamefont {Lodahl}},\ }\bibfield  {title} {\enquote {\bibinfo {title}
  {Strongly modified plasmon–matter interaction with mesoscopic quantum
  emitters},}\ }\href {\doibase 10.1038/nphys1870} {\bibfield  {journal}
  {\bibinfo  {journal} {Nature Physics}\ }\textbf {\bibinfo {volume} {7}},\
  \bibinfo {pages} {215--218} (\bibinfo {year} {2011})}\BibitemShut {NoStop}%
\bibitem [{\citenamefont {Jacob}\ and\ \citenamefont
  {Shalaev}(2011)}]{jacob_plasmonics_2011}%
  \BibitemOpen
  \bibfield  {author} {\bibinfo {author} {\bibfnamefont {Zubin}\ \bibnamefont
  {Jacob}}\ and\ \bibinfo {author} {\bibfnamefont {Vladimir~M.}\ \bibnamefont
  {Shalaev}},\ }\bibfield  {title} {\enquote {\bibinfo {title} {Plasmonics goes
  quantum},}\ }\href {\doibase 10.1126/science.1211736} {\bibfield  {journal}
  {\bibinfo  {journal} {{Science}}\ }\textbf {\bibinfo {volume} {334}},\
  \bibinfo {pages} {463--464} (\bibinfo {year} {2011})}\BibitemShut {NoStop}%
\bibitem [{\citenamefont {Tame}\ \emph {et~al.}(2013)\citenamefont {Tame},
  \citenamefont {{McEnery}}, \citenamefont {Özdemir}, \citenamefont {Lee},
  \citenamefont {Maier},\ and\ \citenamefont {Kim}}]{tame_quantum_2013}%
  \BibitemOpen
  \bibfield  {author} {\bibinfo {author} {\bibfnamefont {M.~S.}\ \bibnamefont
  {Tame}}, \bibinfo {author} {\bibfnamefont {K.~R.}\ \bibnamefont {{McEnery}}},
  \bibinfo {author} {\bibfnamefont {Ş.~K.}\ \bibnamefont {Özdemir}}, \bibinfo
  {author} {\bibfnamefont {J.}~\bibnamefont {Lee}}, \bibinfo {author}
  {\bibfnamefont {S.~A.}\ \bibnamefont {Maier}}, \ and\ \bibinfo {author}
  {\bibfnamefont {M.~S.}\ \bibnamefont {Kim}},\ }\bibfield  {title} {\enquote
  {\bibinfo {title} {Quantum plasmonics},}\ }\href {\doibase 10.1038/nphys2615}
  {\bibfield  {journal} {\bibinfo  {journal} {Nature Physics}\ }\textbf
  {\bibinfo {volume} {9}},\ \bibinfo {pages} {329--340} (\bibinfo {year}
  {2013})}\BibitemShut {NoStop}%
\bibitem [{\citenamefont {Berini}\ and\ \citenamefont
  {De~Leon}(2012)}]{berini_surface_2012}%
  \BibitemOpen
  \bibfield  {author} {\bibinfo {author} {\bibfnamefont {Pierre}\ \bibnamefont
  {Berini}}\ and\ \bibinfo {author} {\bibfnamefont {Israel}\ \bibnamefont
  {De~Leon}},\ }\bibfield  {title} {\enquote {\bibinfo {title} {Surface
  plasmon–polariton amplifiers and lasers},}\ }\href {\doibase
  10.1038/nphoton.2011.285} {\bibfield  {journal} {\bibinfo  {journal} {Nature
  Photonics}\ }\textbf {\bibinfo {volume} {6}},\ \bibinfo {pages} {16--24}
  (\bibinfo {year} {2012})}\BibitemShut {NoStop}%
\bibitem [{\citenamefont {Chikkaraddy}\ \emph {et~al.}(2016)\citenamefont
  {Chikkaraddy}, \citenamefont {de~Nijs}, \citenamefont {Benz}, \citenamefont
  {Barrow}, \citenamefont {Scherman}, \citenamefont {Rosta}, \citenamefont
  {Demetriadou}, \citenamefont {Fox}, \citenamefont {Hess},\ and\ \citenamefont
  {Baumberg}}]{Chikkaraddy2016}%
  \BibitemOpen
  \bibfield  {author} {\bibinfo {author} {\bibfnamefont {Rohit}\ \bibnamefont
  {Chikkaraddy}}, \bibinfo {author} {\bibfnamefont {Bart}\ \bibnamefont
  {de~Nijs}}, \bibinfo {author} {\bibfnamefont {Felix}\ \bibnamefont {Benz}},
  \bibinfo {author} {\bibfnamefont {Steven~J.}\ \bibnamefont {Barrow}},
  \bibinfo {author} {\bibfnamefont {Oren~A.}\ \bibnamefont {Scherman}},
  \bibinfo {author} {\bibfnamefont {Edina}\ \bibnamefont {Rosta}}, \bibinfo
  {author} {\bibfnamefont {Angela}\ \bibnamefont {Demetriadou}}, \bibinfo
  {author} {\bibfnamefont {Peter}\ \bibnamefont {Fox}}, \bibinfo {author}
  {\bibfnamefont {Ortwin}\ \bibnamefont {Hess}}, \ and\ \bibinfo {author}
  {\bibfnamefont {Jeremy~J.}\ \bibnamefont {Baumberg}},\ }\bibfield  {title}
  {\enquote {\bibinfo {title} {Single-molecule strong coupling at room
  temperature in plasmonic nanocavities},}\ }\href {\doibase
  10.1038/nature17974} {\bibfield  {journal} {\bibinfo  {journal} {Nature}\
  }\textbf {\bibinfo {volume} {535}},\ \bibinfo {pages} {127--130} (\bibinfo
  {year} {2016})}\BibitemShut {NoStop}%
\bibitem [{\citenamefont {Benz}\ \emph {et~al.}(2016)\citenamefont {Benz},
  \citenamefont {Schmidt}, \citenamefont {Dreismann}, \citenamefont
  {Chikkaraddy}, \citenamefont {Zhang}, \citenamefont {Demetriadou},
  \citenamefont {Carnegie}, \citenamefont {Ohadi}, \citenamefont {de~Nijs},
  \citenamefont {Esteban}, \citenamefont {Aizpurua},\ and\ \citenamefont
  {Baumberg}}]{Benz2016}%
  \BibitemOpen
  \bibfield  {author} {\bibinfo {author} {\bibfnamefont {F.}~\bibnamefont
  {Benz}}, \bibinfo {author} {\bibfnamefont {M.~K.}\ \bibnamefont {Schmidt}},
  \bibinfo {author} {\bibfnamefont {A.}~\bibnamefont {Dreismann}}, \bibinfo
  {author} {\bibfnamefont {R.}~\bibnamefont {Chikkaraddy}}, \bibinfo {author}
  {\bibfnamefont {Y.}~\bibnamefont {Zhang}}, \bibinfo {author} {\bibfnamefont
  {A.}~\bibnamefont {Demetriadou}}, \bibinfo {author} {\bibfnamefont
  {C.}~\bibnamefont {Carnegie}}, \bibinfo {author} {\bibfnamefont
  {H.}~\bibnamefont {Ohadi}}, \bibinfo {author} {\bibfnamefont
  {B.}~\bibnamefont {de~Nijs}}, \bibinfo {author} {\bibfnamefont
  {R.}~\bibnamefont {Esteban}}, \bibinfo {author} {\bibfnamefont
  {J.}~\bibnamefont {Aizpurua}}, \ and\ \bibinfo {author} {\bibfnamefont
  {J.~J.}\ \bibnamefont {Baumberg}},\ }\bibfield  {title} {\enquote {\bibinfo
  {title} {{Single-molecule optomechanics in ``picocavities''}},}\ }\href
  {\doibase 10.1126/science.aah5243} {\bibfield  {journal} {\bibinfo  {journal}
  {Science}\ }\textbf {\bibinfo {volume} {354}},\ \bibinfo {pages} {726--729}
  (\bibinfo {year} {2016})}\BibitemShut {NoStop}%
\bibitem [{\citenamefont {Morse}\ and\ \citenamefont
  {Feshbach}(1954)}]{morse_methods_1954}%
  \BibitemOpen
  \bibfield  {author} {\bibinfo {author} {\bibfnamefont {Philip~M.}\
  \bibnamefont {Morse}}\ and\ \bibinfo {author} {\bibfnamefont {Herman}\
  \bibnamefont {Feshbach}},\ }\bibfield  {title} {\enquote {\bibinfo {title}
  {Methods of theoretical physics},}\ }\href {\doibase 10.1119/1.1933765}
  {\bibfield  {journal} {\bibinfo  {journal} {American Journal of Physics}\
  }\textbf {\bibinfo {volume} {22}},\ \bibinfo {pages} {410--413} (\bibinfo
  {year} {1954})}\BibitemShut {NoStop}%
\bibitem [{\citenamefont {Kristensen}\ \emph {et~al.}(2012)\citenamefont
  {Kristensen}, \citenamefont {Van~Vlack},\ and\ \citenamefont
  {Hughes}}]{kristensen_generalized_2012}%
  \BibitemOpen
  \bibfield  {author} {\bibinfo {author} {\bibfnamefont {P.~T.}\ \bibnamefont
  {Kristensen}}, \bibinfo {author} {\bibfnamefont {C.}~\bibnamefont
  {Van~Vlack}}, \ and\ \bibinfo {author} {\bibfnamefont {S.}~\bibnamefont
  {Hughes}},\ }\bibfield  {title} {{\selectlanguage {en}\enquote {\bibinfo
  {title} {Generalized effective mode volume for leaky optical cavities},}\
  }}\href {\doibase 10.1364/OL.37.001649} {\bibfield  {journal} {\bibinfo
  {journal} {Optics Letters}\ }\textbf {\bibinfo {volume} {37}},\ \bibinfo
  {pages} {1649} (\bibinfo {year} {2012})}\BibitemShut {NoStop}%
\bibitem [{\citenamefont {Kristensen}\ and\ \citenamefont
  {Hughes}(2014)}]{kristensen_modes_2014}%
  \BibitemOpen
  \bibfield  {author} {\bibinfo {author} {\bibfnamefont {Philip~Tr{\o}st}\
  \bibnamefont {Kristensen}}\ and\ \bibinfo {author} {\bibfnamefont {Stephen}\
  \bibnamefont {Hughes}},\ }\bibfield  {title} {{\selectlanguage {en}\enquote
  {\bibinfo {title} {Modes and {Mode} {Volumes} of {Leaky} {Optical} {Cavities}
  and {Plasmonic} {Nanoresonators}},}\ }}\href {\doibase 10.1021/ph400114e}
  {\bibfield  {journal} {\bibinfo  {journal} {ACS Photonics}\ }\textbf
  {\bibinfo {volume} {1}},\ \bibinfo {pages} {2--10} (\bibinfo {year}
  {2014})}\BibitemShut {NoStop}%
\bibitem [{\citenamefont {Lai}\ \emph {et~al.}(1990)\citenamefont {Lai},
  \citenamefont {Leung}, \citenamefont {Young}, \citenamefont {Barber},\ and\
  \citenamefont {Hill}}]{lai_time-independent_1990}%
  \BibitemOpen
  \bibfield  {author} {\bibinfo {author} {\bibfnamefont {H.~M.}\ \bibnamefont
  {Lai}}, \bibinfo {author} {\bibfnamefont {P.~T.}\ \bibnamefont {Leung}},
  \bibinfo {author} {\bibfnamefont {K.}~\bibnamefont {Young}}, \bibinfo
  {author} {\bibfnamefont {P.~W.}\ \bibnamefont {Barber}}, \ and\ \bibinfo
  {author} {\bibfnamefont {S.~C.}\ \bibnamefont {Hill}},\ }\bibfield  {title}
  {\enquote {\bibinfo {title} {Time-independent perturbation for leaking
  electromagnetic modes in open systems with application to resonances in
  microdroplets},}\ }\href {\doibase 10.1103/PhysRevA.41.5187} {\bibfield
  {journal} {\bibinfo  {journal} {Phys. Rev. A}\ }\textbf {\bibinfo {volume}
  {41}},\ \bibinfo {pages} {5187--5198} (\bibinfo {year} {1990})}\BibitemShut
  {NoStop}%
\bibitem [{\citenamefont {Leung}\ \emph
  {et~al.}(1994{\natexlab{a}})\citenamefont {Leung}, \citenamefont {Liu},\ and\
  \citenamefont {Young}}]{leung_completeness_1994}%
  \BibitemOpen
  \bibfield  {author} {\bibinfo {author} {\bibfnamefont {P.~T.}\ \bibnamefont
  {Leung}}, \bibinfo {author} {\bibfnamefont {S.~Y.}\ \bibnamefont {Liu}}, \
  and\ \bibinfo {author} {\bibfnamefont {K.}~\bibnamefont {Young}},\ }\bibfield
   {title} {{\selectlanguage {en}\enquote {\bibinfo {title} {Completeness and
  orthogonality of quasinormal modes in leaky optical cavities},}\ }}\href
  {\doibase 10.1103/PhysRevA.49.3057} {\bibfield  {journal} {\bibinfo
  {journal} {Physical Review A}\ }\textbf {\bibinfo {volume} {49}},\ \bibinfo
  {pages} {3057--3067} (\bibinfo {year} {1994}{\natexlab{a}})}\BibitemShut
  {NoStop}%
\bibitem [{\citenamefont {Leung}\ \emph
  {et~al.}(1994{\natexlab{b}})\citenamefont {Leung}, \citenamefont {Liu},
  \citenamefont {Tong},\ and\ \citenamefont
  {Young}}]{leung_time-independent_1994}%
  \BibitemOpen
  \bibfield  {author} {\bibinfo {author} {\bibfnamefont {P.~T.}\ \bibnamefont
  {Leung}}, \bibinfo {author} {\bibfnamefont {S.~Y.}\ \bibnamefont {Liu}},
  \bibinfo {author} {\bibfnamefont {S.~S.}\ \bibnamefont {Tong}}, \ and\
  \bibinfo {author} {\bibfnamefont {K.}~\bibnamefont {Young}},\ }\bibfield
  {title} {{\selectlanguage {en}\enquote {\bibinfo {title} {Time-independent
  perturbation theory for quasinormal modes in leaky optical cavities},}\
  }}\href {\doibase 10.1103/PhysRevA.49.3068} {\bibfield  {journal} {\bibinfo
  {journal} {Physical Review A}\ }\textbf {\bibinfo {volume} {49}},\ \bibinfo
  {pages} {3068--3073} (\bibinfo {year} {1994}{\natexlab{b}})}\BibitemShut
  {NoStop}%
\bibitem [{\citenamefont {Leung}\ and\ \citenamefont
  {Pang}(1996)}]{leung_completeness_1996}%
  \BibitemOpen
  \bibfield  {author} {\bibinfo {author} {\bibfnamefont {P.~T.}\ \bibnamefont
  {Leung}}\ and\ \bibinfo {author} {\bibfnamefont {K.~M.}\ \bibnamefont
  {Pang}},\ }\bibfield  {title} {{\selectlanguage {EN}\enquote {\bibinfo
  {title} {Completeness and time-independent perturbation of
  morphology-dependent resonances in dielectric spheres},}\ }}\href {\doibase
  10.1364/JOSAB.13.000805} {\bibfield  {journal} {\bibinfo  {journal} {JOSAB}\
  }\textbf {\bibinfo {volume} {13}},\ \bibinfo {pages} {805--817} (\bibinfo
  {year} {1996})}\BibitemShut {NoStop}%
\bibitem [{\citenamefont {Lee}\ \emph {et~al.}(1999)\citenamefont {Lee},
  \citenamefont {Leung},\ and\ \citenamefont {Pang}}]{lee_dyadic_1999}%
  \BibitemOpen
  \bibfield  {author} {\bibinfo {author} {\bibfnamefont {{KM}}~\bibnamefont
  {Lee}}, \bibinfo {author} {\bibfnamefont {{PT}}~\bibnamefont {Leung}}, \ and\
  \bibinfo {author} {\bibfnamefont {{KM}}~\bibnamefont {Pang}},\ }\bibfield
  {title} {\enquote {\bibinfo {title} {Dyadic formulation of
  morphology-dependent resonances. i. completeness relation},}\ }\href
  {\doibase 10.1364/JOSAB.16.001409} {\bibfield  {journal} {\bibinfo  {journal}
  {JOSAB}\ }\textbf {\bibinfo {volume} {16}},\ \bibinfo {pages} {1409--1417}
  (\bibinfo {year} {1999})}\BibitemShut {NoStop}%
\bibitem [{\citenamefont {Sauvan}\ \emph {et~al.}(2013)\citenamefont {Sauvan},
  \citenamefont {Hugonin}, \citenamefont {Maksymov},\ and\ \citenamefont
  {Lalanne}}]{sauvan_theory_2013}%
  \BibitemOpen
  \bibfield  {author} {\bibinfo {author} {\bibfnamefont {C.}~\bibnamefont
  {Sauvan}}, \bibinfo {author} {\bibfnamefont {J.~P.}\ \bibnamefont {Hugonin}},
  \bibinfo {author} {\bibfnamefont {I.~S.}\ \bibnamefont {Maksymov}}, \ and\
  \bibinfo {author} {\bibfnamefont {P.}~\bibnamefont {Lalanne}},\ }\bibfield
  {title} {{\selectlanguage {en}\enquote {\bibinfo {title} {Theory of the
  {Spontaneous} {Optical} {Emission} of {Nanosize} {Photonic} and {Plasmon}
  {Resonators}},}\ }}\href {\doibase 10.1103/PhysRevLett.110.237401} {\bibfield
   {journal} {\bibinfo  {journal} {Physical Review Letters}\ }\textbf {\bibinfo
  {volume} {110}},\ \bibinfo {pages} {237401} (\bibinfo {year}
  {2013})}\BibitemShut {NoStop}%
\bibitem [{\citenamefont {Bai}\ \emph {et~al.}(2013)\citenamefont {Bai},
  \citenamefont {Perrin}, \citenamefont {Sauvan}, \citenamefont {Hugonin},\
  and\ \citenamefont {Lalanne}}]{bai_efficient_2013-1}%
  \BibitemOpen
  \bibfield  {author} {\bibinfo {author} {\bibfnamefont {Q.}~\bibnamefont
  {Bai}}, \bibinfo {author} {\bibfnamefont {M.}~\bibnamefont {Perrin}},
  \bibinfo {author} {\bibfnamefont {C.}~\bibnamefont {Sauvan}}, \bibinfo
  {author} {\bibfnamefont {J.-P.}\ \bibnamefont {Hugonin}}, \ and\ \bibinfo
  {author} {\bibfnamefont {P.}~\bibnamefont {Lalanne}},\ }\bibfield  {title}
  {{\selectlanguage {EN}\enquote {\bibinfo {title} {Efficient and intuitive
  method for the analysis of light scattering by a resonant nanostructure},}\
  }}\href {\doibase 10.1364/OE.21.027371} {\bibfield  {journal} {\bibinfo
  {journal} {Optics Express}\ }\textbf {\bibinfo {volume} {21}},\ \bibinfo
  {pages} {27371--27382} (\bibinfo {year} {2013})}\BibitemShut {NoStop}%
\bibitem [{\citenamefont {Zschiedrich}\ \emph {et~al.}(2018)\citenamefont
  {Zschiedrich}, \citenamefont {Binkowski}, \citenamefont {Nikolay},
  \citenamefont {Benson}, \citenamefont {Kewes},\ and\ \citenamefont
  {Burger}}]{PhysRevA.98.043806}%
  \BibitemOpen
  \bibfield  {author} {\bibinfo {author} {\bibfnamefont {Lin}\ \bibnamefont
  {Zschiedrich}}, \bibinfo {author} {\bibfnamefont {Felix}\ \bibnamefont
  {Binkowski}}, \bibinfo {author} {\bibfnamefont {Niko}\ \bibnamefont
  {Nikolay}}, \bibinfo {author} {\bibfnamefont {Oliver}\ \bibnamefont
  {Benson}}, \bibinfo {author} {\bibfnamefont {G\"unter}\ \bibnamefont
  {Kewes}}, \ and\ \bibinfo {author} {\bibfnamefont {Sven}\ \bibnamefont
  {Burger}},\ }\bibfield  {title} {\enquote {\bibinfo {title}
  {Riesz-projection-based theory of light-matter interaction in dispersive
  nanoresonators},}\ }\href {\doibase 10.1103/PhysRevA.98.043806} {\bibfield
  {journal} {\bibinfo  {journal} {Phys. Rev. A}\ }\textbf {\bibinfo {volume}
  {98}},\ \bibinfo {pages} {043806} (\bibinfo {year} {2018})}\BibitemShut
  {NoStop}%
\bibitem [{\citenamefont {Lalanne}\ \emph {et~al.}(2018)\citenamefont
  {Lalanne}, \citenamefont {Yan}, \citenamefont {Vynck}, \citenamefont
  {Sauvan},\ and\ \citenamefont {Hugonin}}]{lalanne_light_2018}%
  \BibitemOpen
  \bibfield  {author} {\bibinfo {author} {\bibfnamefont {Philippe}\
  \bibnamefont {Lalanne}}, \bibinfo {author} {\bibfnamefont {Wei}\ \bibnamefont
  {Yan}}, \bibinfo {author} {\bibfnamefont {Kevin}\ \bibnamefont {Vynck}},
  \bibinfo {author} {\bibfnamefont {Christophe}\ \bibnamefont {Sauvan}}, \ and\
  \bibinfo {author} {\bibfnamefont {Jean-Paul}\ \bibnamefont {Hugonin}},\
  }\bibfield  {title} {\enquote {\bibinfo {title} {Light interaction with
  photonic and plasmonic resonances},}\ }\href {\doibase
  10.1002/lpor.201700113} {\bibfield  {journal} {\bibinfo  {journal} {Laser \&
  Photonics Reviews}\ }\textbf {\bibinfo {volume} {12}},\ \bibinfo {pages}
  {1700113} (\bibinfo {year} {2018})}\BibitemShut {NoStop}%
\bibitem [{\citenamefont {Kristensen}\ \emph {et~al.}(2019)\citenamefont
  {Kristensen}, \citenamefont {Herrmann}, \citenamefont {Intravaia},\ and\
  \citenamefont {Busch}}]{1910.05412}%
  \BibitemOpen
  \bibfield  {author} {\bibinfo {author} {\bibfnamefont {Philip~Trøst}\
  \bibnamefont {Kristensen}}, \bibinfo {author} {\bibfnamefont {Kathrin}\
  \bibnamefont {Herrmann}}, \bibinfo {author} {\bibfnamefont {Francesco}\
  \bibnamefont {Intravaia}}, \ and\ \bibinfo {author} {\bibfnamefont {Kurt}\
  \bibnamefont {Busch}},\ }\href@noop {} {\enquote {\bibinfo {title} {Modeling
  electromagnetic resonators using quasinormal modes},}\ } (\bibinfo {year}
  {2019}),\ \Eprint {http://arxiv.org/abs/arXiv:1910.05412} {arXiv:1910.05412}
  \BibitemShut {NoStop}%
\bibitem [{\citenamefont {Muljarov}\ \emph {et~al.}(2010)\citenamefont
  {Muljarov}, \citenamefont {Langbein},\ and\ \citenamefont
  {Zimmermann}}]{muljarov_brillouin-wigner_2010}%
  \BibitemOpen
  \bibfield  {author} {\bibinfo {author} {\bibfnamefont {E.~A.}\ \bibnamefont
  {Muljarov}}, \bibinfo {author} {\bibfnamefont {W.}~\bibnamefont {Langbein}},
  \ and\ \bibinfo {author} {\bibfnamefont {R.}~\bibnamefont {Zimmermann}},\
  }\bibfield  {title} {\enquote {\bibinfo {title} {Brillouin-wigner
  perturbation theory in open electromagnetic systems},}\ }\href {\doibase
  10.1209/0295-5075/92/50010} {\bibfield  {journal} {\bibinfo  {journal}
  {{EPL}}\ }\textbf {\bibinfo {volume} {92}},\ \bibinfo {pages} {50010}
  (\bibinfo {year} {2010})}\BibitemShut {NoStop}%
\bibitem [{\citenamefont {Ge}\ \emph {et~al.}(2014)\citenamefont {Ge},
  \citenamefont {Kristensen}, \citenamefont {Young},\ and\ \citenamefont
  {Hughes}}]{ge_quasinormal_2014}%
  \BibitemOpen
  \bibfield  {author} {\bibinfo {author} {\bibfnamefont {Rong-Chun}\
  \bibnamefont {Ge}}, \bibinfo {author} {\bibfnamefont {Philip~Tr{\o}st}\
  \bibnamefont {Kristensen}}, \bibinfo {author} {\bibfnamefont {Jeff~F}\
  \bibnamefont {Young}}, \ and\ \bibinfo {author} {\bibfnamefont {Stephen}\
  \bibnamefont {Hughes}},\ }\bibfield  {title} {\enquote {\bibinfo {title}
  {Quasinormal mode approach to modelling light-emission and propagation in
  nanoplasmonics},}\ }\href {\doibase 10.1088/1367-2630/16/11/113048}
  {\bibfield  {journal} {\bibinfo  {journal} {New Journal of Physics}\ }\textbf
  {\bibinfo {volume} {16}},\ \bibinfo {pages} {113048} (\bibinfo {year}
  {2014})}\BibitemShut {NoStop}%
\bibitem [{\citenamefont {Kamandar~Dezfouli}\ \emph {et~al.}(2017)\citenamefont
  {Kamandar~Dezfouli}, \citenamefont {Gordon},\ and\ \citenamefont
  {Hughes}}]{2017PRA_hybrid}%
  \BibitemOpen
  \bibfield  {author} {\bibinfo {author} {\bibfnamefont {Mohsen}\ \bibnamefont
  {Kamandar~Dezfouli}}, \bibinfo {author} {\bibfnamefont {Reuven}\ \bibnamefont
  {Gordon}}, \ and\ \bibinfo {author} {\bibfnamefont {Stephen}\ \bibnamefont
  {Hughes}},\ }\bibfield  {title} {\enquote {\bibinfo {title} {Modal theory of
  modified spontaneous emission of a quantum emitter in a hybrid plasmonic
  photonic-crystal cavity system},}\ }\href {\doibase
  10.1103/PhysRevA.95.013846} {\bibfield  {journal} {\bibinfo  {journal} {Phys.
  Rev. A}\ }\textbf {\bibinfo {volume} {95}},\ \bibinfo {pages} {013846}
  (\bibinfo {year} {2017})}\BibitemShut {NoStop}%
\bibitem [{\citenamefont {Fernández-Domínguez}\ \emph
  {et~al.}(2018)\citenamefont {Fernández-Domínguez}, \citenamefont
  {Bozhevolnyi},\ and\ \citenamefont
  {Mortensen}}]{fernandez-dominguez_plasmon-enhanced_2018}%
  \BibitemOpen
  \bibfield  {author} {\bibinfo {author} {\bibfnamefont {Antonio~I.}\
  \bibnamefont {Fernández-Domínguez}}, \bibinfo {author} {\bibfnamefont
  {Sergey~I.}\ \bibnamefont {Bozhevolnyi}}, \ and\ \bibinfo {author}
  {\bibfnamefont {N.~Asger}\ \bibnamefont {Mortensen}},\ }\bibfield  {title}
  {\enquote {\bibinfo {title} {Plasmon-enhanced generation of nonclassical
  light},}\ }\href {\doibase 10.1021/acsphotonics.8b00852} {\bibfield
  {journal} {\bibinfo  {journal} {{ACS} Photonics}\ }\textbf {\bibinfo {volume}
  {5}},\ \bibinfo {pages} {3447--3451} (\bibinfo {year} {2018})}\BibitemShut
  {NoStop}%
\bibitem [{\citenamefont {Ho}\ \emph {et~al.}(1998)\citenamefont {Ho},
  \citenamefont {Leung}, \citenamefont {Maassen van~den Brink},\ and\
  \citenamefont {Young}}]{ho_second_1998}%
  \BibitemOpen
  \bibfield  {author} {\bibinfo {author} {\bibfnamefont {K.~C.}\ \bibnamefont
  {Ho}}, \bibinfo {author} {\bibfnamefont {P.~T.}\ \bibnamefont {Leung}},
  \bibinfo {author} {\bibfnamefont {Alec}\ \bibnamefont {Maassen van~den
  Brink}}, \ and\ \bibinfo {author} {\bibfnamefont {K.}~\bibnamefont {Young}},\
  }\bibfield  {title} {\enquote {\bibinfo {title} {Second quantization of open
  systems using quasinormal modes},}\ }\href {\doibase
  10.1103/PhysRevE.58.2965} {\bibfield  {journal} {\bibinfo  {journal} {Phys.
  Rev. E}\ }\textbf {\bibinfo {volume} {58}},\ \bibinfo {pages} {2965--2978}
  (\bibinfo {year} {1998})}\BibitemShut {NoStop}%
\bibitem [{\citenamefont {Severini}\ \emph {et~al.}(2004)\citenamefont
  {Severini}, \citenamefont {Settimi}, \citenamefont {Sibilia}, \citenamefont
  {Bertolotti}, \citenamefont {Napoli},\ and\ \citenamefont
  {Messina}}]{severini_second_2004}%
  \BibitemOpen
  \bibfield  {author} {\bibinfo {author} {\bibfnamefont {S.}~\bibnamefont
  {Severini}}, \bibinfo {author} {\bibfnamefont {A.}~\bibnamefont {Settimi}},
  \bibinfo {author} {\bibfnamefont {C.}~\bibnamefont {Sibilia}}, \bibinfo
  {author} {\bibfnamefont {M.}~\bibnamefont {Bertolotti}}, \bibinfo {author}
  {\bibfnamefont {A.}~\bibnamefont {Napoli}}, \ and\ \bibinfo {author}
  {\bibfnamefont {A.}~\bibnamefont {Messina}},\ }\bibfield  {title} {\enquote
  {\bibinfo {title} {Second quantization and atomic spontaneous emission inside
  one-dimensional photonic crystals via a quasinormal-modes approach},}\ }\href
  {\doibase 10.1103/PhysRevE.70.056614} {\bibfield  {journal} {\bibinfo
  {journal} {Phys. Rev. E}\ }\textbf {\bibinfo {volume} {70}},\ \bibinfo
  {pages} {056614} (\bibinfo {year} {2004})}\BibitemShut {NoStop}%
\bibitem [{\citenamefont {Franke}\ \emph {et~al.}(2019)\citenamefont {Franke},
  \citenamefont {Hughes}, \citenamefont {Dezfouli}, \citenamefont {Kristensen},
  \citenamefont {Busch}, \citenamefont {Knorr},\ and\ \citenamefont
  {Richter}}]{franke_quantization_2018}%
  \BibitemOpen
  \bibfield  {author} {\bibinfo {author} {\bibfnamefont {Sebastian}\
  \bibnamefont {Franke}}, \bibinfo {author} {\bibfnamefont {Stephen}\
  \bibnamefont {Hughes}}, \bibinfo {author} {\bibfnamefont {Mohsen~Kamandar}\
  \bibnamefont {Dezfouli}}, \bibinfo {author} {\bibfnamefont {Philip~Tr\o{}st}\
  \bibnamefont {Kristensen}}, \bibinfo {author} {\bibfnamefont {Kurt}\
  \bibnamefont {Busch}}, \bibinfo {author} {\bibfnamefont {Andreas}\
  \bibnamefont {Knorr}}, \ and\ \bibinfo {author} {\bibfnamefont {Marten}\
  \bibnamefont {Richter}},\ }\bibfield  {title} {\enquote {\bibinfo {title}
  {Quantization of quasinormal modes for open cavities and plasmonic cavity
  quantum electrodynamics},}\ }\href {\doibase 10.1103/PhysRevLett.122.213901}
  {\bibfield  {journal} {\bibinfo  {journal} {Phys. Rev. Lett.}\ }\textbf
  {\bibinfo {volume} {122}},\ \bibinfo {pages} {213901} (\bibinfo {year}
  {2019})}\BibitemShut {NoStop}%
\bibitem [{\citenamefont {Martin}(2006)}]{martin2006multiple}%
  \BibitemOpen
  \bibfield  {author} {\bibinfo {author} {\bibfnamefont {Paul~A}\ \bibnamefont
  {Martin}},\ }\href@noop {} {\emph {\bibinfo {title} {Multiple scattering:
  interaction of time-harmonic waves with N obstacles}}},\ \bibinfo {number}
  {107}\ (\bibinfo  {publisher} {Cambridge University Press},\ \bibinfo {year}
  {2006})\BibitemShut {NoStop}%
\bibitem [{\citenamefont {Kamandar~Dezfouli}\ and\ \citenamefont
  {Hughes}(2018)}]{kamandar_dezfouli_regularized_2018}%
  \BibitemOpen
  \bibfield  {author} {\bibinfo {author} {\bibfnamefont {Mohsen}\ \bibnamefont
  {Kamandar~Dezfouli}}\ and\ \bibinfo {author} {\bibfnamefont {Stephen}\
  \bibnamefont {Hughes}},\ }\bibfield  {title} {{\selectlanguage {en}\enquote
  {\bibinfo {title} {Regularized quasinormal modes for plasmonic resonators and
  open cavities},}\ }}\href {\doibase 10.1103/PhysRevB.97.115302} {\bibfield
  {journal} {\bibinfo  {journal} {Physical Review B}\ }\textbf {\bibinfo
  {volume} {97}},\ \bibinfo {pages} {115302} (\bibinfo {year}
  {2018})}\BibitemShut {NoStop}%
\bibitem [{\citenamefont {Barth}\ \emph {et~al.}(1992)\citenamefont {Barth},
  \citenamefont {McLeod},\ and\ \citenamefont {Ziolkowski}}]{Neartofar_1992}%
  \BibitemOpen
  \bibfield  {author} {\bibinfo {author} {\bibfnamefont {M.J.}\ \bibnamefont
  {Barth}}, \bibinfo {author} {\bibfnamefont {R.R.}\ \bibnamefont {McLeod}}, \
  and\ \bibinfo {author} {\bibfnamefont {R.W.}\ \bibnamefont {Ziolkowski}},\
  }\bibfield  {title} {\enquote {\bibinfo {title} {A near and far-field
  projection algorithm for finite-difference time-domain codes},}\ }\href
  {\doibase 10.1163/156939392x00995} {\bibfield  {journal} {\bibinfo  {journal}
  {Journal of Electromagnetic Waves and Applications}\ }\textbf {\bibinfo
  {volume} {6}},\ \bibinfo {pages} {5--18} (\bibinfo {year}
  {1992})}\BibitemShut {NoStop}%
\bibitem [{\citenamefont {Schneider}()}]{schneider_understanding_nodate}%
  \BibitemOpen
  \bibfield  {author} {\bibinfo {author} {\bibfnamefont {John~B}\ \bibnamefont
  {Schneider}},\ }\href@noop {} {\emph {\bibinfo {title} {Understanding the
  Finite-Difference Time-Domain Method, \url{www.eecs.wsu.edu/~schneidj/ufdtd,
  2010}}}}\BibitemShut {NoStop}%
\bibitem [{\citenamefont {Hughes}\ \emph {et~al.}(2019)\citenamefont {Hughes},
  \citenamefont {Franke}, \citenamefont {Gustin}, \citenamefont
  {Kamandar~Dezfouli}, \citenamefont {Knorr},\ and\ \citenamefont
  {Richter}}]{Hughes_SPS_2019}%
  \BibitemOpen
  \bibfield  {author} {\bibinfo {author} {\bibfnamefont {Stephen}\ \bibnamefont
  {Hughes}}, \bibinfo {author} {\bibfnamefont {Sebastian}\ \bibnamefont
  {Franke}}, \bibinfo {author} {\bibfnamefont {Chris}\ \bibnamefont {Gustin}},
  \bibinfo {author} {\bibfnamefont {Mohsen}\ \bibnamefont {Kamandar~Dezfouli}},
  \bibinfo {author} {\bibfnamefont {Andreas}\ \bibnamefont {Knorr}}, \ and\
  \bibinfo {author} {\bibfnamefont {Marten}\ \bibnamefont {Richter}},\
  }\bibfield  {title} {\enquote {\bibinfo {title} {Theory and limits of
  on-demand single-photon sources using plasmonic resonators: A quantized
  quasinormal mode approach},}\ }\href {\doibase 10.1021/acsphotonics.9b00849}
  {\bibfield  {journal} {\bibinfo  {journal} {{ACS} Photonics}\ }\textbf
  {\bibinfo {volume} {6}},\ \bibinfo {pages} {2168--2180} (\bibinfo {year}
  {2019})}\BibitemShut {NoStop}%
\bibitem [{\citenamefont {Kristensen}\ \emph {et~al.}(2015)\citenamefont
  {Kristensen}, \citenamefont {Ge},\ and\ \citenamefont
  {Hughes}}]{Kristensen2015}%
  \BibitemOpen
  \bibfield  {author} {\bibinfo {author} {\bibfnamefont {Philip~Tr\o{}st}\
  \bibnamefont {Kristensen}}, \bibinfo {author} {\bibfnamefont {Rong-Chun}\
  \bibnamefont {Ge}}, \ and\ \bibinfo {author} {\bibfnamefont {Stephen}\
  \bibnamefont {Hughes}},\ }\bibfield  {title} {\enquote {\bibinfo {title}
  {Normalization of quasinormal modes in leaky optical cavities and plasmonic
  resonators},}\ }\href {\doibase 10.1103/PhysRevA.92.053810} {\bibfield
  {journal} {\bibinfo  {journal} {Phys. Rev. A}\ }\textbf {\bibinfo {volume}
  {92}},\ \bibinfo {pages} {053810} (\bibinfo {year} {2015})}\BibitemShut
  {NoStop}%
\bibitem [{\citenamefont {Kristensen}\ \emph {et~al.}(2017)\citenamefont
  {Kristensen}, \citenamefont {de~Lasson}, \citenamefont {Heuck}, \citenamefont
  {Gregersen},\ and\ \citenamefont {Mork}}]{Kristensen_coupled_modes_2017}%
  \BibitemOpen
  \bibfield  {author} {\bibinfo {author} {\bibfnamefont {Philip~Trost}\
  \bibnamefont {Kristensen}}, \bibinfo {author} {\bibfnamefont
  {Jakob~Rosenkrantz}\ \bibnamefont {de~Lasson}}, \bibinfo {author}
  {\bibfnamefont {Mikkel}\ \bibnamefont {Heuck}}, \bibinfo {author}
  {\bibfnamefont {Niels}\ \bibnamefont {Gregersen}}, \ and\ \bibinfo {author}
  {\bibfnamefont {Jesper}\ \bibnamefont {Mork}},\ }\bibfield  {title} {\enquote
  {\bibinfo {title} {On the theory of coupled modes in optical cavity-waveguide
  structures},}\ }\href {\doibase 10.1109/jlt.2017.2714263} {\bibfield
  {journal} {\bibinfo  {journal} {Journal of Lightwave Technology}\ }\textbf
  {\bibinfo {volume} {35}},\ \bibinfo {pages} {4247--4259} (\bibinfo {year}
  {2017})}\BibitemShut {NoStop}%
\bibitem [{\citenamefont {Anger}\ \emph {et~al.}(2006)\citenamefont {Anger},
  \citenamefont {Bharadwaj},\ and\ \citenamefont {Novotny}}]{Anger2006}%
  \BibitemOpen
  \bibfield  {author} {\bibinfo {author} {\bibfnamefont {Pascal}\ \bibnamefont
  {Anger}}, \bibinfo {author} {\bibfnamefont {Palash}\ \bibnamefont
  {Bharadwaj}}, \ and\ \bibinfo {author} {\bibfnamefont {Lukas}\ \bibnamefont
  {Novotny}},\ }\bibfield  {title} {\enquote {\bibinfo {title} {Enhancement and
  quenching of single-molecule fluorescence},}\ }\href {\doibase
  10.1103/PhysRevLett.96.113002} {\bibfield  {journal} {\bibinfo  {journal}
  {Phys. Rev. Lett.}\ }\textbf {\bibinfo {volume} {96}},\ \bibinfo {pages}
  {113002} (\bibinfo {year} {2006})}\BibitemShut {NoStop}%
\bibitem [{\citenamefont {Colom}\ \emph {et~al.}(2018)\citenamefont {Colom},
  \citenamefont {McPhedran}, \citenamefont {Stout},\ and\ \citenamefont
  {Bonod}}]{PhysRevB.98.085418}%
  \BibitemOpen
  \bibfield  {author} {\bibinfo {author} {\bibfnamefont {R\'emi}\ \bibnamefont
  {Colom}}, \bibinfo {author} {\bibfnamefont {Ross}\ \bibnamefont {McPhedran}},
  \bibinfo {author} {\bibfnamefont {Brian}\ \bibnamefont {Stout}}, \ and\
  \bibinfo {author} {\bibfnamefont {Nicolas}\ \bibnamefont {Bonod}},\
  }\bibfield  {title} {\enquote {\bibinfo {title} {Modal expansion of the
  scattered field: Causality, nondivergence, and nonresonant contribution},}\
  }\href {\doibase 10.1103/PhysRevB.98.085418} {\bibfield  {journal} {\bibinfo
  {journal} {Phys. Rev. B}\ }\textbf {\bibinfo {volume} {98}},\ \bibinfo
  {pages} {085418} (\bibinfo {year} {2018})}\BibitemShut {NoStop}%
\bibitem [{\citenamefont {Schelkunoff}(1936)}]{FieldEquiv}%
  \BibitemOpen
  \bibfield  {author} {\bibinfo {author} {\bibfnamefont {S.~A.}\ \bibnamefont
  {Schelkunoff}},\ }\bibfield  {title} {{\selectlanguage {en}\enquote {\bibinfo
  {title} {ome equivalence theorems of electromagnetics and their application
  to radiation problems},}\ }}\href {\doibase
  10.1002/j.1538-7305.1936.tb00720.x} {\bibfield  {journal} {\bibinfo
  {journal} {Bell Labs Technical Journal}\ }\textbf {\bibinfo {volume} {15}},\
  \bibinfo {pages} {92} (\bibinfo {year} {1936})}\BibitemShut {NoStop}%
\bibitem [{\citenamefont {Dung}\ \emph {et~al.}(1998)\citenamefont {Dung},
  \citenamefont {Knöll},\ and\ \citenamefont
  {Welsch}}]{dung_three-dimensional_1998}%
  \BibitemOpen
  \bibfield  {author} {\bibinfo {author} {\bibfnamefont {Ho~Trung}\
  \bibnamefont {Dung}}, \bibinfo {author} {\bibfnamefont {Ludwig}\ \bibnamefont
  {Knöll}}, \ and\ \bibinfo {author} {\bibfnamefont {Dirk-Gunnar}\
  \bibnamefont {Welsch}},\ }\bibfield  {title} {\enquote {\bibinfo {title}
  {Three-dimensional quantization of the electromagnetic field in dispersive
  and absorbing inhomogeneous dielectrics},}\ }\href {\doibase
  10.1103/PhysRevA.57.3931} {\bibfield  {journal} {\bibinfo  {journal} {Phys.
  Rev. A}\ }\textbf {\bibinfo {volume} {57}},\ \bibinfo {pages} {3931--3942}
  (\bibinfo {year} {1998})}\BibitemShut {NoStop}%
\bibitem [{\citenamefont {Suttorp}\ and\ \citenamefont
  {Wonderen}(2004)}]{suttorp_fano_2004}%
  \BibitemOpen
  \bibfield  {author} {\bibinfo {author} {\bibfnamefont {L.~G.}\ \bibnamefont
  {Suttorp}}\ and\ \bibinfo {author} {\bibfnamefont {A.~J.~van}\ \bibnamefont
  {Wonderen}},\ }\bibfield  {title} {\enquote {\bibinfo {title} {Fano
  diagonalization of a polariton model for an inhomogeneous absorptive
  dielectric},}\ }\href {\doibase 10.1209/epl/i2004-10131-8} {\bibfield
  {journal} {\bibinfo  {journal} {{EPL}}\ }\textbf {\bibinfo {volume} {67}},\
  \bibinfo {pages} {766} (\bibinfo {year} {2004})}\BibitemShut {NoStop}%
\bibitem [{\citenamefont {Cirac}(1992)}]{PhysRevA.46.4354}%
  \BibitemOpen
  \bibfield  {author} {\bibinfo {author} {\bibfnamefont {J.~I.}\ \bibnamefont
  {Cirac}},\ }\bibfield  {title} {\enquote {\bibinfo {title} {Interaction of a
  two-level atom with a cavity mode in the bad-cavity limit},}\ }\href
  {\doibase 10.1103/PhysRevA.46.4354} {\bibfield  {journal} {\bibinfo
  {journal} {Phys. Rev. A}\ }\textbf {\bibinfo {volume} {46}},\ \bibinfo
  {pages} {4354--4362} (\bibinfo {year} {1992})}\BibitemShut {NoStop}%
\bibitem [{\citenamefont {{COMSOL Inc.}}()}]{comsol}%
  \BibitemOpen
  \bibfield  {author} {\bibinfo {author} {\bibnamefont {{COMSOL Inc.}}},\
  }\href {www.comsol.com} {\enquote {\bibinfo {title} {Comsol multiphysics v
  5.4},}\ }\bibinfo {howpublished} {\url{www.comsol.com}}\BibitemShut {NoStop}%
\bibitem [{\citenamefont {Sasada}\ \emph {et~al.}(2011)\citenamefont {Sasada},
  \citenamefont {Hatano},\ and\ \citenamefont
  {Ordonez}}]{doi:10.1143/JPSJ.80.104707}%
  \BibitemOpen
  \bibfield  {author} {\bibinfo {author} {\bibfnamefont {Keita}\ \bibnamefont
  {Sasada}}, \bibinfo {author} {\bibfnamefont {Naomichi}\ \bibnamefont
  {Hatano}}, \ and\ \bibinfo {author} {\bibfnamefont {Gonzalo}\ \bibnamefont
  {Ordonez}},\ }\bibfield  {title} {\enquote {\bibinfo {title} {Resonant
  spectrum analysis of the conductance of an open quantum system and three
  types of fano parameter},}\ }\href {\doibase 10.1143/JPSJ.80.104707}
  {\bibfield  {journal} {\bibinfo  {journal} {Journal of the Physical Society
  of Japan}\ }\textbf {\bibinfo {volume} {80}},\ \bibinfo {pages} {104707}
  (\bibinfo {year} {2011})}\BibitemShut {NoStop}%
\bibitem [{\citenamefont {Thakkar}\ \emph {et~al.}(2017)\citenamefont
  {Thakkar}, \citenamefont {Rea}, \citenamefont {Smith}, \citenamefont
  {Heylman}, \citenamefont {Quillin}, \citenamefont {Knapper}, \citenamefont
  {Horak}, \citenamefont {Masiello},\ and\ \citenamefont
  {Goldsmith}}]{thakkar_sculpting_2017}%
  \BibitemOpen
  \bibfield  {author} {\bibinfo {author} {\bibfnamefont {Niket}\ \bibnamefont
  {Thakkar}}, \bibinfo {author} {\bibfnamefont {Morgan~T.}\ \bibnamefont
  {Rea}}, \bibinfo {author} {\bibfnamefont {Kevin~C.}\ \bibnamefont {Smith}},
  \bibinfo {author} {\bibfnamefont {Kevin~D.}\ \bibnamefont {Heylman}},
  \bibinfo {author} {\bibfnamefont {Steven~C.}\ \bibnamefont {Quillin}},
  \bibinfo {author} {\bibfnamefont {Kassandra~A.}\ \bibnamefont {Knapper}},
  \bibinfo {author} {\bibfnamefont {Erik~H.}\ \bibnamefont {Horak}}, \bibinfo
  {author} {\bibfnamefont {David~J.}\ \bibnamefont {Masiello}}, \ and\ \bibinfo
  {author} {\bibfnamefont {Randall~H.}\ \bibnamefont {Goldsmith}},\ }\bibfield
  {title} {\enquote {\bibinfo {title} {Sculpting fano resonances to control
  photonic–plasmonic hybridization},}\ }\href {\doibase
  10.1021/acs.nanolett.7b03332} {\bibfield  {journal} {\bibinfo  {journal}
  {Nano Letters}\ }\textbf {\bibinfo {volume} {17}},\ \bibinfo {pages}
  {6927--6934} (\bibinfo {year} {2017})}\BibitemShut {NoStop}%
\bibitem [{\citenamefont {Barth}\ \emph {et~al.}(2010)\citenamefont {Barth},
  \citenamefont {Schietinger}, \citenamefont {Fischer}, \citenamefont {Becker},
  \citenamefont {Nüsse}, \citenamefont {Aichele}, \citenamefont {Löchel},
  \citenamefont {Sönnichsen},\ and\ \citenamefont
  {Benson}}]{barth_nanoassembled_2010}%
  \BibitemOpen
  \bibfield  {author} {\bibinfo {author} {\bibfnamefont {Michael}\ \bibnamefont
  {Barth}}, \bibinfo {author} {\bibfnamefont {Stefan}\ \bibnamefont
  {Schietinger}}, \bibinfo {author} {\bibfnamefont {Sabine}\ \bibnamefont
  {Fischer}}, \bibinfo {author} {\bibfnamefont {Jan}\ \bibnamefont {Becker}},
  \bibinfo {author} {\bibfnamefont {Nils}\ \bibnamefont {Nüsse}}, \bibinfo
  {author} {\bibfnamefont {Thomas}\ \bibnamefont {Aichele}}, \bibinfo {author}
  {\bibfnamefont {Bernd}\ \bibnamefont {Löchel}}, \bibinfo {author}
  {\bibfnamefont {Carsten}\ \bibnamefont {Sönnichsen}}, \ and\ \bibinfo
  {author} {\bibfnamefont {Oliver}\ \bibnamefont {Benson}},\ }\bibfield
  {title} {\enquote {\bibinfo {title} {Nanoassembled plasmonic-photonic hybrid
  cavity for tailored light-matter coupling},}\ }\href {\doibase
  10.1021/nl903555u} {\bibfield  {journal} {\bibinfo  {journal} {Nano Letters}\
  }\textbf {\bibinfo {volume} {10}},\ \bibinfo {pages} {891--895} (\bibinfo
  {year} {2010})}\BibitemShut {NoStop}%
\bibitem [{\citenamefont {Doeleman}\ \emph {et~al.}(2016)\citenamefont
  {Doeleman}, \citenamefont {Verhagen},\ and\ \citenamefont
  {Koenderink}}]{doeleman_antennacavity_2016}%
  \BibitemOpen
  \bibfield  {author} {\bibinfo {author} {\bibfnamefont {Hugo~M.}\ \bibnamefont
  {Doeleman}}, \bibinfo {author} {\bibfnamefont {Ewold}\ \bibnamefont
  {Verhagen}}, \ and\ \bibinfo {author} {\bibfnamefont {A.~Femius}\
  \bibnamefont {Koenderink}},\ }\bibfield  {title} {\enquote {\bibinfo {title}
  {Antenna–cavity hybrids: Matching polar opposites for purcell enhancements
  at any linewidth},}\ }\href {\doibase 10.1021/acsphotonics.6b00453}
  {\bibfield  {journal} {\bibinfo  {journal} {{ACS} Photonics}\ }\textbf
  {\bibinfo {volume} {3}},\ \bibinfo {pages} {1943--1951} (\bibinfo {year}
  {2016})}\BibitemShut {NoStop}%
\bibitem [{\citenamefont {Palstra}\ \emph {et~al.}(2019)\citenamefont
  {Palstra}, \citenamefont {Doeleman},\ and\ \citenamefont
  {Koenderink}}]{palstra_hybrid_2019}%
  \BibitemOpen
  \bibfield  {author} {\bibinfo {author} {\bibfnamefont {Isabelle~M.}\
  \bibnamefont {Palstra}}, \bibinfo {author} {\bibfnamefont {Hugo~M.}\
  \bibnamefont {Doeleman}}, \ and\ \bibinfo {author} {\bibfnamefont
  {A.~Femius}\ \bibnamefont {Koenderink}},\ }\bibfield  {title} {\enquote
  {\bibinfo {title} {Hybrid cavity-antenna systems for quantum optics outside
  the cryostat?}}\ }\href {\doibase 10.1515/nanoph-2019-0062} {\bibfield
  {journal} {\bibinfo  {journal} {Nanophotonics}\ }\textbf {\bibinfo {volume}
  {8}},\ \bibinfo {pages} {1513--1531} (\bibinfo {year} {2019})}\BibitemShut
  {NoStop}%
\bibitem [{\citenamefont {Dezfouli}\ \emph {et~al.}(2019)\citenamefont
  {Dezfouli}, \citenamefont {Gordon},\ and\ \citenamefont
  {Hughes}}]{dezfouli_molecular_2019}%
  \BibitemOpen
  \bibfield  {author} {\bibinfo {author} {\bibfnamefont {Mohsen~Kamandar}\
  \bibnamefont {Dezfouli}}, \bibinfo {author} {\bibfnamefont {Reuven}\
  \bibnamefont {Gordon}}, \ and\ \bibinfo {author} {\bibfnamefont {Stephen}\
  \bibnamefont {Hughes}},\ }\bibfield  {title} {\enquote {\bibinfo {title}
  {Molecular optomechanics in the anharmonic cavity-{QED} regime using hybrid
  metal-dielectric cavity modes},}\ }\href {\doibase
  10.1021/acsphotonics.8b01091} {\bibfield  {journal} {\bibinfo  {journal}
  {{ACS} {PHOTONICS}}\ }\textbf {\bibinfo {volume} {6}},\ \bibinfo {pages}
  {1400--1408} (\bibinfo {year} {2019})}\BibitemShut {NoStop}%
\end{thebibliography}%

\end{document}